\documentclass[twocolumn,astrosymb,trackchanges,twocolappendix]{aastex631_arxiv}

\begin{document}

\title{COBIPULSE: A Systematic Search for Compact Binary Millisecond Pulsars}

\author[0000-0003-0438-4956]{Marco Turchetta}
\affiliation{Department of Physics, Norwegian University of Science and Technology, NO-7491 Trondheim, Norway}
\author[0000-0002-0237-1636]{Manuel Linares}
\affiliation{Department of Physics, Norwegian University of Science and Technology, NO-7491 Trondheim, Norway}
\affiliation{Departament de F{\'i}sica, EEBE, Universitat Polit{\`e}cnica de Catalunya, Av. Eduard Maristany 16, E-08019 Barcelona, Spain}
\author{Karri Koljonen}
\affiliation{Department of Physics, Norwegian University of Science and Technology, NO-7491 Trondheim, Norway}
\author{Jorge Casares}
\affiliation{Instituto de Astrofísica de Canarias, E-38205 La Laguna, Tenerife, Spain}
\affiliation{Departamento de Astrofísica, Universidad de La Laguna, E-38206 La Laguna, Tenerife, Spain}
\author[0000-0003-2446-8882]{Paulo A. Miles-P\'aez}
\affiliation{Centro de Astrobiolog\'ia, CSIC-INTA, Camino Bajo del Castillo s/n, 28692 Villanueva de la Ca\~nada, Madrid, Spain}
\author[0000-0002-4717-5102]{Pablo Rodríguez-Gil}
\affiliation{Instituto de Astrofísica de Canarias, E-38205 La Laguna, Tenerife, Spain}
\affiliation{Departamento de Astrofísica, Universidad de La Laguna, E-38206 La Laguna, Tenerife, Spain}
\author{Tariq Shahbaz}
\affiliation{Instituto de Astrofísica de Canarias, E-38205 La Laguna, Tenerife, Spain}
\affiliation{Departamento de Astrofísica, Universidad de La Laguna, E-38206 La Laguna, Tenerife, Spain}
\author[0000-0001-6841-0725]{Jordan A. Simpson}
\affiliation{Department of Physics, Norwegian University of Science and Technology, NO-7491 Trondheim, Norway}





\begin{abstract}
We report here the results obtained from a systematic optical photometric survey aimed at finding new compact binary millisecond pulsars (also known as ``spiders"): the COmpact BInary PULsar SEarch (COBIPULSE). We acquired multi-band optical images over one year around $33$ unidentified \textit{Fermi}-LAT sources, selected as pulsar candidates based on their curved GeV spectra and steady $\gamma$-ray emission. We present the discovery of four optical variables coinciding with the Fermi sources 3FGL~J0737.2$-$3233, 3FGL~J2117.6$+$3725 (two systems in this field) and 3FGL~J2221.6$+$6507, which we propose as new candidate spider systems. Indeed, they all show optical flux modulation consistent with orbital periods of $0.3548(5) \ \mathrm{d}$, $0.25328(6) \ \mathrm{d}$, $0.441961(2) \ \mathrm{d}$, and $0.165(4) \ \mathrm{d}$, respectively, with amplitudes $\gtrsim 0.3 \ \mathrm{mag}$ and colors compatible with companion star temperatures of $5000$--$6000 \ \mathrm{K}$. These properties are consistent with the ``redback" sub-class of spider pulsars. If confirmed as a millisecond pulsar, 3FGL~J0737.2$-$3233 will be the closest known spider to Earth ($D=659_{-20}^{+16} \ \mathrm{pc}$, from \textit{Gaia}-DR3 parallax).
We searched and did not find any X-ray sources matching our four candidates, placing $3\sigma$ upper limits of $\sim10^{31}$--$10^{32} \ \mathrm{erg} \ \mathrm{s}^{-1}$ ($0.3$--$10 \ \mathrm{keV}$) on their soft X-ray luminosities. We also present and discuss other multi-wavelength information on our spider candidates, from infrared to X-rays.

\end{abstract}



\section{Introduction} \label{sec:intro}
Millisecond pulsars (MSPs) are ``recycled" neutron stars with spin periods $\simeq 1$--$30 \ \mathrm{ms}$, which they reach after a Gyr-long phase of accretion from a low-mass companion star \citep{1982CSci...51.1096R, 1991PhR...203....1B}. This low-mass X-ray binary (LMXB) stage persists until the mass transfer rate decreases and allows the activation of a radio MSP, powered by its rotational energy \citep{2013A&A...558A..39T}.

About $20$\% of the currently known radio MSPs are found in compact binary systems (according to the ATNF Pulsar Catalog\footnote{\url{https://www.atnf.csiro.au/research/pulsar/psrcat/}}), with orbital periods $\lesssim1 \ \mathrm{d}$.
Given such tight orbits, the wind of relativistic particles 
powered by the loss of rotational energy of the pulsar can strongly irradiate and strip matter off the companion star. This gives a proposed formation channel for isolated MSPs, if the pulsar has enough time to fully consume its companion \citep{1988Natur.334..227V}. Such destructive effect also inspired cannibalistic arachnid nicknames for these systems: \textit{redback} (RB) and \textit{black widow} (BW) MSPs, with low-mass non-degenerate companions
($M_{2}\simeq 0.3$--$0.7 \ \mathrm{M}_{\sun}$) and very low-mass semi-degenerate companions
($M_{2}\sim 0.01 \ \mathrm{M}_{\sun}$)
for RBs and BWs, respectively
\citep{1988Natur.333..237F, roberts2012surrounded}. Both sub-types are collectively known as \textit{spiders}.

The recycling scenario for MSP formation outlined above was confirmed by the discovery of
accreting MSPs within LMXBs \citep{Wijnands98}, as well as three particular RBs known as transitional MSPs \citep{2009Sci...324.1411A,2013Natur.501..517P,2014MNRAS.441.1825B}. The latter change between the disk/LMXB and radio pulsar states on timescales of a few weeks, suggesting RBs to be the evolutionary link between these two phases. Generally, spiders provide a unique opportunity to probe the intrabinary shock between pulsar and companion winds \citep{2018ApJ...869..120W}, where charged particles emit synchrotron radiation which dominates the X-ray band \citep{2014ApJ...783...69G,2014ApJ...795...72L}. Furthermore, these systems represent promising environments to find massive neutron stars \citep{2020mbhe.confE..23L}, as they underwent a long phase of mass accretion to reach millisecond spin periods.

Until 2008, spiders were rarely detected in blind radio surveys\footnote{\url{https://www.astro.umd.edu/~eferrara/pulsars/GalacticMSPs.txt}} due to unknown orbital acceleration and eclipses of the pulsations occurring during a large fraction of the orbit.
The radio pulses are absorbed and dispersed by the material stripped off the companion star \citep{2001ApJ...561L..89D}.
Since its launch in 2008, the \textit{Fermi Large Area Telescope} (\textit{Fermi}-LAT) has dramatically boosted the discovery of MSPs in general, and spiders in particular. The third \textit{Fermi}-LAT pulsar catalog reports at least 119 new MSPs, detected either as $\gamma$-ray or radio MSPs via radio searches targeting unidentified \textit{Fermi}-LAT sources \citep{2023ApJ...958..191S}. Among these, 57 spiders have been discovered in the Galactic field, while 17 more systems with no detected pulsations have been classified as spider candidates based on their optical light curve shape and multi-wavelength phenomenology \citep{nedreaas2024spidercat}.

Spiders can also be discovered through their variable optical counterparts, dominated by the companion flux and showing orbital modulation \citep[e.g.][]{2017MNRAS.465.4602L,swihart2021discovery,2023ApJ...943..103A}. Optical light curves of all currently known BWs exhibit a single bright maximum per orbit with sharp minima and $\approx 1 \ \mathrm{mag}$-large peak-to-peak amplitudes, due to the pulsar irradiating flux overwhelming the intrinsic emission of the cold companion, with typical base temperatures of $T_\mathrm{b}\simeq1000$--$3000 \ \mathrm{K}$. RB companions, on the other hand, are more massive and hotter than BW companions, with $T_\mathrm{b}\simeq4000$--$6000 \ \mathrm{K}$. About half of the known RBs show little or no signs of irradiation and ellipsoidal modulation produced by the visibility of the tidally distorted companion star, with sinusoidal optical light curves showing two maxima per orbit \citep{2023MNRAS.525.2565T}. Such systems exhibit smaller peak-to-peak amplitudes $\simeq 0.3 \ \mathrm{mag}$ with respect to spiders in the irradiation regime \citep{2024MNRAS.528.4337D}.


Despite ongoing efforts, the latest \textit{Fermi}-LAT catalog \citep[4FGL-DR4,][]{2023arXiv230712546B} lists more than $2100$ unassociated sources. Many of these might be spiders still ``hidden" in our Galaxy: based on population synthesis models \citep{2018ApJ...863..199G} we predict the detection of at least $\approx70$ new MSPs in the next five years.
In this context, searching for spiders through optical observations of unassociated \textit{Fermi}-LAT sources constitutes a promising strategy. Indeed, there are many cases of spiders detected as MSPs only years after the discovery of their optical variable counterpart, by pointing radio facilities towards their precise sky location (see e.g. \citealt{2015ApJ...812L..24R} and \citealt{2021MNRAS.502..915C} for PSR~J2039$-$5617, \citealt{2017MNRAS.465.4602L} and \citealt{Perez_2023} for PSR~J0212$+$5321, \citealt{2023ApJ...943..103A} and \citealt{10.1093/mnras/stae211} for PSR~J1910$-$5320).

In this paper we present the COmpact BInary PULsar SEarch (COBIPULSE), a systematic optical photometric survey targeting $33$ unidentified $\gamma$-ray sources selected as promising pulsar candidates from the \textit{Fermi}-LAT 3FGL catalog \citep{2015ApJS..218...23A}. In Section \ref{sec:data} we describe our selection criteria for the \textit{Fermi} fields of view, data acquisition, and how we performed photometry and periodicity searches for every source detected in the field. In Section \ref{sec:results} we report the results of our survey, with periodograms and folded light curves of the optical variables classified as spider candidates. In Section \ref{sec:discussion} we discuss separately each discovered candidate and search for radio, IR and X-ray counterparts to our optical candidates from known catalogs. We summarize our main findings in Section \ref{sec:conclusions}.

\section{Observations and data analysis} \label{sec:data}
\begin{figure}[t!]
\centering
\includegraphics[width=\columnwidth]{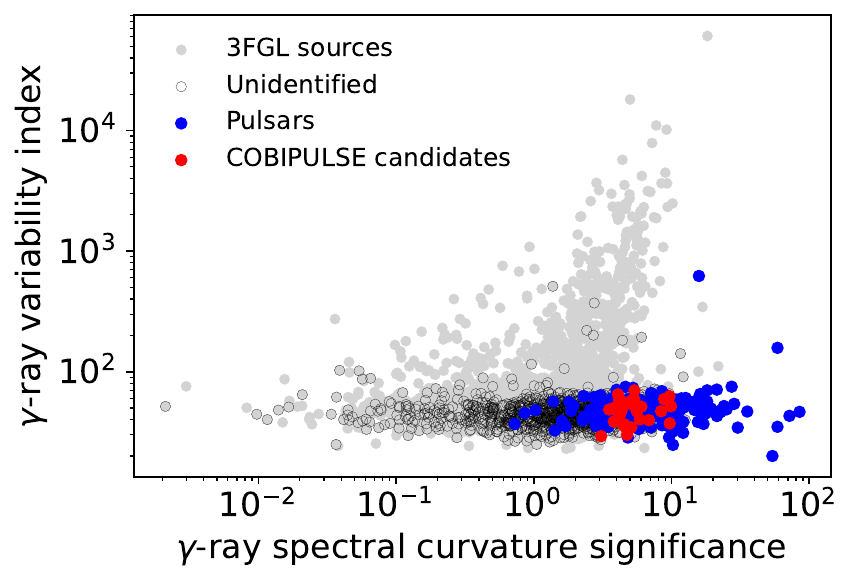}
\caption{$\gamma$-ray variability versus $\gamma$-ray curvature significance plot, where all \textit{Fermi}-LAT 3FGL sources are shown as gray filled circles, unidentified objects as black empty circles, identified pulsars as blue filled circles, and COBIPULSE candidates as red filled circles.}
\label{fig:varvscurv}
\end{figure}
\begin{table*}
\raggedright
\scriptsize
    \caption{Log of COBIPULSE observations.}
    \setlength{\tabcolsep}{7.0pt}
    \begin{tabular}{lccccccc}
    \hline\hline
        Field & Telescope & Instrument & Date & Time & Images in \textit{g'} & Images in \textit{r'} & Images in \textit{i'}\\
        3FGL source & (diameter) & (configuration) & (evening) & (UT) & ($\mathrm{nr}\times\mathrm{exp.}\ \mathrm{time}$) & ($\mathrm{nr}\times\mathrm{exp.}\ \mathrm{time}$) & ($\mathrm{nr}\times\mathrm{exp.}\ \mathrm{time}$)\\ 
        \hline
        {J0238.0$+$5237}& STELLA-1.2m & WiFSIP-1x1 & 2015-09-04 & 01:22$-$04:57 & $9\times300 \ \mathrm{s}$ & $18\times300 \ \mathrm{s}$ & $9\times300 \ \mathrm{s}$\\
        {J0312.1$-$0921}& STELLA-1.2m & WiFSIP-1x1 & 2015-11-14 & 00:01$-$03:33 & $7\times300 \ \mathrm{s}$ & $17\times300 \ \mathrm{s}$ & $7\times300 \ \mathrm{s}$\\
        {J0318.1$+$0252}& STELLA-1.2m & WiFSIP-1x1 & 2015-11-11 & 21:36$-$23:32 & $4\times300 \ \mathrm{s}$ & $9\times300 \ \mathrm{s}$ & $4\times300 \ \mathrm{s}$\\
        {J0336.1$+$7500}& STELLA-1.2m & WiFSIP-1x1 & 2015-11-07 & 21:12$-$23:09 & $6\times300 \ \mathrm{s}$ & $12\times300 \ \mathrm{s}$ & $5\times300 \ \mathrm{s}$\\
        {}& {} & {} & 2015-11-25 & 20:18$-$23:56 & $6\times300 \ \mathrm{s}$ & $11\times300 \ \mathrm{s}$ & $8\times300 \ \mathrm{s}$\\
        {J0545.6$+$6019}& STELLA-1.2m & WiFSIP-1x1 & 2015-09-17 & 02:11$-$05:46 & $9\times300 \ \mathrm{s}$ & $18\times300 \ \mathrm{s}$ & $9\times300 \ \mathrm{s}$\\
        {J0758.6$-$1451}& STELLA-1.2m & WiFSIP-1x1 & 2015-11-28 & 01:56$-$05:33 & $9\times300 \ \mathrm{s}$ & $18\times300 \ \mathrm{s}$ & $9\times300 \ \mathrm{s}$\\
        {}& {} & {} & 2016-01-16 & 00:44$-$04:17 & $8\times300 \ \mathrm{s}$ & $17\times300 \ \mathrm{s}$ & $8\times300 \ \mathrm{s}$\\
        {J0953.7$-$1510}& STELLA-1.2m & WiFSIP-1x1 & 2015-12-19 & 02:37$-$04:07 & $4\times300 \ \mathrm{s}$ & $9\times300 \ \mathrm{s}$ & $4\times300 \ \mathrm{s}$\\
        {}& {} & {} & 2016-01-20 & 03:13$-$04:17 & $4\times300 \ \mathrm{s}$ & $7\times300 \ \mathrm{s}$ & $3\times300 \ \mathrm{s}$\\
        {J1120.6$+$0713}& STELLA-1.2m & WiFSIP-1x1 & 2015-12-21 & 02:47$-$06:25 & $9\times300 \ \mathrm{s}$ & $18\times300 \ \mathrm{s}$ & $8\times300 \ \mathrm{s}$\\
        {}& {} & {} & 2016-01-22 & 00:43$-$04:23 & $9\times300 \ \mathrm{s}$ & $18\times300 \ \mathrm{s}$ & $9\times300 \ \mathrm{s}$\\
        {J1543.5$-$0244}& STELLA-1.2m & WiFSIP-1x1 & 2016-03-01 & 02:49$-$06:30 & $8\times300 \ \mathrm{s}$ & $17\times300 \ \mathrm{s}$ & $8\times300 \ \mathrm{s}$\\
        {}& {} & {} & 2016-03-17 & 03:04$-$05:29 & $4\times300 \ \mathrm{s}$ & $7\times300 \ \mathrm{s}$ & $8\times300 \ \mathrm{s}$\\
        {}& {} & {} & 2016-04-01 & 01:46$-$05:31 & $8\times300 \ \mathrm{s}$ & $18\times300 \ \mathrm{s}$ & $9\times300 \ \mathrm{s}$\\
        {J1601.9$+$2306}& STELLA-1.2m & WiFSIP-1x1 & 2015-08-26 & 21:04$-$00:39 & $9\times300 \ \mathrm{s}$ & $18\times300 \ \mathrm{s}$ & $9\times300 \ \mathrm{s}$\\
        {}& {} & {} & 2016-02-18 & 02:59$-$04:31 & $5\times300 \ \mathrm{s}$ & $10\times300 \ \mathrm{s}$ & $4\times300 \ \mathrm{s}$\\
        {}& {} & {} & 2016-03-06 & 02:11$-$03:18 & $4\times300 \ \mathrm{s}$ & $7\times300 \ \mathrm{s}$ & $3\times300 \ \mathrm{s}$\\
        {J1625.1$-$0021}& STELLA-1.2m & WiFSIP-1x1 & 2016-03-29 & 02:05$-$05:49 & $9\times300 \ \mathrm{s}$ & $16\times300 \ \mathrm{s}$ & $9\times300 \ \mathrm{s}$\\
        {}& {} & {} & 2016-04-19 & 00:08$-$03:53 & $9\times300 \ \mathrm{s}$ & $18\times300 \ \mathrm{s}$ & $9\times300 \ \mathrm{s}$\\
        {J1730.6$-$0357}& STELLA-1.2m & WiFSIP-1x1 & 2016-04-10 & 02:00$-$05:44 & $8\times300 \ \mathrm{s}$ & $18\times300 \ \mathrm{s}$ & $9\times300 \ \mathrm{s}$\\
        {}& {} & {} & 2016-04-26 & 01:22$-$05:02 & $9\times300 \ \mathrm{s}$ & $18\times300 \ \mathrm{s}$ & $9\times300 \ \mathrm{s}$\\
        {J2103.7$-$1113}& STELLA-1.2m & WiFSIP-1x1 & 2015-11-14 & 21:23$-$22:50 & $4\times300 \ \mathrm{s}$ & $8\times300 \ \mathrm{s}$ & $3\times300 \ \mathrm{s}$\\
        {J2212.5$+$0703}& STELLA-1.2m & WiFSIP-1x1 & 2015-08-06 & 01:16$-$02:20 & $4\times300 \ \mathrm{s}$ & $6\times300 \ \mathrm{s}$ & $3\times300 \ \mathrm{s}$\\
        {}& {} & {} & 2015-08-12 & 00:36$-$04:11 & $9\times300 \ \mathrm{s}$ & $17\times300 \ \mathrm{s}$ & $9\times300 \ \mathrm{s}$\\
        {J2221.6$+$6507}& STELLA-1.2m & WiFSIP-1x1 & 2015-11-01 & 21:31$-$01:02 & $9\times300 \ \mathrm{s}$ & $18\times300 \ \mathrm{s}$ & $9\times300 \ \mathrm{s}$\\
        {}& {} & {} & 2015-11-17 & 20:37$-$00:10 & $9\times300 \ \mathrm{s}$ & $18\times300 \ \mathrm{s}$ & $9\times300 \ \mathrm{s}$\\
        {J2233.1$+$6542}& STELLA-1.2m & WiFSIP-1x1 & 2015-09-05 & 01:48$-$05:24 & $9\times300 \ \mathrm{s}$ & $18\times300 \ \mathrm{s}$ & $9\times300 \ \mathrm{s}$\\
        {J2310.1$-$0557}& STELLA-1.2m & WiFSIP-1x1 & 2015-10-11 & 02:22$-$03:51 & $4\times300 \ \mathrm{s}$ & $9\times300 \ \mathrm{s}$ & $4\times300 \ \mathrm{s}$\\
        {}& {} & {} & 2015-11-10 & 21:39$-$00:26 & $7\times300 \ \mathrm{s}$ & $12\times300 \ \mathrm{s}$ & $6\times300 \ \mathrm{s}$\\
        {J1225.9$+$2953}& INT-2.5m & WFC-1x1 & 2015-06-05 & 23:07$-$00:35 & $12\times60 \ \mathrm{s}$ & $24\times60 \ \mathrm{s}$ & $12\times60 \ \mathrm{s}$\\
        {J1630.2$-$1052}& INT-2.5m & WFC-1x1 & 2015-06-03 & 01:00$-$02:42 & $7\times180 \ \mathrm{s}$ & $14\times180 \ \mathrm{s}$ & $7\times180 \ \mathrm{s}$\\
        {J1827.7$+$1141}& INT-2.5m & WFC-1x1 & 2015-06-03 & 03:00$-$04:51 & $12\times180 \ \mathrm{s}$ & $22\times180 \ \mathrm{s}$ & $12\times180 \ \mathrm{s}$\\
        {J1829.2$+$3229}& INT-2.5m & WFC-1x1 & 2015-06-04 & 00:41$-$02:32 & $12\times60 \ \mathrm{s}$ & $29\times60 \ \mathrm{s}$ & $15\times60 \ \mathrm{s}$\\
        {J1950.2$+$1215}& INT-2.5m & WFC-1x1 & 2015-06-05 & 03:31$-$04:36 & $9\times60 \ \mathrm{s}$ & $18\times60 \ \mathrm{s}$ & $9\times60 \ \mathrm{s}$\\
        {J2117.6$+$3725}& INT-2.5m & WFC-1x1 & 2015-06-04 & 02:58$-$05:20 & $19\times60 \ \mathrm{s}$ & $38\times60 \ \mathrm{s}$ & $19\times60 \ \mathrm{s}$\\
        {J0514.6$-$4406}& LCO-1m & Sinistro-1x1 & 2015-11-02 & 05:11$-$08:31 & $9\times300 \ \mathrm{s}$ & $18\times300 \ \mathrm{s}$ & $9\times300 \ \mathrm{s}$\\
        {}& {} & {} & 2015-11-04 & 04:00$-$07:20 & $9\times300 \ \mathrm{s}$ & $18\times300 \ \mathrm{s}$ & $8\times300 \ \mathrm{s}$\\
        {}& {} & {} & 2015-11-05 & 05:12$-$08:32 & $9\times300 \ \mathrm{s}$ & $18\times300 \ \mathrm{s}$ & $9\times300 \ \mathrm{s}$\\
        {}& {} & {} & 2015-11-06 & 05:09$-$05:43 & $3\times300 \ \mathrm{s}$ & $5\times300 \ \mathrm{s}$ & $2\times300 \ \mathrm{s}$\\
        {}& {} & {} & 2015-11-07 & 04:50$-$08:10 & $9\times300 \ \mathrm{s}$ & $18\times300 \ \mathrm{s}$ & $9\times300 \ \mathrm{s}$\\
        {}& {} & {} & 2015-11-20 & 04:25$-$07:45 & $9\times300 \ \mathrm{s}$ & $18\times300 \ \mathrm{s}$ & $8\times300 \ \mathrm{s}$\\
        {J0737.2$-$3233}& LCO-1m & Sinistro-1x1 & 2015-11-09 & 05:14$-$08:34 & $9\times300 \ \mathrm{s}$ & $18\times300 \ \mathrm{s}$ & $9\times300 \ \mathrm{s}$\\
        {}& {} & {} & 2015-11-12 & 05:09$-$08:29 & $9\times300 \ \mathrm{s}$ & $18\times300 \ \mathrm{s}$ & $9\times300 \ \mathrm{s}$\\
        {}& {} & {} & 2015-11-21 & 03:47$-$07:07 & $9\times300 \ \mathrm{s}$ & $18\times300 \ \mathrm{s}$ & $9\times300 \ \mathrm{s}$\\
        {J0933.9$-$6232}& LCO-1m & Sinistro-1x1 & 2015-12-04 & 04:53$-$08:13 & $9\times300 \ \mathrm{s}$ & $17\times300 \ \mathrm{s}$ & $9\times300 \ \mathrm{s}$\\
        {}& {} & {} & 2015-12-08 & 04:55$-$06:43 & $6\times300 \ \mathrm{s}$ & $11\times300 \ \mathrm{s}$ & $5\times300 \ \mathrm{s}$\\
        {}& {} & {} & 2015-12-26 & 03:51$-$06:47 & $8\times300 \ \mathrm{s}$ & $18\times300 \ \mathrm{s}$ & $9\times300 \ \mathrm{s}$\\
        {J1035.7$-$6720}& LCO-1m & Sinistro-1x1 & 2015-12-19 & 05:06$-$08:26 & $9\times300 \ \mathrm{s}$ & $18\times300 \ \mathrm{s}$ & $9\times300 \ \mathrm{s}$\\
        {}& {} & {} & 2015-12-23 & 04:39$-$07:36 & $9\times300 \ \mathrm{s}$ & $18\times300 \ \mathrm{s}$ & $8\times300 \ \mathrm{s}$\\
        {}& {} & {} & 2015-12-27 & 04:50$-$08:10 & $8\times300 \ \mathrm{s}$ & $17\times300 \ \mathrm{s}$ & $8\times300 \ \mathrm{s}$\\
        {}& {} & {} & 2016-01-08 & 03:45$-$06:41 & $9\times300 \ \mathrm{s}$ & $18\times300 \ \mathrm{s}$ & $8\times300 \ \mathrm{s}$\\
        {J1119.9$-$2204}& LCO-1m & Sinistro-1x1 & 2016-02-19 & 02:01$-$05:21 & $9\times300 \ \mathrm{s}$ & $18\times300 \ \mathrm{s}$ & $9\times300 \ \mathrm{s}$\\
        {}& {} & {} & 2016-03-28 & 23:54$-$02:50 & $9\times300 \ \mathrm{s}$ & $17\times300 \ \mathrm{s}$ & $8\times300 \ \mathrm{s}$\\
        {J1231.6$-$5113}& LCO-1m & Sinistro-1x1 & 2016-02-19 & 06:00$-$09:20 & $9\times300 \ \mathrm{s}$ & $18\times300 \ \mathrm{s}$ & $9\times300 \ \mathrm{s}$\\
        {}& {} & {} & 2016-02-21 & 04:09$-$07:29 & $9\times300 \ \mathrm{s}$ & $18\times300 \ \mathrm{s}$ & $9\times300 \ \mathrm{s}$\\
        {J1744.1$-$7619}& LCO-0.4m & SBIG-2x2 & 2016-06-13 & 09:46$-$12:09 & $11\times150 \ \mathrm{s}$ & $17\times150 \ \mathrm{s}$ & $12\times150 \ \mathrm{s}$\\
        {J1757.7$-$6030}& LCO-0.4m & SBIG-2x2 & 2016-06-12 & 10:24$-$12:35 & $11\times150 \ \mathrm{s}$ & $19\times150 \ \mathrm{s}$ & $11\times150 \ \mathrm{s}$\\
        {J1946.4$-$5403}& LCO-0.4m & SBIG-2x2 & 2016-06-07 & 12:18$-$13:50 & $2\times150 \ \mathrm{s}$ & $3\times150 \ \mathrm{s}$ & $1\times150 \ \mathrm{s}$\\
        {}& {} & {} & 2016-06-12 & 16:13$-$19:41 & $8\times150 \ \mathrm{s}$ & $8\times150 \ \mathrm{s}$ & $4\times150 \ \mathrm{s}$\\
        {J2039.9$-$5618}& LCO-0.4m & SBIG-2x2 & 2016-07-28 & 14:31$-$17:57 & $17\times150 \ \mathrm{s}$ & $34\times150 \ \mathrm{s}$ & $17\times150 \ \mathrm{s}$\\
        {}& {} & {} & 2016-08-06 & 11:38$-$14:58 & $13\times150 \ \mathrm{s}$ & $24\times150 \ \mathrm{s}$ & $12\times150 \ \mathrm{s}$\\
        \hline
    \end{tabular}
    \label{tab:observationlog}
\end{table*}
\subsection{Pulsar candidates from the \textit{Fermi}-3FGL catalog} \label{subsec:gammaselection}
We built our sample of candidates for COBIPULSE based on the two main properties shown by pulsars in the $\gamma$-ray \textit{Fermi}-LAT energy band: they are steady $\gamma$-ray emitters with curved spectra in the 0.1--300 GeV energy range \citep{Ackermann_2012}. These were quantified in the 3FGL catalog by the variability index and the spectral curvature significance, respectively (for details see Eqs. (3) and (4) in \citealt{2012ApJS..199...31N}). We therefore selected, among the $1010$ 3FGL unidentified sources available in 2015, pulsar candidates that fulfilled all the following criteria\footnote{The selection and optical observations of these \textit{Fermi} fields were performed in 2015, thus we used the 3FGL catalog \citep{2015ApJS..218...23A} back then instead of the latest 4FGL-DR4 \citep{2023arXiv230712546B}.}:
\begin{enumerate}
\item They have spectral curvature significance higher than $3\sigma$.
\item Their variability index is lower than $100$.
\item They are located at Galactic latitudes $|b|>5^{\circ}$ (to avoid contamination from the diffuse $\gamma$-ray emission close to the Galactic plane).
\item The semi-major axis of the \textit{Fermi} $95\%$ error ellipse is $<30'$ (to fit them in the STELLA, INT and LCO fields of view; see Section \ref{subsec:optphotometry}).
\end{enumerate}
This led us to the $33$ COBIPULSE candidates selected for this study, listed in Table \ref{tab:observationlog}. 
As we can see from Figure \ref{fig:varvscurv}, COBIPULSE candidates (shown as red circles) are all located in the bottom right region of the $\gamma$-ray variability-curvature diagram,  along with the majority of the known pulsar population (reported in blue).

\subsection{Observations and optical photometry} \label{subsec:optphotometry}
We acquired optical images of the $33$ selected fields with the $1.2$-$\mathrm{m}$ STELLA, $2.5$-$\mathrm{m}$ INT, $1$-$\mathrm{m}$ LCO, and $0.4$-$\mathrm{m}$ LCO telescopes, using the WiFSIP, WFC, Sinistro, and SBIG cameras mounted on each of them, respectively. The observations were taken over 14 months, from 2015 June 3 to 2016 August 6, alternating the SDSS \textit{g'}, \textit{r'} and \textit{i'} optical filters across the nights. We chose to cycle the exposures through the sequence \textit{g'}-\textit{r'}-\textit{i'}-\textit{r'} to improve the sampling in \textit{r'} and investigate the source variability in this band, while obtaining color information from \textit{g'} and \textit{i'}. We report in Table \ref{tab:observationlog} details of the instrumental setups used for our observations.

We processed STELLA/WiFSIP, LCO/Sinistro, and LCO/SBIG data using their respective data-processing pipelines, which included tasks such as bad-pixel masking, bias subtraction and flat-field correction (see \citealt{2013PASP..125.1031B}, \citealt{2016SPIE.9910E..0NW},  and \citealt{2018SPIE10707E..0KM} for more details). Concerning the INT data, we performed data reduction separately for each of the four CCDs in the WFC mosaic camera using standard routines in \textsc{IRAF}\footnote{\url{https://iraf-community.github.io/}}. For INT/WFC, we performed our search only with CCD n.4 data, which contains the entire $95\%$ error ellipse for each of the 6 \textit{Fermi}-3FGL fields observed with this telescope.

We combined the images acquired from a given field into a single median image for each of the optical bands, \textit{g'}, \textit{r'} and \textit{i'}, to obtain a deeper source detection sensitivity (see the \textit{r'}-band fields in Appendix \ref{sec:appA}, Figures \ref{fig:STELLAFoVs}-\ref{fig:LCOFoVs_cont1}). We identified all the sources in the field having a signal-to-noise ratio $\geq2$ in the combined frames using the \textsc{SEP}\footnote{\url{https://github.com/kbarbary/sep}} package \citep{2016zndo....159035B}, based on the \textsc{SExtractor} software \citep{1996A&AS..117..393B}. Using this detection threshold, our survey is sensitive down to magnitudes $\textit{r'}\simeq19$, $21$, $21$ and $22$ for $0.4$-$\mathrm{m}$ LCO, $1$-$\mathrm{m}$ LCO, $1.2$-$\mathrm{m}$ STELLA, and $2.5$-$\mathrm{m}$ INT, respectively. This allows us to detect potential redback systems, which typically cover the magnitude range $\textit{r'}\simeq14-20$. However, we stress that the faintest sources identified in the median images might be lost or barely detected in some of the individual frames, affecting the extraction of their optical light curves. We detected between about 200 and 22000 sources, depending on each field and band (Table \ref{tab:results}).

Next, we performed circular aperture photometry systematically using \textsc{SEP} on all the identified stars. We experimented with different aperture radii, and concluded that $1.2\times$ the average full-width at half maximum (FWHM) is optimal to minimize sky noise contribution and photometry contamination between nearby stars in the most crowded fields. We used three different sets of comparison stars (one for each optical filter \textit{g'}, \textit{r'} and \textit{i'}), selected as the most stable sources in the field using the \textsc{astrosource}\footnote{\url{https://github.com/zemogle/astrosource}} package \citep{2021JOSS....6.2641F}, all showing variability with rms amplitudes of $\simeq0.001\mathrm{-}0.01\ \mathrm{mag}$\footnote{By summing together the counts measured from $N$ reference stars, we improve the signal-to-noise ratio on the apparent magnitudes of our targets by a factor of $\sqrt{N}$ \citep{1992PASP..104..435H}.}. We also restricted the selection of comparison stars to sources in the magnitude range $\simeq14\mathrm{-}16$, with brightness similar to or greater than our targets of interest (the brightest known spider PSR~J0212$+$5321 has $\textit{r'}\simeq14.3$, \citealt{2017MNRAS.465.4602L}), and far from the saturation regime of the detectors. 
\newpage
\subsection{Variable selection and periodicity search} \label{subsec:variablesandperiods}
We computed the average differential magnitude $\Delta m$\footnote{The differential magnitude is defined as the difference between the apparent magnitudes of target and comparison stars, determined from the ratio of target to comparison measured counts.} and its standard deviation $\sigma$ for each source detected in the \textit{r'}-band, where the data sampling is best (twice the \textit{g'} and \textit{i'}-band cadence, Sec. \ref{subsec:optphotometry}). The corresponding $\sigma$-$\Delta m$ plots are shown in Appendix \ref{sec:appB} (Figures \ref{fig:sigmavsdm}-\ref{fig:sigmavsdm_cont5}). For each COBIPULSE field we considered all the sources fulfilling the criteria described below for further study:
\begin{enumerate}
    \item They lie inside a square centered on the \textit{Fermi}-3FGL coordinates, with sides twice the semi-major axis of the 95\% error ellipse.
    \item Their magnitude standard deviation $\sigma$ is larger than the median value of $\sigma$ computed over the sources lying in the same differential magnitude bin $\Delta m$ of width $0.1 \ \mathrm{mag}$. Since sources showing peak-to-peak amplitudes $\geq0.1 \ \mathrm{mag}$ typically corresponded to $\sigma\geq2\times$ the median, we chose $1\times$ the median as a conservative $\sigma$-threshold to not miss any variable. When we have less than $5$ sources detected in the corresponding differential magnitude bin, we skip this filtering step and use only criterion n.1, as the statistics are too poor to establish whether a source is variable or not.
\end{enumerate}
The numbers for these sources of interest are reported in the $5^{\mathrm{th}}$ column of Table \ref{tab:results}, labeled as ``photometric variables". We detected between 30 and 1700 photometric variables per field.

We then performed Lomb-Scargle (LS; \citealt{1976Ap&SS..39..447L}; \citealt{1982ApJ...263..835S}) and phase-dispersion minimization periodicity searches (PDM; \citealt{1978ApJ...224..953S}) systematically on the \textit{r'}-band light curve of each source selected in the above step. The inspected grid of periods spans the range $0.02\mathrm{-}2.5 \ \mathrm{d}$, to accommodate typical spider orbital periods (see Table 1 in \citealt{2021JCAP...02..030L}), using a period resolution of $\simeq2 \ \mathrm{min}$. For each case, we folded the \textit{r'}-band light curves on the periods corresponding to the maximum and minimum of the LS and PDM periodograms, respectively. 
After checking for possible spurious variability introduced by stars close to the target, we considered all the sources showing sinusoid-like folded light curves, sharp flux minima\footnote{Optical light curves of spiders typically have a sinusoidal shape in the ellipsoidal regime and sharp minima with larger amplitudes in the irradiated regime (see Section \ref{sec:intro}).} or monotonically varying light curves in the \textit{r'}-band. Also, the \textit{g'} and \textit{i'}-band light curves have been folded with the best photometric period $P_{\mathrm{best}}$, that we chose from the method (LS or PDM) yielding the smoothest light curve shape. We classified as ``smooth variables" all the sources investigated above showing the same trend consistently in their \textit{g'}, \textit{r'} and \textit{i'}-band light curves, finding between 0 and 5 smooth variables per field (numbers reported in the $6^\mathrm{th}$ column of Table \ref{tab:results}). The latter selection step excludes undesired variables from our search, like spurious/artificial variables, flares and planetary transits.

We evaluated the significance of the detected photometric periods by using a Fisher randomization test \citep{1985AJ.....90.2317L}. This method consists in permuting the original sequence of magnitudes $m_{i}$ measured at the times $t_{i}$ to assign them a different time $t_{j}$, computing then a new periodogram on this shuffled time series. We repeated such process for $N\sim10000$ times for all our smooth variables. Then we considered the number of permutations whose LS/PDM periodogram contained a maximum/minimum which is larger/smaller than the power we found from the unrandomized dataset at any period. This represents the probability that no periodic component is present in the data, known as false-alarm-probability (FAP), with the detection confidence level corresponding to $1-\mathrm{FAP}$.

To estimate the uncertainty in the photometric periods, we generated mock light curves by introducing artificial Gaussian noise, where the magnitude $m'_{i}$ at the $i$-th point was drawn from a Gaussian distribution with mean value $m_{i}$ and standard deviation $\sigma_{m_{i}}$, with $m_{i}$ and $\sigma_{m_{i}}$ indicating the magnitude at the $i$-th point of the original light curve and its measurement error, respectively. We performed the same periodicity search as described above for the mock light curves and recorded the periods corresponding to the maximum/minimum values of their LS/PDM periodograms. We repeated such procedure for $N\sim1000$ different variations of the original light curve, finding $P_{\mathrm{best},i}$ for each one of them. We assumed the standard deviation $\sigma_{P_{\mathrm{best}}}$ of $P_{\mathrm{best},i}$ as the error on the photometric period estimated from the original data.

\subsection{Supplementary data from the ZTF optical survey} \label{subsec:ZTFdata}
To expand the light curves of the northern sky sources we deemed as ``smooth variables" in the previous section, we used the Zwicky Transient Facility (ZTF) survey data from the latest public release \citep[2018 March--2023 November,][]{2019PASP..131a8002B}. For 13 of our fields which are covered by ZTF (declination $\delta>-31^{\circ}$), we downloaded data for the \textit{g'} and \textit{r'} bands from the IRSA light curve service\footnote{\url{https://irsa.ipac.caltech.edu/docs/program_interface/ztf_lightcurve_api.html}}, extracted using circular apertures of radius $0.5\times$FWHM of their stellar point-spread function (PSF) \citep{2019PASP..131a8002B}. Then, we applied the same periodicity search methods explained in Section \ref{subsec:variablesandperiods}.

The much longer time span of ZTF with respect to the COBIPULSE survey allowed us to confirm or refute as variables the northern sources showing monotonic variability or poor data sampling in our light curves. Furthermore, this analysis was useful to check and possibly increase the accuracy of the photometric periods when full phase coverage was missing in our data (see Section \ref{subsec:J2117res}). Among the northern sources previously referred to as ``smooth variables", we thus classified ``periodic variables" as those displaying a periodic pattern in the ZTF light curves consistent with the COBIPULSE light curves (numbers are reported in the $7^{\mathrm{th}}$ column of Table \ref{tab:results}). Concerning the southern sky variable sources picked out in Section \ref{subsec:variablesandperiods}, we have been able to claim all of them as ``periodic variables" based solely on our survey, as they were observed across multiple nights covering more than one orbit.

\subsection{Classification of spider candidates} \label{subsec:spiderclass}
We use the following multi-wavelength scheme to classify a source as a potential spider candidate (numbers reported in the $8^{\mathrm{th}}$ column of Table \ref{tab:results}):
\begin{enumerate}
    \item {\it Unknown source}: It lacks a previous firm classification or published optical light curve.
    \item {\it $\gamma$-rays}: It is located inside the corresponding \textit{Fermi}-3FGL 95\% error ellipse. An additional match with the latest 4FGL associated source would strengthen the spider identification.
    \item {\it Optical}: Its multi-band optical light curves show a sinusoidal shape or sharp flux minima (see Section \ref{subsec:variablesandperiods}), with peak-to-peak amplitudes $\geq 0.1 \ \mathrm{mag}$. This threshold value
    takes into account also possible face-on inclinations. 
\end{enumerate}

Furthermore, we searched for matching X-ray sources in the \textit{Chandra}, \textit{Swift}, \textit{XMM-Newton} and \textit{eROSITA} latest catalogs \citep[][respectively]{2020AAS...23515405E,Evans_2020,2020A&A...641A.136W,2024A&A...682A..34M} using a search radius of $3''$ around the optical location. Detecting an X-ray source strengthens the spider association, as in these systems we usually observe X-ray emission produced by the intra-binary shock (see Section \ref{sec:intro}). However, the distance might be poorly constrained (see Section \ref{sec:discussion}) and some of the known spiders are faint X-ray emitters ($L_{\mathrm{X}} < 10^{31} \ \mathrm{erg} \ \mathrm{s}^{-1}$). Together with the different depth of X-ray observations in different fields, this implies that the absence of an X-ray counterpart does not rule out the optical source as a spider candidate.

\section{Results} \label{sec:results}
\begin{table*}
\raggedright
    \caption{Photometry and periodicity search results.}
    \setlength{\tabcolsep}{4.0pt}
    \begin{tabular}{lccccccc}
    \hline\hline
        Field & Sources in \textit{g'} & Sources in \textit{r'} & Sources in \textit{i'} & Phot. variables & Smooth variables & Per. variables & Spider candidates\\
        3FGL & (nr) & (nr) & (nr) & (nr) & (nr) & (nr) & (nr)\\ 
        \hline
    \multicolumn{8}{c}{STELLA/WiFSIP ($1.2$$\mathrm{m}$)}\\
    \hline
        {J0238.0$+$5237}& 4081 & 5997 & 6245 & 204 & 1 & 1 &--\\
        {J0312.1$-$0921}& 351 & 564 & 479 & 44 & -- & -- &--\\
        {J0318.1$+$0252}& 302 & 220 & 177 & 36 & 3 & -- &--\\
        {J0336.1$+$7500}& 2600 & 2576 & 2862 & 80 & -- & -- &--\\
        {J0545.6$+$6019}& 2189 & 3100 & 3563 & 83 & -- & -- &--\\
        {J0758.6$-$1451}& 2399 & 3702 & 3167 & 1293 & 3 & 3 &--\\
        {J0953.7$-$1510}& 1011 & 1445 & 1180 & 93 & 1 & 1 &--\\
        {J1120.6$+$0713}& 386 & 513 & 536 & 43 & -- & -- &--\\
        {J1543.5$-$0244}& 1550 & 1141 & 1585 & 322 & 1 & 1 &--\\
        {J1601.9$+$2306}& 979 & 778 & 628 & 150 & -- & -- &--\\
        {J1625.1$-$0021}& 626 & 820 & 446 & 33 & -- & -- &--\\
        {J1730.6$-$0357}& 5266 & 7499 & 6432 & 450 & 1 & -- &--\\
        {J2103.7$-$1113}& 609 & 930 & 1397 & 70 & -- & -- &--\\
        {J2212.5$+$0703}& 1956 & 3828 & 3011 & 175 & 1 & 1 &--\\
        {J2221.6$+$6507}& 3802 & 6210 & 6050 & 1169 & 3 & 3 & 1\\
        {J2233.1$+$6542}& 900 & 1755 & 3123 & 202 & 1 & 1 &--\\
        {J2310.1$-$0557}& 704 & 790 & 1402 & 189 & 1 & -- &--\\
        \hline
        \multicolumn{8}{c}{INT/WFC ($2.5$-$\mathrm{m}$)}\\
    \hline
        {J1225.9$+$2953}& 715 & 752 & 513 & 56 & -- & -- &--\\
        {J1630.2$-$1052}& 1430 & 2186 & 2831 & 861 & 4 & -- &--\\
        {J1827.7$+$1141}& 6470 & 8506 & 9998 & 874 & -- & -- &--\\
        {J1829.2$+$3229}& 778 & 1538 & 1757 & 579 & -- & -- &--\\
        {J1950.2$+$1215}& 5497 & 8100 & 7723 & 1660 & 1 & 1 &--\\
        {J2117.6$+$3725}& 3852 & 4953 & 5870 & 673 & 2 & 2 & 2\\
        \hline
        \multicolumn{8}{c}{LCO/Sinistro ($1$-$\mathrm{m}$)}\\
    \hline
        {J0514.6$-$4406}& 1741 & 1915 & 3194 & 578 & 1 & 1 &--\\
        {J0737.2$-$3233}& 12034 & 15061 & 15603 & 1588 & 1 & 1 & 1\\
        {J0933.9$-$6232}& 15847 & 20725 & 22368 & 340 & -- & -- &--\\
        {J1035.7$-$6720}& 12027 & 12914 & 21331 & 213 & -- & -- &--\\
        {J1119.9$-$2204}\tablenotemark{a}& 890 & 1778 & 1634 & 74 & -- & -- & --\\
        {J1231.6$-$5113}& 7958 & 12182 & 15572 & 439 & 5 & 5 &--\\
        \hline
        \multicolumn{8}{c}{LCO/SBIG ($0.4$-$\mathrm{m}$)}\\
    \hline
        {J1744.1$-$7619}& 1045 & 1711 & 1440 & 30 & -- & -- &--\\
        {J1757.7$-$6030}& 1628 & 3227 & 2469 & 114 & -- & -- &--\\
        {J1946.4$-$5403}& 1491 & 1939 & 1699 & 115 & -- & -- &--\\
        {J2039.9$-$5618}\tablenotemark{b}& 1189 & 1607 & 937 & 44 & -- & -- &--\\
        \hline
    \end{tabular}
    \tablenotetext{a}{Already identified as a MSP candidate evolving in a MSP-He white dwarf binary by \cite{2022ApJ...926..201S}. In Section \ref{subsec:J1119andJ2039disc} we explain the non-detection of its optical counterpart as a variable from our survey.}
    \tablenotetext{b}{Already identified as a redback candidate by \cite{2015ApJ...812L..24R} and \cite{2015ApJ...814...88S}, and confirmed as a $\gamma$-ray MSP by \cite{2021MNRAS.502..915C}. We did not detect this optical variable in our data (see Section \ref{subsec:J1119andJ2039disc} for details).}
    \label{tab:results}
\end{table*}
\begin{table*}
\raggedright
    \caption{Optical location, \textit{g'}, \textit{r'} and \textit{i'}-band magnitudes ranges and best estimate of the photometric period for each spider candidate.}
    \setlength{\tabcolsep}{5.2pt}
    \begin{tabular}{lccccccc}
    \hline\hline
        Name & R.A. (J2000)\tablenotemark{a} & Decl. (J2000)\tablenotemark{a} & Error radius\tablenotemark{b} & \textit{g'} band & \textit{r'} band & \textit{i'} band & Photometric period\\
        3FGL & (h:m:s) & ($^{\circ}$: $'$: $''$) & ($''$) & (mag) & (mag) & (mag) & (d)\\ 
        \hline
    \hline
        J0737.2$-$3233 & 07:36:56.22 & $-$32:32:55.3 & 0.8 & [17.3-17.6] & [16.4-16.6] & [15.9-16.0] & 0.3548(5)\\
        J2117.6$+$3725-A & 21:17:56.30 & $+$37:26:44.6 & 0.9 & [19.1-19.5] & [18.2-18.7] & [17.8-18.1] & 0.12664(3)\\
        J2117.6$+$3725-B & 21:18:10.14 & $+$37:27:25.7 & 0.9 & [19.0-19.4] & [18.5-18.9] & [18.3-18.7] & 0.2209805(8)\\
        J2221.6$+$6507 & 22:22:32.80 & $+$65:00:21.0 & 1.0 & [18.9-19.5] & [17.8-18.4] & [17.1-17.8] & 0.165(4)\\
        \hline
    \end{tabular}
    \tablenotetext{a}{The equatorial coordinates have been obtained using our astrometry-corrected combined image in the \textit{r'}-band of the corresponding field of view.}
    \tablenotetext{b}{The error radius on the optical location has been estimated as FWHM/2 of the corresponding source profile.}
    \label{tab:candresults}
\end{table*}
\begin{figure}[ht!]
\centering
\includegraphics[width=\columnwidth]{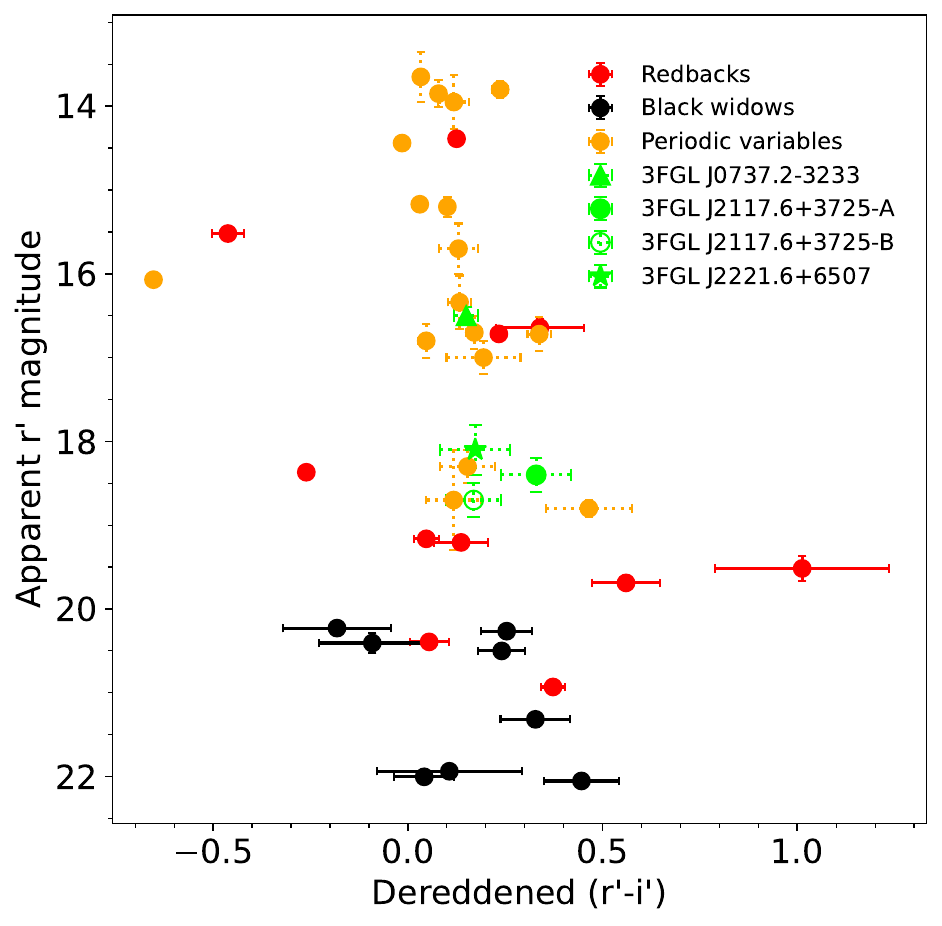}
\caption{Apparent \textit{r'} magnitude versus intrinsic (dereddened) $(\textit{r}'-\textit{i}')$ color. RBs and BWs are shown as red and black circles, respectively, with corresponding values taken from \textit{Pan-STARRS}~1 catalog magnitudes. The periodic variables and spider candidates identified in this work are marked with orange circles and different green symbols for each system, respectively. For both variables and candidates we plot the average value computed from their light curves, with dotted lines indicating their magnitude range on the $y$-axis and the standard deviation of their color on the $x$-axis.}
\label{fig:apprvsri}
\end{figure}

We report the main results obtained from our analysis in Table \ref{tab:results}, for each \textit{Fermi}-3FGL pulsar candidate. This wide-field optical survey led to the identification of 21 periodic variables in total, 3 of them newly discovered\footnote{We checked and did not find any variable star classification or optical light curve in the literature for these sources from \textit{Catalina Sky Survey} \citep[CSS;][]{catalinasouth_2017}, \textit{ATLAS} \citep{2018AJ....156..241H}, \textit{ZTF} \citep{Chen_2020}, and \textit{Gaia}-Data Release 3 \citep[DR3;][]{refId0,2022gdr3.reptE..10R} variables catalog.}. By following the classification method outlined in Section \ref{subsec:spiderclass}, we identified 4 spider candidates in three different 3FGL fields: 3FGL~J0737.2$-$3233, 3FGL~J2117.6$+$3725-A, 3FGL~J2117.6$+$3725-B and 3FGL~J2221.6$+$6507 (see Sections \ref{subsec:J0737res}, \ref{subsec:J2117res}, and \ref{subsec:J2221res} for details). We summarize their main properties in Table \ref{tab:candresults}. The corresponding periodograms and multi-band optical light curves are shown in the left and upper-right panels, respectively, of Figures \ref{fig:J0737_var1}, \ref{fig:J2117_var1}, \ref{fig:J2117_var2}, and \ref{fig:J2221_var2}. We report in Appendix \ref{sec:appC} detailed information and optical light curves for all the other 17 periodic variables found in this survey, identified either as eclipsing binaries, pulsating stars or W UMa binaries (see Table \ref{tab:periodicresults}, for a review about this type of optical variables see \citealt{Chambliss_1992}).

In Figure \ref{fig:apprvsri}, we plot the average apparent magnitude in the \textit{r'}-band as a function of the mean dereddened $(\textit{r}'-\textit{i}')$ color shown in the light curve from our spider candidates (highlighted in green symbols), together with all the other 17 COBIPULSE periodic variables (shown as orange circles). For comparison, we also include in the same magnitude-color diagram the optical counterparts of all currently known spider systems identified from the \textit{Pan-STARRS}~1 catalog \citep[PS1;][]{2016arXiv161205560C} by \cite{nedreaas2024spidercat} (12 RBs and 9 BWs, shown as red and black circles, respectively). All the $(\textit{g}'-\textit{r}')$ and $(\textit{r}'-\textit{i}')$ observed colors were corrected into intrinsic (dereddened) colors by using the corresponding color excess value $E(g-r)$ computed for each source from the 3D dust map of \cite{2019ApJ...887...93G}.

As clearly shown in Figure \ref{fig:apprvsri}, the COBIPULSE survey is sensitive enough to detect periodicities from sources as faint as $\textit{r'}\simeq19$. This indicates that our search is best suited to discover new RB systems. Variable light curves from optically fainter BWs ($\textit{r'}>20$) are more difficult to detect with the present survey. Indeed, our candidates (marked in green) are all located in the RB region of the magnitude-color diagram of Figure \ref{fig:apprvsri} ($\textit{r'}\simeq14-20$), suggesting a likely RB nature (see Section \ref{sec:discussion} for a detailed discussion). 

\subsection{3FGL~J0737.2$-$3233} \label{subsec:J0737res}
\begin{figure}[t!]
\centering
\includegraphics[width=\columnwidth]{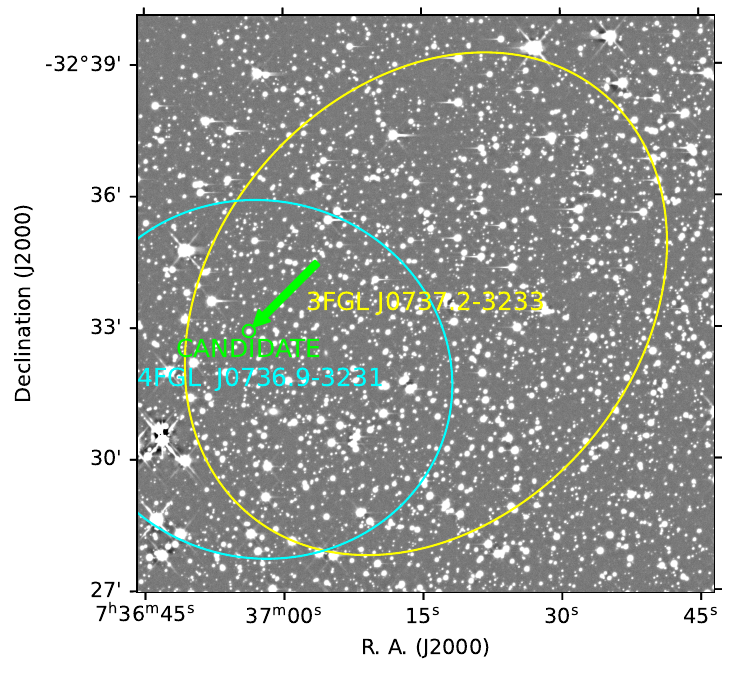}
\caption{3FGL~J0737.2$-$3233 field of view observed with LCO/Sinistro. 3FGL and 4FGL 95\% error ellipses are shown in yellow and cyan, respectively, while the spider candidate is highlighted in green.}
\label{fig:J0737_FoV}
\end{figure}
\begin{figure}[t!]
\centering
\includegraphics[width=\columnwidth]{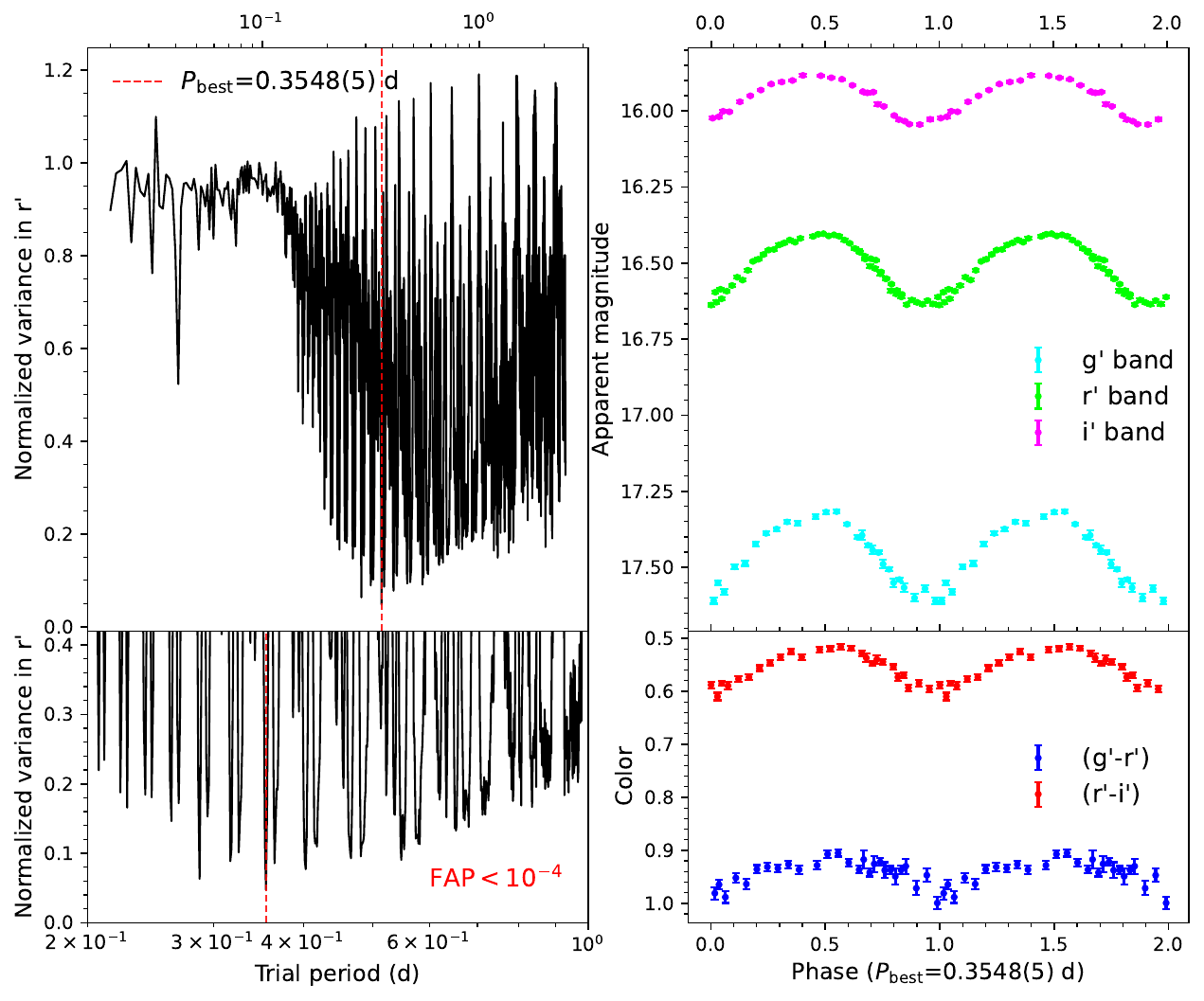}
\caption{\textit{Top left panel}: PDM periodogram of the \textit{r'}-band light curve of J0737. \textit{Bottom left panel}: Zoom of the deepest minimum in the periodogram found at $P_{\mathrm{best}}=0.3548 \pm0.0005 \ \mathrm{d}$. \textit{Top right panel}: Light curves of J0737 in the \textit{g'}, \textit{r'} and \textit{i'} optical bands folded on $P_{\mathrm{best}}$ and reference epoch (minimum light) $T_{0}=57336.219 \ \mathrm{MJD}$, with two cycles shown for displaying purposes. \textit{Bottom right panel}: Observed color curves of J0737 folded on $P_{\mathrm{best}}$.}
\label{fig:J0737_var1}
\end{figure}

We identify an optical variable and spider candidate (hereafter referred to as J0737) inside both the 95\% error ellipses of 3FGL~J0737.2$-$3233 and 4FGL~J0736.9$-$3231 (see Figure \ref{fig:J0737_FoV}).

We find from the global minimum of its PDM periodogram a photometric period $P_{\mathrm{best}}=0.3548\pm0.0005 \ \mathrm{d}$ (as shown in the left panels of Figure \ref{fig:J0737_var1}), with a FAP probability $<10^{-4}$ corresponding to a confidence level more than $3\sigma$ (see Section \ref{subsec:variablesandperiods})\footnote{Here we do not find any of the permuted time series yielding a PDM minimum smaller than the original minimum. Therefore, we place a lower limit on the detection significance. The same consideration applies also for the spider candidates in Section \ref{subsec:J2117res} and \ref{subsec:J2221res}.}. We fold the \textit{g'}, \textit{r'} and \textit{i'} light curves and color curves on this period using as reference epoch the time $T_{0}=57336.219\pm0.002 \ \mathrm{MJD}$, corresponding to the brightness minimum in the \textit{r'}-band\footnote{All the light curves and color curves presented hereafter in this paper have been folded using the time of flux minimum in the \textit{r'}-band as the reference epoch, as this is supposed to correspond to the companion's inferior conjunction for spider systems.} (see right panels of Figure \ref{fig:J0737_var1}). We find the same periodic modulation in all three bands, with peak-to-peak amplitudes of $0.3$, $0.2$ and $0.2 \ \mathrm{mag}$ in \textit{g'}, \textit{r'} and \textit{i'} filters, respectively. Additionally, the color curves clearly show a periodic trend, especially $(\textit{r}'-\textit{i}')$, peaking around the same phase of maximum flux.

We corrected the observed colors into intrinsic (dereddened) colors, and computed their mean and standard deviation along one full cycle, obtaining $(\textit{g}'-\textit{r}')=0.37\pm0.06 \ \mathrm{mag}$ and $(\textit{r}'-\textit{i}')=0.15\pm0.05 \ \mathrm{mag}$. These two colors have been matched to the spectral templates for low mass stars provided by \cite{allard11}, which turn out to be compatible with effective temperatures of $T_\mathrm{eff}=6100\pm200 \ \mathrm{K}$ and $T_\mathrm{eff}=5500\pm400 \ \mathrm{K}$ for $(\textit{g}'-\textit{r}')$ and $(\textit{r}'-\textit{i}')$, respectively. Hereinafter, we use this method to estimate the effective temperatures from the optical colors.

Given the shape of its light and color curve shape, in Section \ref{subsec:J0737disc} we discuss J0737 as a likely irradiated RB system observed at a low orbital inclination.
\subsection{3FGL~J2117.6$+$3725} \label{subsec:J2117res}
\begin{figure}[t!]
\centering
\includegraphics[width=\columnwidth]{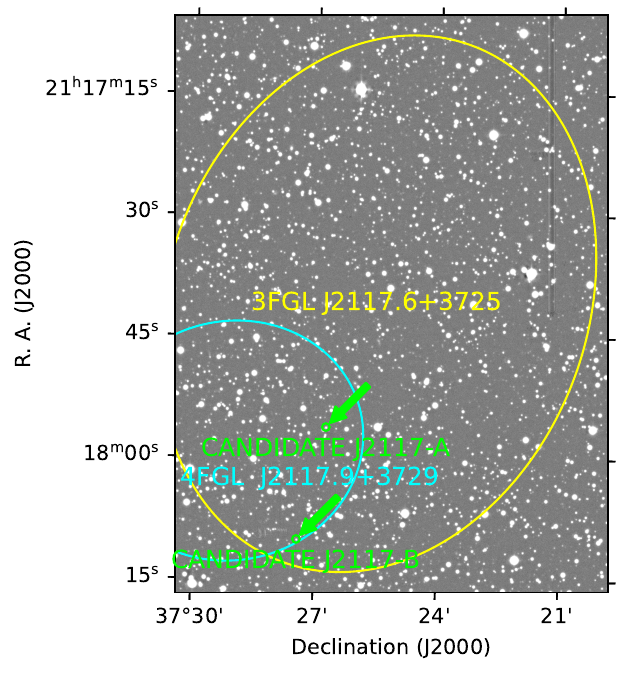}
\caption{3FGL~J2117.6$+$3725 field of view observed with INT/WFC. 3FGL and 4FGL 95\% error ellipses are shown in yellow and cyan, respectively, while the two spider candidates are highlighted in green.}
\label{fig:J2117_FoV}
\end{figure}
\begin{figure}[t!]
\centering
\includegraphics[width=\columnwidth]{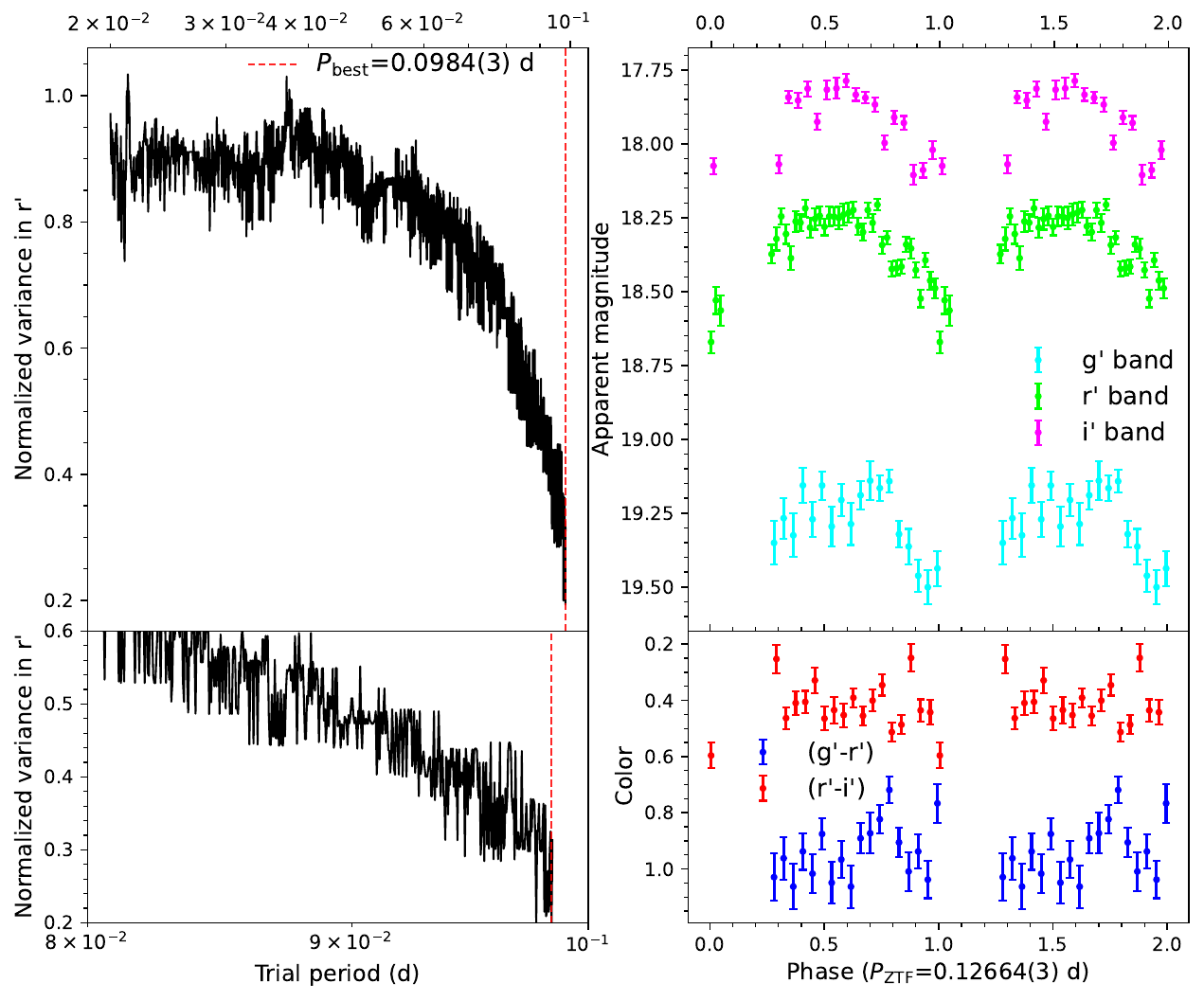}
\caption{\textit{Top left panel}: PDM periodogram performed on the \textit{r'}-band light curve of J2117-A.  \textit{Bottom left panel}: Zoom of the minimum in the periodogram found at $P_{\mathrm{best}}=0.0984\pm0.0003 \ \mathrm{d}$. \textit{Top right panel}: Light curves of J2117-A in the \textit{g'}, \textit{r'} and \textit{i'} optical bands folded at the photometric period $P_{\mathrm{ZTF}}=0.12664 \ \mathrm{d}$ and reference epoch $T_{0}=59025.415 \ \mathrm{MJD}$, estimated from the periodicity search on ZTF data, where two cycles are shown for displaying purposes. \textit{Bottom right panel}: Observed color curves of J2117-A folded at $P_{\mathrm{ZTF}}$.}
\label{fig:J2117_var1}
\end{figure}
\begin{figure}[t!]
\centering
\includegraphics[width=\columnwidth]{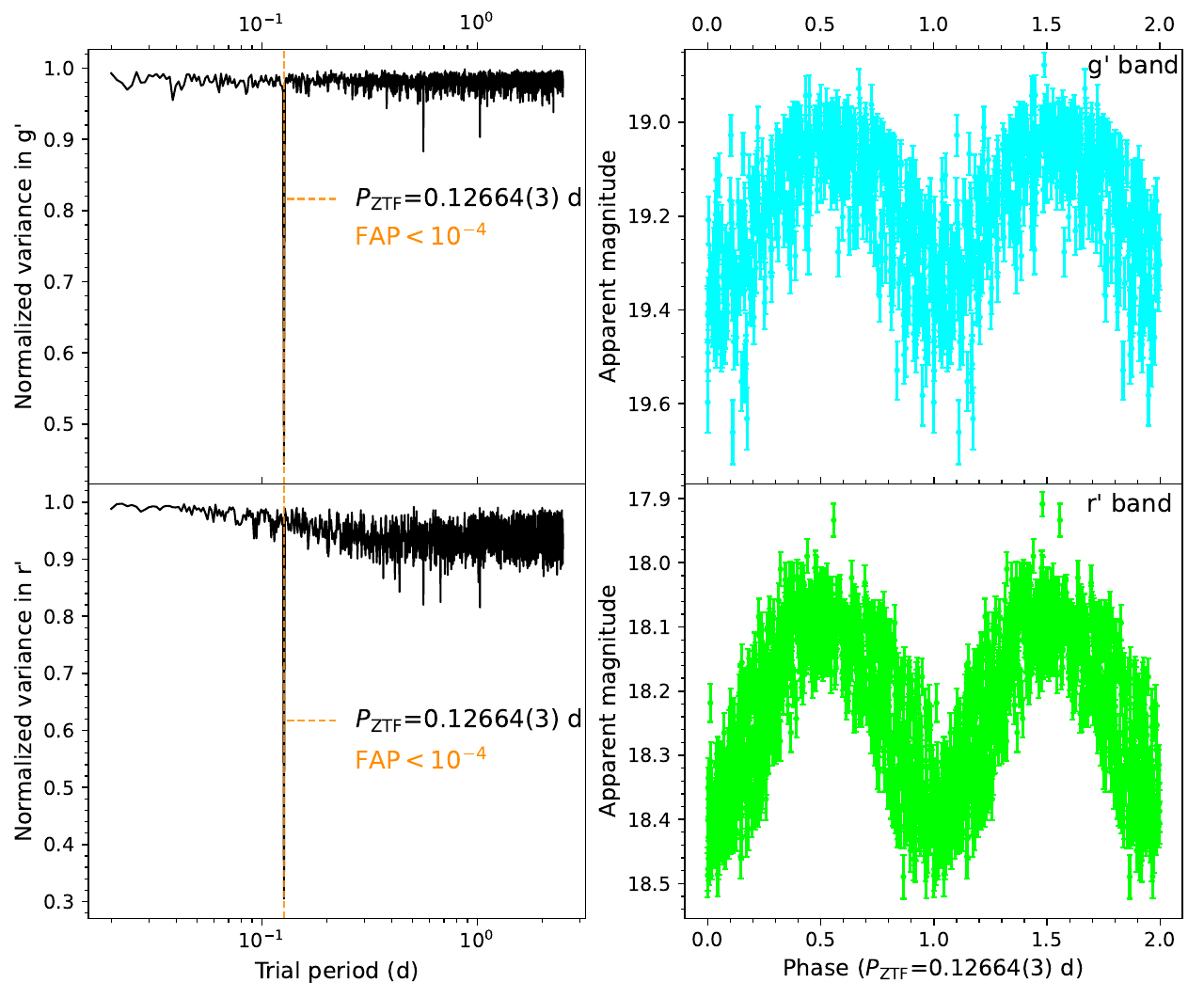}
\caption{\textit{Left panels}: PDM periodograms performed on the \textit{g'} and \textit{r'}-band light curves of J2117-A acquired with ZTF. \textit{Right panels}: ZTF optical light curves of J2117-A folded at the photometric period $P_{\mathrm{ZTF}}=0.12664 \ \mathrm{d}$ and reference epoch $T_{0}=59025.415 \ \mathrm{MJD}$.}
\label{fig:J2117_var1_ZTF}
\end{figure}
\begin{figure}[t!]
\centering
\includegraphics[width=\columnwidth]{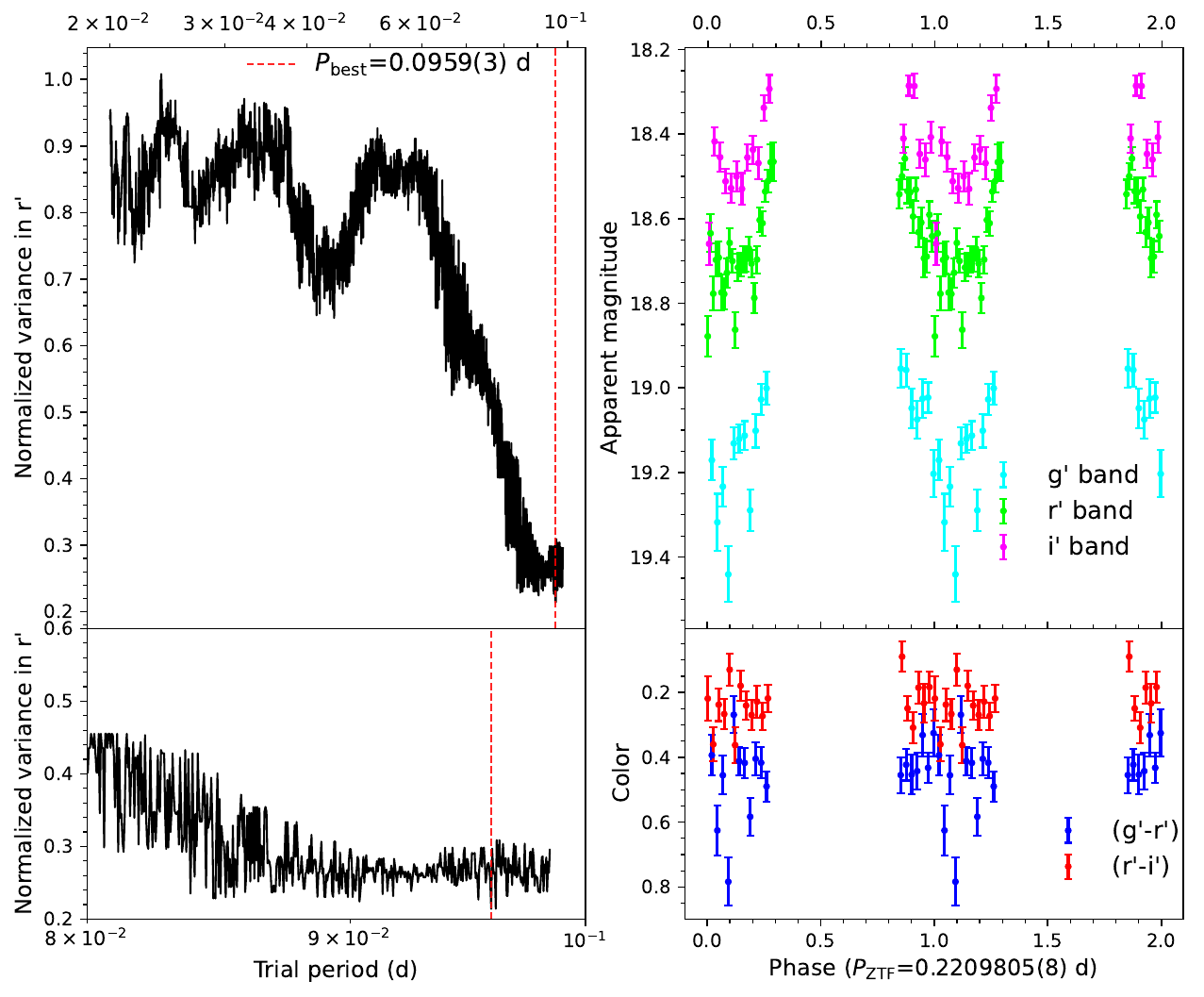}
\caption{\textit{Left top panel}: PDM periodogram performed on the \textit{r'}-band light curve of J2117-B.  \textit{Left bottom panel}: Zoom of the minimum in the periodogram found at $P_{\mathrm{best}}=0.0959\pm0.0003 \ \mathrm{d}$. \textit{Right top panel}: Light curves of J2117-B in the \textit{g'}, \textit{r'} and \textit{i'} optical bands folded at the photometric period $P_{\mathrm{ZTF}}=0.2209805 \ \mathrm{d}$ and reference epoch $T_{0}=58745.308 \ \mathrm{MJD}$, estimated from the periodicity search on ZTF data, where two cycles are shown for displaying purposes. \textit{Right bottom panel}: Observed color curves of J2117-B folded at $P_{\mathrm{ZTF}}$.}
\label{fig:J2117_var2}
\end{figure}
\begin{figure}[t!]
\centering
\includegraphics[width=\columnwidth]{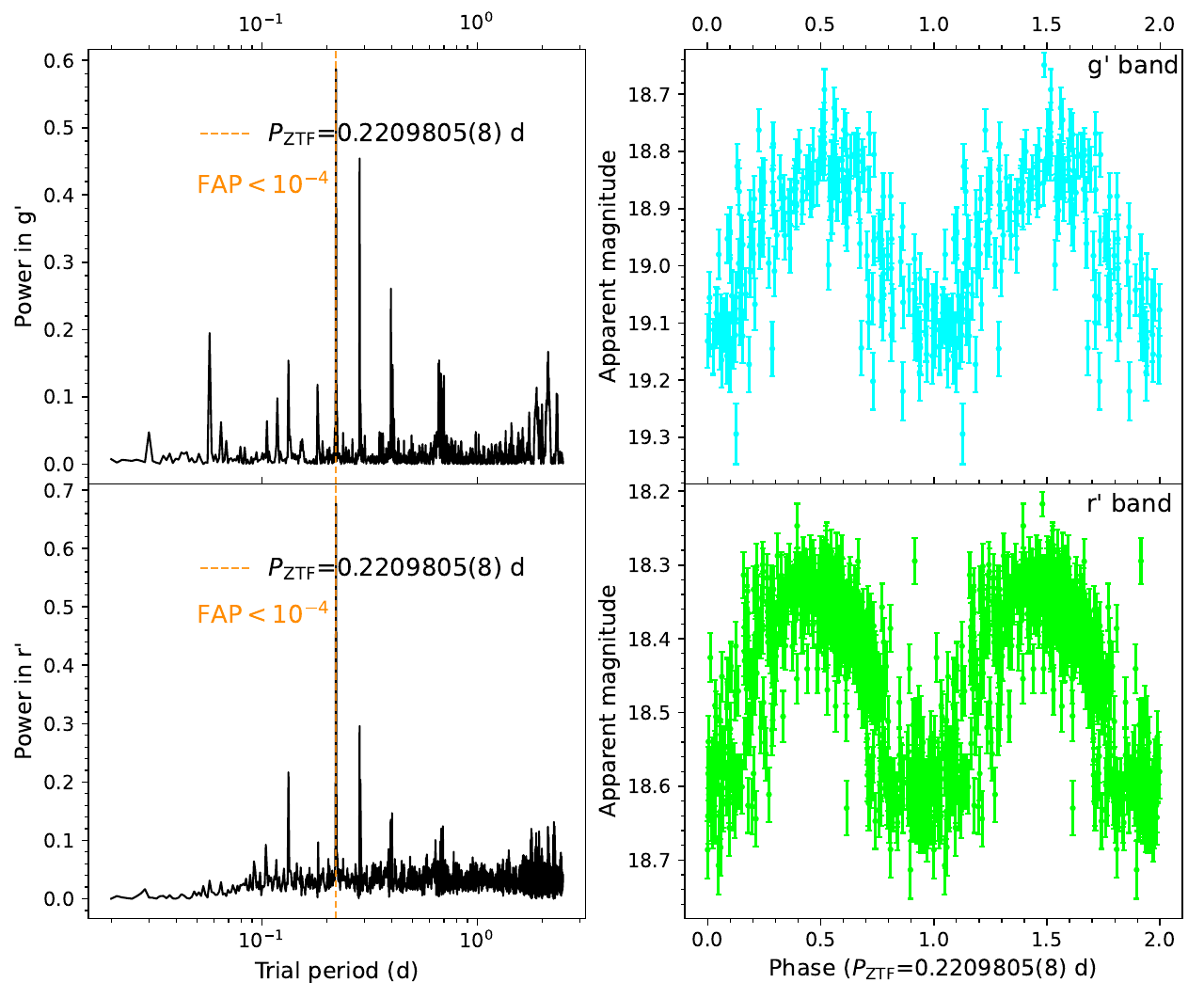}
\caption{\textit{Left panels}: LS periodograms performed on the \textit{g'} and \textit{r'}-band light curves of J2117-B acquired with ZTF. \textit{Right panels}: ZTF optical light curves of J2117-B folded at the photometric period $P_{\mathrm{ZTF}}=0.2209805 \ \mathrm{d}$ and reference epoch $T_{0}=58745.308 \ \mathrm{MJD}$.}
\label{fig:J2117_var2_ZTF}
\end{figure}

We identify two optical variables (hereafter referred to as J2117-A and J2117-B, respectively) inside both 95\% error regions of 3FGL~J2117.6$+$3725 and 4FGL~J2117.9$+$3729, as illustrated in Figure \ref{fig:J2117_FoV}.

The periods corresponding to the minima of the PDM periodograms of the INT light curves are $P_{\mathrm{best}}=0.0984\pm0.0003 \ \mathrm{d}$ and $P_{\mathrm{best}}=0.0959\pm0.0003 \ \mathrm{d}$ for J2117-A and J2117-B, respectively (see left panels of Figures \ref{fig:J2117_var1} and \ref{fig:J2117_var2}). However, the two periodograms show that the variance is monotonically decreasing towards the right edge of the period grid. This indicates that the observing time window does not cover a full cycle of the two sources, likely indicating incorrect estimates of their periods. Therefore, we instead obtain the periods from PDM and LS periodograms of the ZTF optical light curves (see Section \ref{subsec:ZTFdata}). The photometric periods for J2117-A and J2117-B are $P_{\mathrm{ZTF}}=0.12664\pm0.00003 \ \mathrm{d}$ and $P_{\mathrm{ZTF}}=0.2209805\pm0.0000008 \ \mathrm{d}$, respectively (shown in Figures \ref{fig:J2117_var1_ZTF} and \ref{fig:J2117_var2_ZTF}), both detected with a FAP probability $<10^{-4}$ and therefore a confidence level of more than $3\sigma$. Here, we used as reference epochs $T_{0}=59025.415\pm0.002 \ \mathrm{MJD}$ and $T_{0}=58745.308\pm0.002 \ \mathrm{MJD}$, respectively.

These two variables show periodic modulation consistent across different optical bands with amplitudes of $0.3$--$0.4 \ \mathrm{mag}$. We estimated dereddened colors of $(\textit{g}'-\textit{r}')=0.8\pm0.1 \ \mathrm{mag}$ and $(\textit{r}'-\textit{i}')=0.3\pm0.1 \ \mathrm{mag}$ for J2117-A, corresponding to temperatures of $T_\mathrm{eff}=4950\pm250 \ \mathrm{K}$ and $T_\mathrm{eff}=4650\pm350 \ \mathrm{K}$, respectively. Applying the same procedure, we obtain for J2117-B $(\textit{g}'-\textit{r}')=0.4\pm0.2 \ \mathrm{mag}$ and $(\textit{r}'-\textit{i}')=0.17\pm0.09 \ \mathrm{mag}$, with respective temperatures of $T_\mathrm{eff}=6300\pm600 \ \mathrm{K}$ and $T_\mathrm{eff}=5700\pm900 \ \mathrm{K}$.

In Section \ref{subsec:J2117disc} we discuss these two variables as potential optical counterparts to RB MSPs dominated by ellipsoidal modulation rather than irradiation, as they show small-amplitude modulations and constant colors.

\subsection{3FGL~J2221.6$+$6507} \label{subsec:J2221res}
\begin{figure}[t!]
\centering
\includegraphics[width=\columnwidth]{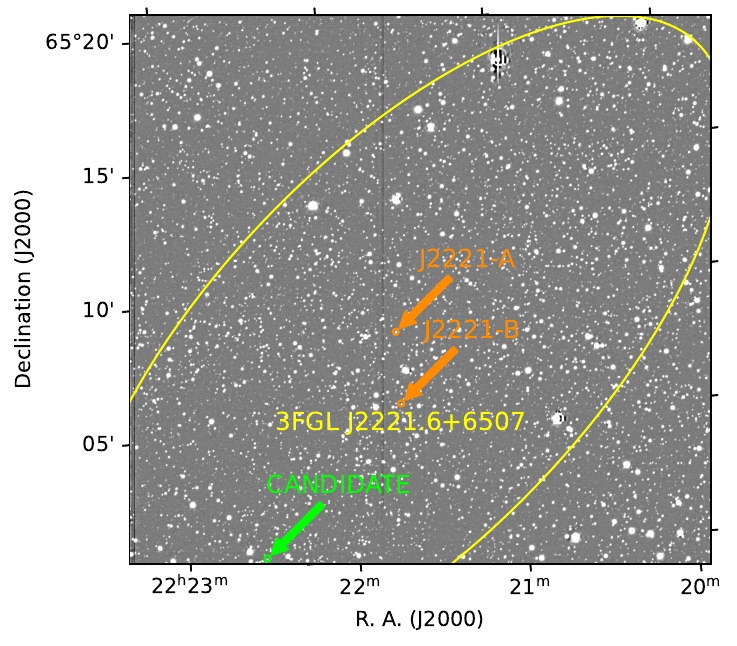}
\caption{3FGL~J2221.6$+$6507 field of view observed with STELLA/WiFSIP. 3FGL 95\% error ellipse is shown in yellow, while the spider candidate is highlighted in green and the two other periodic variables are reported in orange.}
\label{fig:J2221_FoV}
\end{figure}
\begin{figure}[t!]
\centering
\includegraphics[width=\columnwidth]{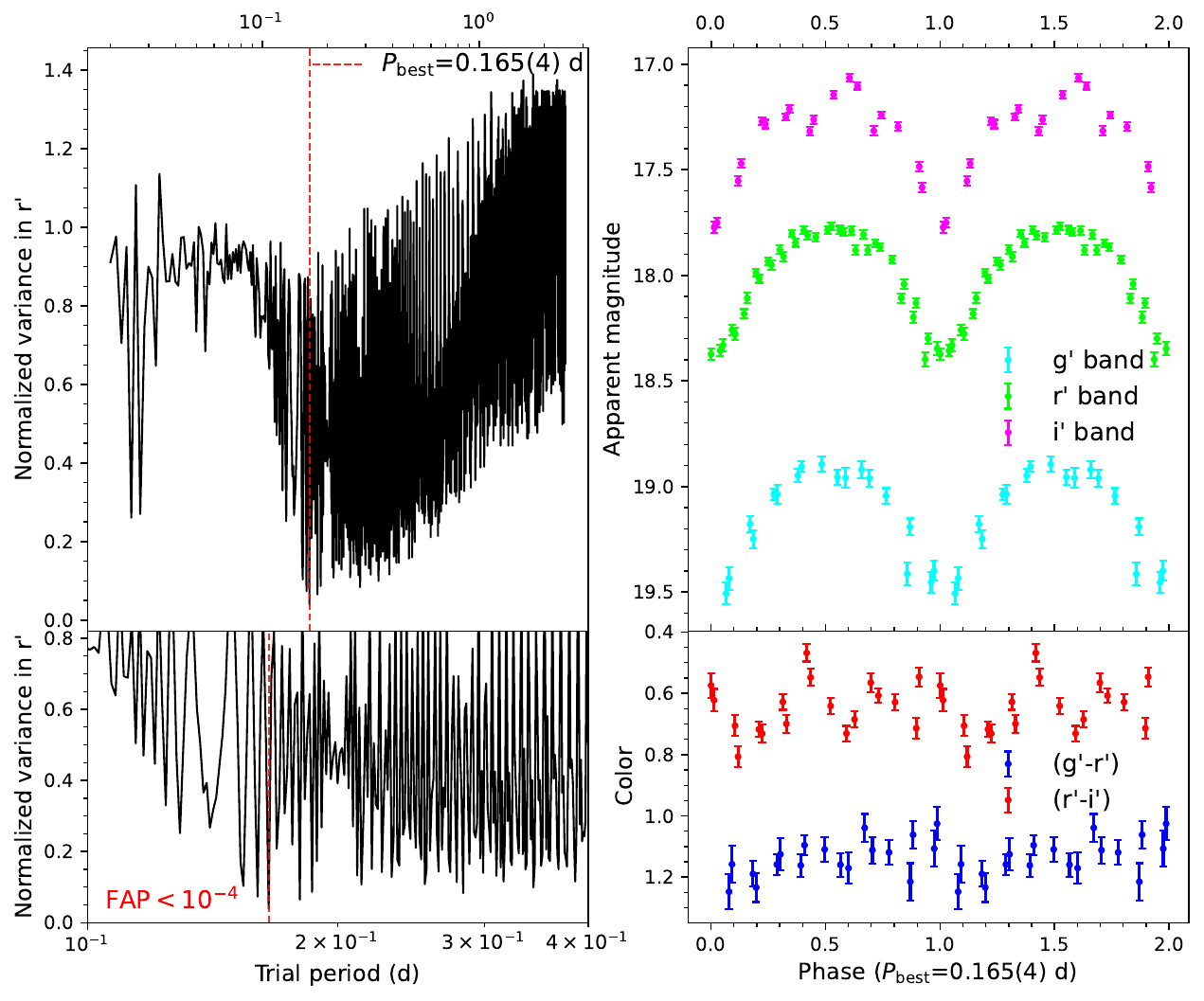}
\caption{\textit{Top left panel}: PDM periodogram performed on the \textit{r'}-band light curve of J2221. \textit{Left bottom panel}: Zoom of the deepest minimum in the periodogram found at $P_{\mathrm{best}}=0.165 \pm0.004 \ \mathrm{d}$. \textit{Top right panel}: Light curves of J2221 in the \textit{g'}, \textit{r'} and \textit{i'} optical bands folded on $P_{\mathrm{best}}$ and reference epoch $T_{0}=57343.939 \ \mathrm{MJD}$, with two cycles shown for displaying purposes. \textit{Bottom right panel}: Observed color curves of J2221 folded on $P_{\mathrm{best}}$.}
\label{fig:J2221_var2}
\end{figure}

We detected three optical variables inside the 95\% error ellipse of 3FGL~J2221.6$+$6507. Two of these are identified as W UMa eclipsing binary stars in the ZTF catalog \citep{Chen_2020} (see Appendix \ref{sec:appC}, reported in orange as J2221-A and J2221-B in Figure \ref{fig:J2221_FoV}), while we classify the other as a spider candidate (henceforth called J2221, green in Figure \ref{fig:J2221_FoV}).
Here, the location of the 3FGL $\gamma$-ray source should be taken with caution, as it is quite close to the Galactic plane (Galactic latitude $b=6.7^{\circ}$) and is lacking a 4FGL source association (see discussion in Section \ref{subsec:J2221disc}).

We found from the PDM periodogram a periodicity at $P_{\mathrm{best}}=0.165\pm0.004 \ \mathrm{d}$ for J2221, which we used to fold our data (see Figure \ref{fig:J2221_var2}). We obtained for this detection a FAP probability $<10^{-4}$, equivalent to a significance level higher than $3\sigma$. The observed periodic modulation is consistent in the \textit{g'}, \textit{r'} and \textit{i'} bands, showing a broad flux maximum and a sharp minimum. We measured peak-to-peak amplitudes of $0.6 \ \mathrm{mag}$ in all three bands. Also the color curves show a complex variable trend especially in $(\textit{r}'-\textit{i}')$, with at least two local maxima at $\phi=0.5$ and $\phi=0.75$.
We estimate for the dereddened colors $(\textit{g}'-\textit{r}')=0.5\pm0.1 \ \mathrm{mag}$ and $(\textit{r}'-\textit{i}')=0.2\pm0.1 \ \mathrm{mag}$, which correspond to temperatures of $T_\mathrm{eff}=5850\pm350 \ \mathrm{K}$ and $T_\mathrm{eff}=5550\pm750 \ \mathrm{K}$, respectively.

\section{Discussion} \label{sec:discussion}

\subsection{3FGL~J0737.2$-$3233} \label{subsec:J0737disc}
We observe a clear periodic pattern in the folded optical light curves of J0737, with consistent color trends, as we can see in Figure \ref{fig:J0737_var1}. This indicates that the star is showing temperature variations along the photometric period $P_{\mathrm{best}}=0.3548 \pm 0.0005 \ \mathrm{d}$, with a mean temperature of $T_{\mathrm{eff}}=6100 \pm 200 \ \mathrm{K}$ estimated from the dereddened $(\textit{g}'-\textit{r}')$ color.

Such features suggest that J0737 is likely a RB system where the companion star is irradiated by the pulsar wind. Thus, the optical light curves have been folded with the presumed orbital period $P_{\mathrm{orb}}=P_{\mathrm{best}}=0.3548 \pm 0.0005 \ \mathrm{d}$, showing one flux maximum at the superior conjunction of the companion ($\phi=0.5$).
Also, the optical colors peak at $\phi=0.5$, consistent with the proposed scenario. However, the light curves exhibit peak-to-peak amplitudes of $0.2$--$0.3\ \mathrm{mag}$, smaller than the $\approx1$-mag modulations typically observed in RBs undergoing an irradiated regime (see e.g. \citealt{2018ApJ...859...54L} for PSR~J2215+5135 and \citealt{2011ApJ...743L..26R} for PSR~J2339$-$0533). A low orbital inclination can explain the small amplitudes of the light curves of J0737, implying that we observe only a fraction of the irradiated inner face of the companion at $\phi=0.5$. This hypothesis is further supported by the small temperature change along the orbit, $\sim200 \ \mathrm{K}$, estimated from a base (non-irradiated) temperature of $T_{\mathrm{b}}=6000 \pm 200 \ \mathrm{K}$ at $\phi=0$ and a day-side (irradiated) temperature of $T_{\mathrm{day}}=6200 \pm 200 \ \mathrm{K}$ at $\phi=0.5$.

We also found the source 5592027220171664128 from \textit{Gaia}-DR3 \citep{brown2016gaia,refId0} coincident with J0737. The measurement of its parallax, $1.50\pm0.05 \ \mathrm{mas}$, is precise enough to obtain an accurate estimate of the geometric distance to this object.
By assuming the distance prior distribution from \cite{2021AJ....161..147B}, based on stellar populations, we obtained a distance of $D=659_{-20}^{+16} \ \mathrm{pc}$\footnote{We computed the median value and the $1\sigma$ lower and upper ends on the geometric distances through \url{http://dc.zah.uni-heidelberg.de/__system__/adql/query/form}.}.
If J0737 is confirmed as a compact MSP, it will be the closest known spider MSP to Earth (considering only parallax distance measurements). Among RB spiders, only PSR~J2339$-$3233 has a similar distance, $D=750 \ \mathrm{pc}$, estimated via dispersion measure by \cite{2014AAS...22314007R} using the electron density model of \cite{2017ApJ...835...29Y}. However, its \textit{Gaia} geometric distance is $1.71 \ \mathrm{kpc}$. In general, dispersion measure distances systematically underestimate the real spider distances due to small-scale inaccuracies of the electron density map of \cite{2017ApJ...835...29Y}, as shown by \cite{koljonen2024}.

RB MSPs are generally characterized by X-ray emission produced via synchrotron radiation in the intrabinary shock region (see Section \ref{sec:intro}), with a typical power-law photon index of $\Gamma\simeq1$--$1.5$ \citep{2014ApJ...795...72L}. In the case of J0737, we did not find any X-ray counterparts in the latest point-source catalogs from \textit{Chandra}, \textit{Swift}, \textit{XMM-Newton} or \textit{eROSITA}. The most constraining X-ray upper limit is obtained from \textit{Swift}/XRT (0.3--10 keV), that observed the corresponding \textit{Fermi} field between 2010 July 31 and 2017 March 6, pointing to J0737 for a total exposure of $10 \ \mathrm{ks}$. Using the \textit{Swift}-SXPS query server at \url{https://www.swift.ac.uk/LSXPS/ulserv.php}, we estimate a $3\sigma$ upper limit of $1.5\times10^{-3} \ \mathrm{ct} \ \mathrm{s}^{-1}$ from our source location. Given the low counts upper limit and the proximity of this source, we assume a photon index $\Gamma=2.5$, larger than typical values for RBs and more suited for low X-ray luminosity spiders \citep{Swihart_2022}. Using also an interstellar hydrogen absorption\footnote{Calculated using the H\,{\sc i} column density tool at \url{https://heasarc.gsfc.nasa.gov/cgi-bin/Tools/w3nh/w3nh.pl}.} of $N_{\mathrm{H}}=3.5\times10^{21} \ \mathrm{cm}^{-2}$, the corresponding upper limit to the X-ray unabsorbed energy flux \footnote{\textit{Swift}-XRT counts converted into 0.3--10 keV flux using the tool located at \url{https://heasarc.gsfc.nasa.gov/cgi-bin/Tools/w3pimms/w3pimms.pl}.} is $1.0\times10^{-13} \ \mathrm{erg} \ \mathrm{cm}^{-2} \ \mathrm{s}^{-1}$. Taking the parallax distance reported earlier as a reference value, we obtained an upper limit to the X-ray (0.3--10 keV) luminosity of $5.4\times10^{30} \ (D/0.7 \ \mathrm{kpc})^2 \ \mathrm{erg} \ \mathrm{s}^{-1}$. For comparison, the X-ray luminosity distribution of Galactic RBs peaks at the higher value of $8\times10^{31} \ \mathrm{erg} \ \mathrm{s}^{-1}$ \citep{2023MNRAS.525.3963K}. J0737 would then be one of the least luminous RBs in X-rays, which are generally brighter than BWs at these wavelengths \citep{2023MNRAS.525.3963K}. However, our upper limit still does not rule out a faint X-ray emission from this system, that could be detected with observations deeper than $10 \ \mathrm{ks}$. 

The match between our optical variable and the pulsar-like \textit{Fermi} unidentified source 3FGL~J0737.2$-$3233, along with its 4FGL association 4FGL~J0736.9$-$3231, as shown in Figure \ref{fig:J0737_FoV}, strengthens the identification of J0737 as a RB MSP candidate (for details, see Section \ref{subsec:gammaselection}). The integrated 0.1--100 GeV energy flux from 4FGL is $(1.0\pm0.1)\times10^{-11}\ \mathrm{erg} \ \mathrm{cm}^{-2} \ \mathrm{s}^{-1}$, which corresponds to a $\gamma$-ray luminosity of $L_{\gamma}=5.3\times10^{32} \ (D/0.7 \ \mathrm{kpc})^2 \ \mathrm{erg} \ \mathrm{s}^{-1}$, taking as reference the parallax distance. This value lies within the range of $\gamma$-ray luminosities, $\sim10^{32}$--$10^{34} \ \mathrm{erg} \ \mathrm{s}^{-1}$, typically observed from MSPs \citep{2023ApJ...958..191S}.

We also identified an infrared counterpart to J0737, with ID \textit{2MASS} J07365621$-$3232551 \citep{2006AJ....131.1163S} and magnitudes $J=14.31 \pm 0.04$, $H=13.63 \pm 0.03$, and $K=13.48 \pm 0.04$. We completed this multi-wavelength search by looking for counterparts in the radio band but did not find any matches from the 1.4 GHz \textit{NRAO VLA Sky Survey} \citep[NVSS,][]{1998AJ....115.1693C} or the 4.0--10.0 GHz radio follow-up on all 3FGL unassociated sources by \cite{2017ApJ...838..139S}. Additionally, the other candidates discussed in Sections \ref{subsec:J2117disc} and \ref{subsec:J2221disc} also lack a radio counterpart.

\subsection{3FGL~J2117.6+3725} \label{subsec:J2117disc}
\begin{figure}[ht!]
\centering
\includegraphics[width=\columnwidth]{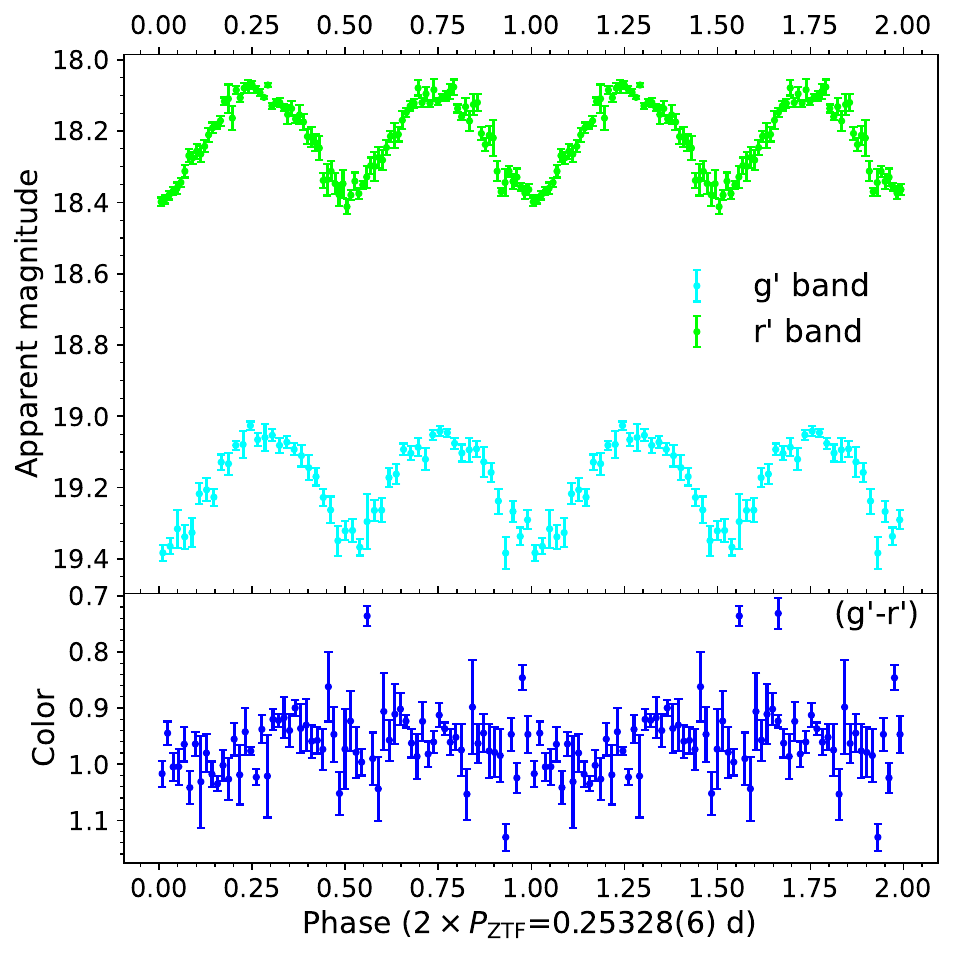}
\caption{\textit{Top panel}: ZTF \textit{g'}- and \textit{r'}-band light curves of J2117-A folded with the assumed orbital period $P_{\mathrm{orb}}=0.25328 \ \mathrm{d}$ and reference epoch $T_{0}=59025.415 \ \mathrm{MJD}$, where two cycles are shown for displaying purposes. \textit{Bottom panel}: Observed $(\textit{g}'-\textit{r}')$ color curve of J2117-A.}
\label{fig:J2117_var1_ZTF_lccol}
\end{figure}
\begin{figure}[ht!]
\centering
\includegraphics[width=\columnwidth]{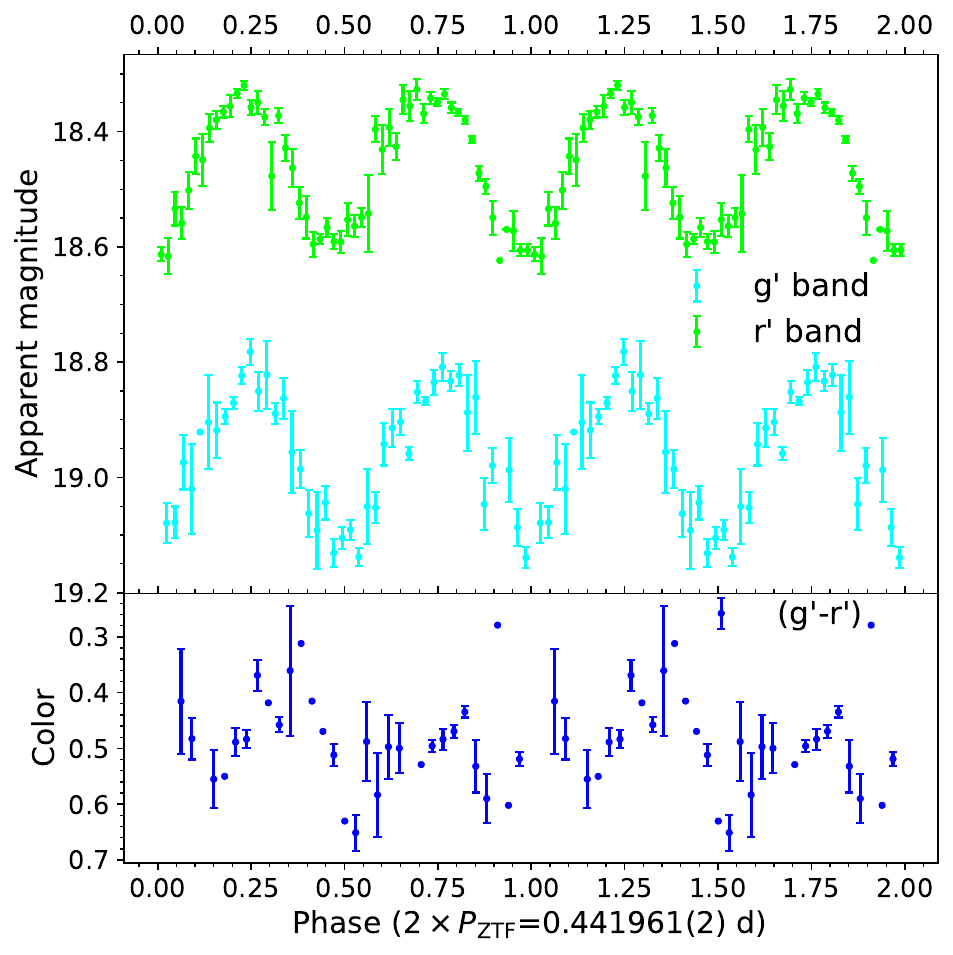}
\caption{\textit{Top panel}: ZTF \textit{g'}- and \textit{r'}-band light curves of J2117-B folded with the assumed orbital period $P_{\mathrm{orb}}=0.441961 \ \mathrm{d}$ and reference epoch $T_{0}=58745.308 \ \mathrm{MJD}$, where two cycles are shown for displaying purposes. \textit{Bottom panel}: Observed $(\textit{g}'-\textit{r}')$ color curve of J2117-B.}
\label{fig:J2117_var2_ZTF_lccol}
\end{figure}

We find a deep minimum and a strong maximum from the PDM and LS periodograms, respectively, of the ZTF optical light curves of J2117-A and J2117-B, corresponding to the two photometric periods $P_{\mathrm{ZTF}}=0.12664 \pm 0.00003$ and $P_{\mathrm{ZTF}}=0.2209805 \pm 0.0000008 \ \mathrm{d}$, respectively (left panels of Figures \ref{fig:J2117_var1_ZTF} and \ref{fig:J2117_var2_ZTF}). The two folded optical light curves show similar periodic shapes, which are consistent in the \textit{g'} and \textit{r'} bands, with peak-to-peak amplitudes of $0.4$ and $0.3 \ \mathrm{mag}$ for J2117-A and J2117-B, respectively, illustrated in the right panels of Figures \ref{fig:J2117_var1_ZTF} and \ref{fig:J2117_var2_ZTF}. We also report folded INT light curves for the two variables in the right panels of Figures \ref{fig:J2117_var1} and \ref{fig:J2117_var2}. Their colors do not show variability consistent with the modulation of the respective light curves, opposite to what is observed in the irradiation case of J0737 (Figure \ref{fig:J0737_var1}). This suggests that both J2117-A and J2117-B do not experience significant temperature variation, with $T_{\mathrm{eff}}=4950 \pm 250 \ \mathrm{K}$ and $T_{\mathrm{eff}}=6300 \pm 600 \ \mathrm{K}$ from $(\textit{g}'-\textit{r}')$ color, respectively.

These observations make both systems compatible with being the optical counterpart of a RB MSP not irradiating its companion star, showing only ellipsoidal modulation. Therefore, we show in Figures \ref{fig:J2117_var1_ZTF_lccol} and \ref{fig:J2117_var2_ZTF_lccol} their ZTF light curves and color curves now folded with the presumed orbital periods of J2117-A ($P_{\mathrm{orb}}=2\times P_{\mathrm{ZTF}}=0.25328 \pm 0.00006 \ \mathrm{d}$) and J2117-B  ($P_{\mathrm{orb}}=2\times P_{\mathrm{ZTF}}=0.441961 \pm 0.000002 \ \mathrm{d}$), obtained as twice their photometric periods. In order to visualize more clearly the periodic pattern, we also rebinned the \textit{g'}- and \textit{r'}-band light curves of J2117-A into $50$ and $100$ phase bins, respectively, while we used $50$ bins for both light curves of J2117-B\footnote{We computed the mean magnitude over all the data points included in the corresponding bin, with their standard deviation as uncertainty.}.

As we can see from Figures \ref{fig:J2117_var1_ZTF_lccol} and \ref{fig:J2117_var2_ZTF_lccol}, both rebinned light curves of J2117-A and J2117-B show two flux minima at $\phi=0.5$ and $\phi=1$, and two maxima at ascending nodes at $\phi=0.25$ and $\phi=0.75$. We do not observe any clear variable trend in the $(g'-r')$ colors of these two systems, which are constant within the errors along their orbits.
Both cases closely resemble the optical light curve and color curve shapes typically observed in non-irradiated RBs dominated by ellipsoidal modulation (e.g. PSR~J2039$-$5618, \citealt{2015ApJ...814...88S}; PSR~J2129$-$0429, \citealt{2016ApJ...816...74B} and PSR~J1622$-$0315, \citealt{2023MNRAS.525.2565T}).
In the following, we will thereby consider both J2117-A and J2117-B for our multi-wavelength discussion.

Our candidates J2117-A and J2117-B match two optical sources detected and flagged as variable by \textit{Gaia}, with IDs 1963955610938800896 and 1963954236549278592, respectively \citep{refId0}. According to their machine learning algorithm \citep{2022gdr3.reptE..10R}, the \textit{Gaia} collaboration classifies both these variables as ``eclipsing binaries of beta Persei type" \citep{Chambliss_1992}. It is interesting to note that the same \textit{Gaia} class is assigned to the variable optical counterparts of known spiders such as the RB PSR~J0212$+$5321 \citep{Perez_2023}.
Given their respective geometric parallaxes of $0.50 \pm 0.11 \ \mathrm{mas}$ and $0.11 \pm 0.13 \ \mathrm{mas}$, by using the distance prior from \cite{2021AJ....161..147B} we obtain the distance estimates of $D=2.2_{-0.4}^{+0.6} \ \mathrm{kpc}$ and $D=4.5_{-1.2}^{+1.5} \ \mathrm{kpc}$ for J2117-A and J2117-B, respectively.

The latest catalogs from \textit{Chandra}, \textit{Swift}, \textit{XMM-Newton} or \textit{eROSITA} did not provide any X-ray counterparts to our two candidates. \textit{Swift}/XRT observed this field between 2010 September 24 and 2019 October 2 for a total exposure of $3.8 \ \mathrm{ks}$. Here we used the same tools and method adopted in Section \ref{subsec:J0737disc} to derive the upper limits to the X-ray fluxes of J2117-A and J2117-B, assuming a photon index $\Gamma=1.2$ and a hydrogen column density $N_{\mathrm{H}}=1.6\times10^{21} \ \mathrm{cm}^{-2}$. We obtained as $3\sigma$ upper limits to the counts of these two sources $4.1\times10^{-3} \ \mathrm{ct} \ \mathrm{s}^{-1}$ and $2.7\times10^{-3} \ \mathrm{ct} \ \mathrm{s}^{-1}$, which correspond to the upper limits to the unabsorbed energy flux of $2.8\times10^{-13}\ \mathrm{erg} \ \mathrm{cm}^{-2} \ \mathrm{s}^{-1}$ and $1.8\times10^{-13}\ \mathrm{erg} \ \mathrm{cm}^{-2} \ \mathrm{s}^{-1}$ (0.3--10 keV). Using the geometric distances previously estimated for J2117-A and J2117-B, we found as upper limits to the X-ray luminosity $1.6\times10^{32} \ (D/2.2 \ \mathrm{kpc})^2 \ \mathrm{erg} \ \mathrm{s}^{-1}$ and $4.3\times10^{32} \ (D/4.5 \ \mathrm{kpc})^2 \ \mathrm{erg} \ \mathrm{s}^{-1}$, respectively. For comparison, the X-ray luminosity distribution of Galactic RB MSPs peaks at a lower luminosity of $8\times10^{31} \ \mathrm{erg} \ \mathrm{s}^{-1}$ \citep{2023MNRAS.525.3963K}. Therefore, our upper limits do not exclude the presence of the X-ray counterparts to these two RB candidates.

As shown in Figure \ref{fig:J2117_FoV}, both J2117-A and J2117-B match the $\gamma$-ray pulsar-like source 3FGL~J2117.6+3725 and also the close 4FGL~J2117.9+3729, strengthening their spider association. The 0.1--100 GeV energy flux of the 4FGL object is $(4.3\pm1.0)\times10^{-12}\ \mathrm{erg} \ \mathrm{cm}^{-2} \ \mathrm{s}^{-1}$, from which we obtained the respective $\gamma$-ray luminosities of $L_{\gamma}=2.5\times10^{33} \ (D/2.2 \ \mathrm{kpc})^2 \ \mathrm{erg} \ \mathrm{s}^{-1}$ and $L_{\gamma}=1.0\times10^{34} \ (D/4.5 \ \mathrm{kpc})^2 \ \mathrm{erg} \ \mathrm{s}^{-1}$. Both these estimates are compatible with the typical MSPs luminosities $\sim10^{32}$--$10^{34} \ \mathrm{erg} \ \mathrm{s}^{-1}$ measured in $\gamma$-ray wavelengths \citep{2023ApJ...958..191S}.

Furthermore, we identified an infrared counterpart to J2117-A, with ID \textit{2MASS} J21175634$+$3726451 \citep{2006AJ....131.1163S} and magnitudes $J=16.37 \pm 0.13$, $H=15.80 \pm 0.15$, and $K=15.65$\footnote{The uncertainty on the $K$ magnitude is not reported from the \textit{2MASS} catalog in this case due to low quality photometry in this band.}. For the other candidate J2117-B instead we did not find any infrared counterpart from \textit{2MASS} or \textit{AllWISE}.

\subsection{3FGL~J2221.6+6507} \label{subsec:J2221disc}
The folded optical light curves of J2221, reported in Figure \ref{fig:J2221_var2}, show a periodic shape with a sharp flux minimum. Also, the ($g'-r'$) and ($r'-i'$) colors seem variable, peaking within the broad light curve maximum. These color curves indicate temperature variations along the photometric period $P_{\mathrm{best}}=0.165 \pm 0.004 \ \mathrm{d}$ and a mean temperature of $T_{\mathrm{eff}}=5850 \pm 350 \ \mathrm{K}$.

The properties listed above make J2221 a potential RB candidate. Its light curve shape and peak-to-peak amplitude of $0.6 \ \mathrm{mag}$ suggest a mildly-irradiated companion star. Therefore, the optical light curves have been folded with $P_{\mathrm{orb}}=P_{\mathrm{best}}=0.165 \pm 0.004 \ \mathrm{d}$, the same value as the photometric period, and exhibit one flux maximum at phase $\phi=0.5$. From Figure \ref{fig:J2221_var2}, we can notice a complex variable trend in the color curves showing at least two local maxima (at $\phi=0.5$ and $\phi=0.75$), which is inconsistent with a direct isotropic heating of the companion's inner face. This, combined with the light maximum flatness and asymmetry with respect to the superior conjunction ($\phi=0.5$), may suggest hot spots or asymmetric heating for J2221 companion. Similar behaviors have been found and discussed in previous works for other RBs (e.g. \citealt{2016ApJ...823..105D,2016ApJ...828....7R, 2019ApJ...876....8S, 2021MNRAS.507.2174S}).
We obtain for J2221 a base temperature $T_{\mathrm{b}}=5500\pm300 \ \mathrm{K}$ and a day-side temperature $T_{\mathrm{day}}=6000\pm350 \ \mathrm{K}$ from the ($g'-r'$) colors.

Our candidate is coincident with the variable \textit{Gaia}-DR3 source 2206544469143391872, again ranked in the same category of eclipsing binary as J2117-B from \cite{2022gdr3.reptE..10R} (see Section \ref{subsec:J2117disc} for details). Using the distance prior distribution from \cite{2021AJ....161..147B}, we convert the parallax measurement $0.46\pm0.10 \ \mathrm{mas}$ of J2221 to a geometric distance of $D=2.4_{-0.5}^{+0.6} \ \mathrm{kpc}$.

No X-ray counterparts or archival observations were found for J2221 from \textit{Chandra}, \textit{Swift}, \textit{XMM-Newton} or \textit{eROSITA}. This optical variable is identified inside the 95\% error ellipse of the pulsar-like \textit{Fermi} unidentified source 3FGL~J2221.6+6507, which has a $\gamma$-ray energy flux of $(9.3\pm1.7)\times10^{-12}\ \mathrm{erg} \ \mathrm{cm}^{-2} \ \mathrm{s}^{-1}$ (0.1--100 GeV). Using the previous distance estimate as reference, we derived a luminosity of $L_{\gamma}=6.1\times10^{33} \ (D/2.4 \ \mathrm{kpc})^2 \ \mathrm{erg} \ \mathrm{s}^{-1}$, which is compatible with the range of $\gamma$-ray luminosities for MSPs \citep{2023ApJ...958..191S}. Although this 3FGL object has a detection significance of $7.4\sigma$, J2221 is missing a 4FGL association, and the closest source 4FGL~J2227.1+6455 is located $30'$ away from our optical location. We deem this suspect detection of 3FGL~J2221.6+6507 to be caused by contamination from the Galactic diffuse emission, as this source is quite close to the Galactic plane (Galactic latitude $b=6.7^{\circ}$). Indeed, the model of $\gamma$-ray diffuse background used in the \textit{Fermi}-3FGL catalog represents the main factor of uncertainties in detections nearby the Galactic plane. This is especially the case for faint sources with spectral energy flux densities peaking at low energies $<1 \ \mathrm{GeV}$ \citep{2015ApJS..218...23A}, like J2221, which has a peak emission of $3.4\times10^{-12}\ \mathrm{erg} \ \mathrm{cm}^{-2} \ \mathrm{s}^{-1}$ in the 0.3--1 GeV band, dropping down to $1.8\times10^{-12}\ \mathrm{erg} \ \mathrm{cm}^{-2} \ \mathrm{s}^{-1}$ at $>1 \ \mathrm{GeV}$. The lack of this source from the 4FGL catalog is likely due to the global improvement of the diffuse emission model with respect to 3FGL. This could undermine the reliability of J2221 as a spider candidate, but we note that a few confirmed spiders lack 4FGL counterparts (e.g. the RB PSR~J1723$-$2837 or the BW PSR~J1720$-$0533, \citealt{lindseth2023gamma}).

For completeness, we also report the infrared counterpart to our optical variable, identified as \textit{2MASS} 22223274$+$6500207 by \cite{2006AJ....131.1163S} with magnitudes $J=15.49 \pm 0.06$, $H=14.79 \pm 0.07$, and $K=14.56 \pm 0.09$.

\subsection{The MSP candidate 3FGL~J1119.9$-$2204 and the confirmed RB 3FGL~J2039.9$-$5618} \label{subsec:J1119andJ2039disc}
While this work was in progress, MSPs were identified in two Fermi sources observed in our survey, 3FGL~J1119.9$-$2204 \citep{2022ApJ...926..201S} and 3FGL~J2039.9$-$5618 \citep{2015ApJ...812L..24R, 2015ApJ...814...88S}. In the following, we briefly discuss why we do not detect any periodic variable in these two fields.

\cite{2022ApJ...926..201S} identified the likely optical and X-ray counterparts to 3FGL~J1119.9$-$2204, matching also its 4FGL association 4FGL~J1120.0$-$2204. Through optical spectroscopy and radial velocity modeling, they classified this system as a X-ray compact binary, containing a $\approx0.17 \ \mathrm{M}_{\sun}$ companion star at a temperature of $8650 \ \mathrm{K}$, an orbital period of $8.7 \ \mathrm{h}$, and a low inclination of $\sim16^{\circ}$--$19^{\circ}$. Given these parameter estimates and after binary evolution simulations, \cite{2022ApJ...926..201S} argue that 4FGL~J1120.0$-$2204 is in an intermediate stage of evolving into an MSP binary with an extremely low-mass white dwarf companion. They also attribute the lack of orbital variability in the optical band to the nearly face-on orientation, the companion underfilling its Roche lobe, and the absence of heating from the pulsar wind. Their interpretation is in agreement with the non-detection of this source as an optical variable in our survey. However, we still detect it as a stable source in all three optical bands, with mean magnitudes of $g'=15.46\pm0.02$, $r'=15.65\pm0.02$, and $i'=15.73\pm0.02$. The detection of a radio or $\gamma$-ray MSP from the same location would confirm the scenario proposed by \cite{2022ApJ...926..201S}.

3FGL~J2039.9$-$5618 was identified as a RB candidate by \cite{2015ApJ...812L..24R} and \cite{2015ApJ...814...88S}, who found the X-ray and optical variable counterparts to this Fermi source. \cite{2021MNRAS.502..915C} later confirmed this system as a $\gamma$-ray MSP. Its optical light curve shows ellipsoidal modulation with an orbital period of $0.22748\pm0.00043 \ \mathrm{d}$ \citep{2015ApJ...814...88S} and asymmetry between the two peaks, indicating some residual companion heating. \cite{2015ApJ...814...88S} measured mean magnitudes of $g'=19.40\pm0.02$, $r'=18.71\pm0.02$ and $i'=18.59\pm0.02$ for this RB, just above the sensitivity limit of $r'\simeq19$ attained by the 0.4-m LCO telescope used for our observations. Indeed, we marginally detect this source in the $r'$-band (at $r'=18.5\pm0.1$), resulting in scarce and low quality photometry with which we cannot assess its variability.

\section{Conclusions} \label{sec:conclusions}
The COBIPULSE survey is aimed at finding new spider MSPs through their variable optical counterparts. To pursue this goal, we performed multi-band optical observations of 33 promising \textit{Fermi}-3FGL candidates, selected based on their pulsar-like $\gamma$-ray properties. 
This systematic search led to the discovery of four spider candidates, matching 3FGL~J0737.2$-$3233/4FGL~J0736.9$-$3231 (J0737), 3FGL~J2117.6$+$3725/4FGL~J2117.9$+$3729 (J2117-A and J2117-B), and 3FGL~J2221.6$+$6507 (J2221). They all show flux modulations in the sub-day period range, with peak-to-peak amplitudes $\gtrsim0.3 \ \mathrm{mag}$. We classified all four as likely RB MSP candidates, based on their companion temperatures of $5000$--$6000 \ \mathrm{K}$.

We argue that J0737 is a RB candidate system, with a mean companion temperature of $6100\pm200 \ \mathrm{K}$ and an orbital period of $0.3548 \pm 0.0005 \ \mathrm{d}$. Its variable color curves peaking at the same phase $\phi=0.5$ as the light maxima, indicate that the companion star is likely irradiated by the pulsar wind. The small-amplitude modulation of $0.2$--$0.3 \ \mathrm{mag}$ shown by the J0737 light curves can be explained by a low orbital inclination. 
If confirmed as a radio or $\gamma$-ray MSP, this would be the closest known spider to Earth, with a distance of $659_{-20}^{+16} \ \mathrm{pc}$ estimated from the \textit{Gaia} parallax measurement. Also, given our $3\sigma$ upper limit on the X-ray luminosity of $5.4\times10^{30} \ (D/0.7 \ \mathrm{kpc})^2 \ \mathrm{erg} \ \mathrm{s}^{-1}$ (0.3--10 keV), J0737 would be one of the least luminous RBs in X-rays.
%

We identify two optical variables in the field of 3FGL~J2117.6$+$3725 (J2117-A and J2117-B), both exhibiting peak-to-peak amplitudes of $0.3$--$0.4 \ \mathrm{mag}$ and constant colors, indicative of small temperature differences of the companion star along the orbit. 
Thus, we classify both these sources as non-irradiated RB candidates. We estimate for J2117-A and J2117-B respective companion temperatures of $4950 \pm 250 \ \mathrm{K}$ and $6300 \pm 600 \ \mathrm{K}$, and infer orbital periods of $0.25328 \pm 0.00006 \ \mathrm{d}$ and $0.441961 \pm 0.000002 \ \mathrm{d}$.
Archival X-ray observations only allow us to place shallow upper limits to the X-ray luminosity of $1.6\times10^{32} \ (D/2.2 \ \mathrm{kpc})^2 \ \mathrm{erg} \ \mathrm{s}^{-1}$ and $4.3\times10^{32} \ (D/4.5 \ \mathrm{kpc})^2 \ \mathrm{erg} \ \mathrm{s}^{-1}$ for J2117-A and J2117-B, respectively, fully compatible with the presence of an X-ray counterpart not detected yet. 

We deem J2221 as a RB MSP candidate, with a mean companion temperature of $5850\pm350 \ \mathrm{K}$ and an orbital of period $0.165 \pm 0.004 \ \mathrm{d}$. Its variable optical colors peaking within the broad light curve maximum and amplitude of $0.6 \ \mathrm{mag}$ suggest it to be in a mildly-irradiated regime. We did not find any archival X-ray observations for J2221.

Our work significantly impacts the expansion of the spider population by providing precise sky locations for targeted radio and $\gamma$-ray follow-up of four RB candidates, thereby facilitating their detection as MSPs. This is particularly relevant for J0737, J2117-A, and J2117-B, which we consider to be the strongest of the four candidates we discovered, as discussed earlier. The orbital periods we have estimated will also aid in the accelerated pulsation searches from these systems by narrowing down the parameter space. Phase resolved optical spectroscopy will help confirming the nature of these sources and determine their fundamental parameters.  
\section*{acknowledgments}
This project has received funding from the European Research Council (ERC) under the European Union’s Horizon 2020 research and innovation programme (grant agreement No. 101002352, PI: M. Linares). This work was supported by the Agencia Estatal de Investigación del Ministerio de Ciencia e Innovación (MCIN/AEI) and the European Regional Development Fund (ERDF) under grant PID2021--124879NB--I00. We thank M. Kennedy for discussions on periodicity search methods and for suggesting the use of the ZTF public data release to extend our dataset. We also thank M. Satybaldiev for his assistance with the method for estimating periods uncertainties. We acknowledge the \textsc{astrosource} software developers, in particular M. Fitzgerald for a discussion on ensemble photometry. JC acknowledges support by the Spanish Ministry of Science via the Plan de Generacion de conocimiento through grant PID2022-143331NB-100. P.A.M.P. acknowledges support from grant RYC2021-031173-I funded by MCIN/AEI/ 10.13039/501100011033 and by the 'European Union NextGenerationEU/PRTR’. This article makes use of observations made in the Observatorios de Canarias del IAC with the STELLA telescope operated on the island of Tenerife by the Leibniz Institute for Astrophysics Potsdam (AIP) in the Observatorio del Teide, and the Isaac Newton Telescope (INT) operated on the island of La Palma by the Isaac Newton Group of Telescopes (ING) in the Observatorio del Roque de los Muchachos. This work is also based on observations from the Las Cumbres Observatory global telescope network, which were performed using the Sinistro camera at the 1-meter LCOGT network and SBIG camera mounted on the 0.4-meter autonomous telescopes at the LCOGT southern sites (Cerro Tololo-CHI, Siding Spring-AUS). This research has made use of the NASA/IPAC Infrared Science Archive, which is funded by the National Aeronautics and Space Administration and operated by the California Institute of Technology. This work has also made use of data from the Zwicky Transient Facility (ZTF). ZTF is supported by the National Science Foundation under Grants No. AST-1440341 and AST-2034437 and a collaboration including current partners Caltech, IPAC, the Weizmann Institute for Science, the Oskar Klein Center at Stockholm University, the University of Maryland, Deutsches Elektronen-Synchrotron and Humboldt University, the TANGO Consortium of Taiwan, the University of Wisconsin at Milwaukee, Trinity College Dublin, Lawrence Livermore National Laboratories, IN2P3, University of Warwick, Ruhr University Bochum, Northwestern University and former partners the University of Washington, Los Alamos National Laboratories, and Lawrence Berkeley National Laboratories. Operations are conducted by COO, IPAC, and UW.. ZTF is supported by the National Science Foundation under Grants No. AST-1440341 and AST-2034437 and a collaboration including current partners Caltech, IPAC, the Weizmann Institute for Science, the Oskar Klein Center at Stockholm University, the University of Maryland, Deutsches Elektronen-Synchrotron and Humboldt University, the TANGO Consortium of Taiwan, the University of Wisconsin at Milwaukee, Trinity College Dublin, Lawrence Livermore National Laboratories, IN2P3, University of Warwick, Ruhr University Bochum, Northwestern University and former partners the University of Washington, Los Alamos National Laboratories, and Lawrence Berkeley National Laboratories. Operations are conducted by COO, IPAC, and UW. This publication makes use of data products from the Two Micron All Sky Survey, which is a joint project of the University of Massachusetts and the Infrared Processing and Analysis Center/California Institute of Technology, funded by the National Aeronautics and Space Administration and the National Science Foundation. This publication makes use of data products from the Wide-field Infrared Survey Explorer, which is a joint project of the University of California, Los Angeles, and the Jet Propulsion Laboratory/California Institute of Technology, funded by the National Aeronautics and Space Administration. This research has made use of data products from the Pan-STARRS1 Surveys (PS1), which have been made possible through contributions by the Institute for Astronomy at the University of Hawaii, the Pan-STARRS Project Office, the Max-Planck Society and its participating institutes. This work has made use of data products from the European Space Agency (ESA) mission Gaia (\url{https://www.cosmos.esa.int/gaia}), processed by the Gaia Data Processing and Analysis Consortium (DPAC, \url{https://www.cosmos.esa.int/web/gaia/dpac/consortium}). Funding for the DPAC has been provided by national institutions, in particular the institutions participating in the Gaia Multilateral Agreement. This research has made use of data products from the Catalina Sky Survey (CSS), specifically the Catalina Sky Survey Periodic Variables Catalog. NNG05GF22G issued through the Science Mission Directorate Near-Earth Objects Observations Program. This work also used data products from the ATLAS project. The ATLAS project is primarily funded by NASA under grant number NN12AR55G, and it is operated by the Institute for Astronomy at the University of Hawaii. We also used data and/or software provided by the High Energy Astrophysics Science Archive Research Center, which is a service of the Astrophysics Science Division at NASA and Goddard Space Flight Center. This work made use of data supplied by the UK Swift Science Data Centre at the University of Leicester. This research used also data from the 4XMM-DR13s catalog, created from observations obtained with XMM-Newton, an ESA science mission with instruments and contributions directly funded by ESA Member States and NASA.

\section*{data availability}
The raw STELLA, INT and LCO images with bias and flats frames used for data reduction can be obtained by contacting M. Turchetta. ZTF data are public and can be obtained through the IRSA light curve service \url{https://irsa.ipac.caltech.edu/docs/program_interface/ztf_lightcurve_api.html}.

%
\vspace{10mm}
\facilities{STELLA:1.5m/WiFSIP, INT:2.5m/WFC, LCO:1m/Sinistro, LCO:0.4m/SBIG, IRSA, PO:1.2m}

\software{\textsc{IRAF} \citep{1986SPIE..627..733T},
          \textsc{SEP} \citep{2016zndo....159035B}, 
          Source Extractor \citep{1996A&AS..117..393B},
          \textsc{astrosource} \citep{2021JOSS....6.2641F}
          }
\bibliography{sample631}{}

\begin{thebibliography}{}
\expandafter\ifx\csname natexlab\endcsname\relax\def\natexlab#1{#1}\fi
\providecommand{\url}[1]{\href{#1}{#1}}
\providecommand{\dodoi}[1]{doi:~\href{http://doi.org/#1}{\nolinkurl{#1}}}
\providecommand{\doeprint}[1]{\href{http://ascl.net/#1}{\nolinkurl{http://ascl.net/#1}}}
\providecommand{\doarXiv}[1]{\href{https://arxiv.org/abs/#1}{\nolinkurl{https://arxiv.org/abs/#1}}}

\bibitem[{{Acero} {et~al.}(2015){Acero}, {Ackermann}, {Ajello}, {Albert}, {Atwood}, {Axelsson}, {Baldini}, {Ballet}, {Barbiellini}, {Bastieri}, {Belfiore}, {Bellazzini}, {Bissaldi}, {Blandford}, {Bloom}, {Bogart}, {Bonino}, {Bottacini}, {Bregeon}, {Britto}, {Bruel}, {Buehler}, {Burnett}, {Buson}, {Caliandro}, {Cameron}, {Caputo}, {Caragiulo}, {Caraveo}, {Casandjian}, {Cavazzuti}, {Charles}, {Chaves}, {Chekhtman}, {Cheung}, {Chiang}, {Chiaro}, {Ciprini}, {Claus}, {Cohen-Tanugi}, {Cominsky}, {Conrad}, {Cutini}, {D'Ammando}, {de Angelis}, {DeKlotz}, {de Palma}, {Desiante}, {Digel}, {Di Venere}, {Drell}, {Dubois}, {Dumora}, {Favuzzi}, {Fegan}, {Ferrara}, {Finke}, {Franckowiak}, {Fukazawa}, {Funk}, {Fusco}, {Gargano}, {Gasparrini}, {Giebels}, {Giglietto}, {Giommi}, {Giordano}, {Giroletti}, {Glanzman}, {Godfrey}, {Grenier}, {Grondin}, {Grove}, {Guillemot}, {Guiriec}, {Hadasch}, {Harding}, {Hays}, {Hewitt}, {Hill}, {Horan}, {Iafrate}, {Jogler}, {J{\'o}hannesson}, {Johnson}, {Johnson}, {Johnson}, {Johnson}, {Kamae},
  {Kataoka}, {Katsuta}, {Kuss}, {La Mura}, {Landriu}, {Larsson}, {Latronico}, {Lemoine-Goumard}, {Li}, {Li}, {Longo}, {Loparco}, {Lott}, {Lovellette}, {Lubrano}, {Madejski}, {Massaro}, {Mayer}, {Mazziotta}, {McEnery}, {Michelson}, {Mirabal}, {Mizuno}, {Moiseev}, {Mongelli}, {Monzani}, {Morselli}, {Moskalenko}, {Murgia}, {Nuss}, {Ohno}, {Ohsugi}, {Omodei}, {Orienti}, {Orlando}, {Ormes}, {Paneque}, {Panetta}, {Perkins}, {Pesce-Rollins}, {Piron}, {Pivato}, {Porter}, {Racusin}, {Rando}, {Razzano}, {Razzaque}, {Reimer}, {Reimer}, {Reposeur}, {Rochester}, {Romani}, {Salvetti}, {S{\'a}nchez-Conde}, {Saz Parkinson}, {Schulz}, {Siskind}, {Smith}, {Spada}, {Spandre}, {Spinelli}, {Stephens}, {Strong}, {Suson}, {Takahashi}, {Takahashi}, {Tanaka}, {Thayer}, {Thayer}, {Thompson}, {Tibaldo}, {Tibolla}, {Torres}, {Torresi}, {Tosti}, {Troja}, {Van Klaveren}, {Vianello}, {Winer}, {Wood}, {Wood}, {Zimmer}, \& {Fermi-LAT Collaboration}}]{2015ApJS..218...23A}
{Acero}, F., {Ackermann}, M., {Ajello}, M., {et~al.} 2015, \apjs, 218, 23, \dodoi{10.1088/0067-0049/218/2/23}

\bibitem[{Ackermann {et~al.}(2012)Ackermann, Ajello, Allafort, Antolini, Baldini, Ballet, Barbiellini, Bastieri, Bellazzini, Berenji, Blandford, Bloom, Bonamente, Borgland, Bouvier, Brandt, Bregeon, Brigida, Bruel, Buehler, Burnett, Buson, Caliandro, Cameron, Caraveo, Casandjian, Cavazzuti, Cecchi, Çelik, Charles, Chekhtman, Chen, Cheung, Chiang, Ciprini, Claus, Cohen-Tanugi, Conrad, Cutini, de~Angelis, DeCesar, Luca, de~Palma, Dermer, do~e Silva, Drell, Drlica-Wagner, Dubois, Enoto, Favuzzi, Fegan, Ferrara, Focke, Fortin, Fukazawa, Funk, Fusco, Gargano, Gasparrini, Gehrels, Germani, Giglietto, Giordano, Giroletti, Glanzman, Godfrey, Grenier, Grondin, Grove, Guillemot, Guiriec, Gustafsson, Hadasch, Hanabata, Harding, Hayashida, Hays, Healey, Hill, Horan, Hou, Jóhannesson, Johnson, Johnson, Kamae, Katagiri, Kataoka, Kerr, Knödlseder, Kuss, Lande, Latronico, Lee, Lemoine-Goumard, Longo, Loparco, Lott, Lovellette, Lubrano, Madejski, Mazziotta, McEnery, Mehault, Michelson, Mignani, Mitthumsiri, Mizuno, Monte,
  Monzani, Morselli, Moskalenko, Murgia, Nakamori, Naumann-Godo, Nolan, Norris, Nuss, Ohsugi, Okumura, Omodei, Orlando, Ormes, Ozaki, Paneque, Panetta, Parent, Pelassa, Pesce-Rollins, Pierbattista, Piron, Pivato, Porter, Rainò, Rando, Ray, Razzano, Reimer, Reimer, Reposeur, Romani, Sadrozinski, Salvetti, Parkinson, Schalk, Sgrò, Shaw, Siskind, Smith, Spandre, Spinelli, Suson, Takahashi, Tanaka, Thayer, Thayer, Thompson, Tibaldo, Tibolla, Torres, Tosti, Tramacere, Troja, Usher, Vandenbroucke, Vasileiou, Vianello, Vilchez, Vitale, Waite, Wallace, Wang, Winer, Wolff, Wood, Wood, Yang, \& Zimmer}]{Ackermann_2012}
Ackermann, M., Ajello, M., Allafort, A., {et~al.} 2012, The Astrophysical Journal, 753, 83, \dodoi{10.1088/0004-637X/753/1/83}

\bibitem[{{Allard} {et~al.}(2011){Allard}, {Homeier}, \& {Freytag}}]{allard11}
{Allard}, F., {Homeier}, D., \& {Freytag}, B. 2011, in Astronomical Society of the Pacific Conference Series, Vol. 448, 16th Cambridge Workshop on Cool Stars, Stellar Systems, and the Sun, ed. C.~{Johns-Krull}, M.~K. {Browning}, \& A.~A. {West}, 91, \dodoi{10.48550/arXiv.1011.5405}

\bibitem[{{Archibald} {et~al.}(2009){Archibald}, {Stairs}, {Ransom}, {Kaspi}, {Kondratiev}, {Lorimer}, {McLaughlin}, {Boyles}, {Hessels}, {Lynch}, {van Leeuwen}, {Roberts}, {Jenet}, {Champion}, {Rosen}, {Barlow}, {Dunlap}, \& {Remillard}}]{2009Sci...324.1411A}
{Archibald}, A.~M., {Stairs}, I.~H., {Ransom}, S.~M., {et~al.} 2009, Science, 324, 1411, \dodoi{10.1126/science.1172740}

\bibitem[{{Au} {et~al.}(2023){Au}, {Strader}, {Swihart}, {Lin}, {Kong}, {Takata}, {Hui}, {Panurach}, {Molina}, {Aydi}, {Sokolovsky}, \& {Li}}]{2023ApJ...943..103A}
{Au}, K.-Y., {Strader}, J., {Swihart}, S.~J., {et~al.} 2023, \apj, 943, 103, \dodoi{10.3847/1538-4357/acae8a}

\bibitem[{{Bailer-Jones} {et~al.}(2021){Bailer-Jones}, {Rybizki}, {Fouesneau}, {Demleitner}, \& {Andrae}}]{2021AJ....161..147B}
{Bailer-Jones}, C.~A.~L., {Rybizki}, J., {Fouesneau}, M., {Demleitner}, M., \& {Andrae}, R. 2021, \aj, 161, 147, \dodoi{10.3847/1538-3881/abd806}

\bibitem[{{Ballet} {et~al.}(2023){Ballet}, {Bruel}, {Burnett}, {Lott}, \& {The Fermi-LAT collaboration}}]{2023arXiv230712546B}
{Ballet}, J., {Bruel}, P., {Burnett}, T.~H., {Lott}, B., \& {The Fermi-LAT collaboration}. 2023, arXiv e-prints, arXiv:2307.12546, \dodoi{10.48550/arXiv.2307.12546}

\bibitem[{{Barbary} {et~al.}(2016){Barbary}, {Boone}, {McCully}, {Craig}, {Deil}, \& {Rose}}]{2016zndo....159035B}
{Barbary}, K., {Boone}, K., {McCully}, C., {et~al.} 2016, {kbarbary/sep: v1.0.0}, v1.0.0,  Zenodo, \dodoi{10.5281/zenodo.159035}

\bibitem[{{Bassa} {et~al.}(2014){Bassa}, {Patruno}, {Hessels}, {Keane}, {Monard}, {Mahony}, {Bogdanov}, {Corbel}, {Edwards}, {Archibald}, {Janssen}, {Stappers}, \& {Tendulkar}}]{2014MNRAS.441.1825B}
{Bassa}, C.~G., {Patruno}, A., {Hessels}, J.~W.~T., {et~al.} 2014, \mnras, 441, 1825, \dodoi{10.1093/mnras/stu708}

\bibitem[{{Bellm} {et~al.}(2016){Bellm}, {Kaplan}, {Breton}, {Phinney}, {Bhalerao}, {Camilo}, {Dahal}, {Djorgovski}, {Drake}, {Hessels}, {Laher}, {Levitan}, {Lewis}, {Mahabal}, {Ofek}, {Prince}, {Ransom}, {Roberts}, {Russell}, {Sesar}, {Surace}, \& {Tang}}]{2016ApJ...816...74B}
{Bellm}, E.~C., {Kaplan}, D.~L., {Breton}, R.~P., {et~al.} 2016, \apj, 816, 74, \dodoi{10.3847/0004-637X/816/2/74}

\bibitem[{{Bellm} {et~al.}(2019){Bellm}, {Kulkarni}, {Graham}, {Dekany}, {Smith}, {Riddle}, {Masci}, {Helou}, {Prince}, {Adams}, {Barbarino}, {Barlow}, {Bauer}, {Beck}, {Belicki}, {Biswas}, {Blagorodnova}, {Bodewits}, {Bolin}, {Brinnel}, {Brooke}, {Bue}, {Bulla}, {Burruss}, {Cenko}, {Chang}, {Connolly}, {Coughlin}, {Cromer}, {Cunningham}, {De}, {Delacroix}, {Desai}, {Duev}, {Eadie}, {Farnham}, {Feeney}, {Feindt}, {Flynn}, {Franckowiak}, {Frederick}, {Fremling}, {Gal-Yam}, {Gezari}, {Giomi}, {Goldstein}, {Golkhou}, {Goobar}, {Groom}, {Hacopians}, {Hale}, {Henning}, {Ho}, {Hover}, {Howell}, {Hung}, {Huppenkothen}, {Imel}, {Ip}, {Ivezi{\'c}}, {Jackson}, {Jones}, {Juric}, {Kasliwal}, {Kaspi}, {Kaye}, {Kelley}, {Kowalski}, {Kramer}, {Kupfer}, {Landry}, {Laher}, {Lee}, {Lin}, {Lin}, {Lunnan}, {Giomi}, {Mahabal}, {Mao}, {Miller}, {Monkewitz}, {Murphy}, {Ngeow}, {Nordin}, {Nugent}, {Ofek}, {Patterson}, {Penprase}, {Porter}, {Rauch}, {Rebbapragada}, {Reiley}, {Rigault}, {Rodriguez}, {van Roestel}, {Rusholme}, {van
  Santen}, {Schulze}, {Shupe}, {Singer}, {Soumagnac}, {Stein}, {Surace}, {Sollerman}, {Szkody}, {Taddia}, {Terek}, {Van Sistine}, {van Velzen}, {Vestrand}, {Walters}, {Ward}, {Ye}, {Yu}, {Yan}, \& {Zolkower}}]{2019PASP..131a8002B}
{Bellm}, E.~C., {Kulkarni}, S.~R., {Graham}, M.~J., {et~al.} 2019, \pasp, 131, 018002, \dodoi{10.1088/1538-3873/aaecbe}

\bibitem[{{Bertin} \& {Arnouts}(1996)}]{1996A&AS..117..393B}
{Bertin}, E., \& {Arnouts}, S. 1996, \aaps, 117, 393, \dodoi{10.1051/aas:1996164}

\bibitem[{{Bhattacharya} \& {van den Heuvel}(1991)}]{1991PhR...203....1B}
{Bhattacharya}, D., \& {van den Heuvel}, E.~P.~J. 1991, \physrep, 203, 1, \dodoi{10.1016/0370-1573(91)90064-S}

\bibitem[{{Brown} {et~al.}(2013){Brown}, {Baliber}, {Bianco}, {Bowman}, {Burleson}, {Conway}, {Crellin}, {Depagne}, {De Vera}, {Dilday}, {Dragomir}, {Dubberley}, {Eastman}, {Elphick}, {Falarski}, {Foale}, {Ford}, {Fulton}, {Garza}, {Gomez}, {Graham}, {Greene}, {Haldeman}, {Hawkins}, {Haworth}, {Haynes}, {Hidas}, {Hjelstrom}, {Howell}, {Hygelund}, {Lister}, {Lobdill}, {Martinez}, {Mullins}, {Norbury}, {Parrent}, {Paulson}, {Petry}, {Pickles}, {Posner}, {Rosing}, {Ross}, {Sand}, {Saunders}, {Shobbrook}, {Shporer}, {Street}, {Thomas}, {Tsapras}, {Tufts}, {Valenti}, {Vander Horst}, {Walker}, {White}, \& {Willis}}]{2013PASP..125.1031B}
{Brown}, T.~M., {Baliber}, N., {Bianco}, F.~B., {et~al.} 2013, \pasp, 125, 1031, \dodoi{10.1086/673168}

\bibitem[{{Chambers} {et~al.}(2016){Chambers}, {Magnier}, {Metcalfe}, {Flewelling}, {Huber}, {Waters}, {Denneau}, {Draper}, {Farrow}, {Finkbeiner}, {Holmberg}, {Koppenhoefer}, {Price}, {Rest}, {Saglia}, {Schlafly}, {Smartt}, {Sweeney}, {Wainscoat}, {Burgett}, {Chastel}, {Grav}, {Heasley}, {Hodapp}, {Jedicke}, {Kaiser}, {Kudritzki}, {Luppino}, {Lupton}, {Monet}, {Morgan}, {Onaka}, {Shiao}, {Stubbs}, {Tonry}, {White}, {Ba{\~n}ados}, {Bell}, {Bender}, {Bernard}, {Boegner}, {Boffi}, {Botticella}, {Calamida}, {Casertano}, {Chen}, {Chen}, {Cole}, {Deacon}, {Frenk}, {Fitzsimmons}, {Gezari}, {Gibbs}, {Goessl}, {Goggia}, {Gourgue}, {Goldman}, {Grant}, {Grebel}, {Hambly}, {Hasinger}, {Heavens}, {Heckman}, {Henderson}, {Henning}, {Holman}, {Hopp}, {Ip}, {Isani}, {Jackson}, {Keyes}, {Koekemoer}, {Kotak}, {Le}, {Liska}, {Long}, {Lucey}, {Liu}, {Martin}, {Masci}, {McLean}, {Mindel}, {Misra}, {Morganson}, {Murphy}, {Obaika}, {Narayan}, {Nieto-Santisteban}, {Norberg}, {Peacock}, {Pier}, {Postman}, {Primak}, {Rae}, {Rai},
  {Riess}, {Riffeser}, {Rix}, {R{\"o}ser}, {Russel}, {Rutz}, {Schilbach}, {Schultz}, {Scolnic}, {Strolger}, {Szalay}, {Seitz}, {Small}, {Smith}, {Soderblom}, {Taylor}, {Thomson}, {Taylor}, {Thakar}, {Thiel}, {Thilker}, {Unger}, {Urata}, {Valenti}, {Wagner}, {Walder}, {Walter}, {Watters}, {Werner}, {Wood-Vasey}, \& {Wyse}}]{2016arXiv161205560C}
{Chambers}, K.~C., {Magnier}, E.~A., {Metcalfe}, N., {et~al.} 2016, arXiv e-prints, arXiv:1612.05560, \dodoi{10.48550/arXiv.1612.05560}

\bibitem[{Chambliss(1992)}]{Chambliss_1992}
Chambliss, C.~R. 1992, Publications of the Astronomical Society of the Pacific, 104, 663, \dodoi{10.1086/133036}

\bibitem[{Chen {et~al.}(2020)Chen, Wang, Deng, de~Grijs, Yang, \& Tian}]{Chen_2020}
Chen, X., Wang, S., Deng, L., {et~al.} 2020, The Astrophysical Journal Supplement Series, 249, 18, \dodoi{10.3847/1538-4365/ab9cae}

\bibitem[{{Clark} {et~al.}(2021){Clark}, {Nieder}, {Voisin}, {Allen}, {Aulbert}, {Behnke}, {Breton}, {Choquet}, {Corongiu}, {Dhillon}, {Eggenstein}, {Fehrmann}, {Guillemot}, {Harding}, {Kennedy}, {Machenschalk}, {Marsh}, {Mata S{\'a}nchez}, {Mignani}, {Stringer}, {Wadiasingh}, \& {Wu}}]{2021MNRAS.502..915C}
{Clark}, C.~J., {Nieder}, L., {Voisin}, G., {et~al.} 2021, \mnras, 502, 915, \dodoi{10.1093/mnras/staa3484}

\bibitem[{{Condon} {et~al.}(1998){Condon}, {Cotton}, {Greisen}, {Yin}, {Perley}, {Taylor}, \& {Broderick}}]{1998AJ....115.1693C}
{Condon}, J.~J., {Cotton}, W.~D., {Greisen}, E.~W., {et~al.} 1998, \aj, 115, 1693, \dodoi{10.1086/300337}

\bibitem[{{D'Amico} {et~al.}(2001){D'Amico}, {Possenti}, {Manchester}, {Sarkissian}, {Lyne}, \& {Camilo}}]{2001ApJ...561L..89D}
{D'Amico}, N., {Possenti}, A., {Manchester}, R.~N., {et~al.} 2001, \apjl, 561, L89, \dodoi{10.1086/324562}

\bibitem[{{Deneva} {et~al.}(2016){Deneva}, {Ray}, {Camilo}, {Halpern}, {Wood}, {Cromartie}, {Ferrara}, {Kerr}, {Ransom}, {Wolff}, {Chambers}, \& {Magnier}}]{2016ApJ...823..105D}
{Deneva}, J.~S., {Ray}, P.~S., {Camilo}, F., {et~al.} 2016, \apj, 823, 105, \dodoi{10.3847/0004-637X/823/2/105}

\bibitem[{{Dodge} {et~al.}(2024){Dodge}, {Breton}, {Clark}, {Burgay}, {Strader}, {Au}, {Barr}, {Buchner}, {Dhillon}, {Ferrara}, {Freire}, {Griessmeier}, {Kennedy}, {Kramer}, {Li}, {Padmanabh}, {Phosrisom}, {Stappers}, {Swihart}, \& {Thongmeearkom}}]{2024MNRAS.528.4337D}
{Dodge}, O.~G., {Breton}, R.~P., {Clark}, C.~J., {et~al.} 2024, \mnras, 528, 4337, \dodoi{10.1093/mnras/stae211}

\bibitem[{Dodge {et~al.}(2024)Dodge, Breton, Clark, Burgay, Strader, Au, Barr, Buchner, Dhillon, Ferrara, Freire, Griessmeier, Kennedy, Kramer, Li, Padmanabh, Phosrisom, Stappers, Swihart, \& Thongmeearkom}]{10.1093/mnras/stae211}
Dodge, O.~G., Breton, R.~P., Clark, C.~J., {et~al.} 2024, Monthly Notices of the Royal Astronomical Society, 528, 4337, \dodoi{10.1093/mnras/stae211}

\bibitem[{Drake {et~al.}(2017)Drake, Djorgovski, Catelan, Graham, Mahabal, Larson, Christensen, Torrealba, Beshore, McNaught, Garradd, Belokurov, \& Koposov}]{catalinasouth_2017}
Drake, A.~J., Djorgovski, S.~G., Catelan, M., {et~al.} 2017, Monthly Notices of the Royal Astronomical Society, 469, 3688, \dodoi{10.1093/mnras/stx1085}

\bibitem[{{Evans} {et~al.}(2020){Evans}, {Primini}, {Miller}, {Evans}, {Allen}, {Anderson}, {Becker}, {Budynkiewicz}, {Burke}, {Chen}, {Civano}, {D'Abrusco}, {Doe}, {Fabbiano}, {Martinez Galarza}, {Gibbs}, {Glotfelty}, {Graessle}, {Grier}, {Hain}, {Hall}, {Harbo}, {Houck}, {Lauer}, {Laurino}, {Lee}, {McCollough}, {McDowell}, {McLaughlin}, {Morgan}, {Mossman}, {Nguyen}, {Nichols}, {Nowak}, {Paxson}, {Perdikeas}, {Plummer}, {Rots}, {Siemiginowska}, {Sundheim}, {Thong}, {Tibbetts}, {Van Stone}, {Winkelman}, \& {Zografou}}]{2020AAS...23515405E}
{Evans}, I.~N., {Primini}, F.~A., {Miller}, J.~B., {et~al.} 2020, in American Astronomical Society Meeting Abstracts, Vol. 235, American Astronomical Society Meeting Abstracts \#235, 154.05.
\newblock \url{https://ui.adsabs.harvard.edu/abs/2020AAS...23515405E}

\bibitem[{Evans {et~al.}(2020)Evans, Page, Osborne, Beardmore, Willingale, Burrows, Kennea, Perri, Capalbi, Tagliaferri, \& Cenko}]{Evans_2020}
Evans, P.~A., Page, K.~L., Osborne, J.~P., {et~al.} 2020, The Astrophysical Journal Supplement Series, 247, 54, \dodoi{10.3847/1538-4365/ab7db9}

\bibitem[{{Fitzgerald} {et~al.}(2021){Fitzgerald}, {Gomez}, {Salimpour}, {Singleton}, \& {Wibowo}}]{2021JOSS....6.2641F}
{Fitzgerald}, M., {Gomez}, E., {Salimpour}, S., {Singleton}, J., \& {Wibowo}, R. 2021, The Journal of Open Source Software, 6, 2641, \dodoi{10.21105/joss.02641}

\bibitem[{{Fruchter} {et~al.}(1988){Fruchter}, {Stinebring}, \& {Taylor}}]{1988Natur.333..237F}
{Fruchter}, A.~S., {Stinebring}, D.~R., \& {Taylor}, J.~H. 1988, \nat, 333, 237, \dodoi{10.1038/333237a0}

\bibitem[{{Gaia Collaboration}(2016)}]{brown2016gaia}
{Gaia Collaboration}. 2016, Astronomy \& Astrophysics, 595, A2, \dodoi{10.1051/0004-6361/201629512}

\bibitem[{{Gaia Collaboration}(2023)}]{refId0}
---. 2023, A\&A, 674, A1, \dodoi{10.1051/0004-6361/202243940}

\bibitem[{{Gentile} {et~al.}(2014){Gentile}, {Roberts}, {McLaughlin}, {Camilo}, {Hessels}, {Kerr}, {Ransom}, {Ray}, \& {Stairs}}]{2014ApJ...783...69G}
{Gentile}, P.~A., {Roberts}, M.~S.~E., {McLaughlin}, M.~A., {et~al.} 2014, \apj, 783, 69, \dodoi{10.1088/0004-637X/783/2/69}

\bibitem[{{Gonthier} {et~al.}(2018){Gonthier}, {Harding}, {Ferrara}, {Frederick}, {Mohr}, \& {Koh}}]{2018ApJ...863..199G}
{Gonthier}, P.~L., {Harding}, A.~K., {Ferrara}, E.~C., {et~al.} 2018, \apj, 863, 199, \dodoi{10.3847/1538-4357/aad08d}

\bibitem[{{Green} {et~al.}(2019){Green}, {Schlafly}, {Zucker}, {Speagle}, \& {Finkbeiner}}]{2019ApJ...887...93G}
{Green}, G.~M., {Schlafly}, E., {Zucker}, C., {Speagle}, J.~S., \& {Finkbeiner}, D. 2019, \apj, 887, 93, \dodoi{10.3847/1538-4357/ab5362}

\bibitem[{{Heinze} {et~al.}(2018){Heinze}, {Tonry}, {Denneau}, {Flewelling}, {Stalder}, {Rest}, {Smith}, {Smartt}, \& {Weiland}}]{2018AJ....156..241H}
{Heinze}, A.~N., {Tonry}, J.~L., {Denneau}, L., {et~al.} 2018, \aj, 156, 241, \dodoi{10.3847/1538-3881/aae47f}

\bibitem[{{Honeycutt}(1992)}]{1992PASP..104..435H}
{Honeycutt}, R.~K. 1992, \pasp, 104, 435, \dodoi{10.1086/133015}

\bibitem[{{Koljonen} \& {Linares}(2023)}]{2023MNRAS.525.3963K}
{Koljonen}, K. I.~I., \& {Linares}, M. 2023, \mnras, 525, 3963, \dodoi{10.1093/mnras/stad2485}

\bibitem[{Koljonen {et~al.}(2024)Koljonen, Lindseth, Linares, Harding, \& Turchetta}]{koljonen2024}
Koljonen, K. I.~I., Lindseth, S.~S., Linares, M., Harding, A.~K., \& Turchetta, M. 2024, Monthly Notices of the Royal Astronomical Society, 529, 575, \dodoi{10.1093/mnras/stae498}

\bibitem[{{Linares}(2014)}]{2014ApJ...795...72L}
{Linares}, M. 2014, \apj, 795, 72, \dodoi{10.1088/0004-637X/795/1/72}

\bibitem[{{Linares}(2020)}]{2020mbhe.confE..23L}
{Linares}, M. 2020, in Multifrequency Behaviour of High Energy Cosmic Sources - XIII. 3-8 June 2019. Palermo, 23, \dodoi{10.22323/1.362.0023}

\bibitem[{{Linares} \& {Kachelrie{\ss}}(2021)}]{2021JCAP...02..030L}
{Linares}, M., \& {Kachelrie{\ss}}, M. 2021, \jcap, 2021, 030, \dodoi{10.1088/1475-7516/2021/02/030}

\bibitem[{{Linares} {et~al.}(2017){Linares}, {Miles-P{\'a}ez}, {Rodr{\'\i}guez-Gil}, {Shahbaz}, {Casares}, {Fari{\~n}a}, \& {Karjalainen}}]{2017MNRAS.465.4602L}
{Linares}, M., {Miles-P{\'a}ez}, P., {Rodr{\'\i}guez-Gil}, P., {et~al.} 2017, \mnras, 465, 4602, \dodoi{10.1093/mnras/stw3057}

\bibitem[{{Linares} {et~al.}(2018){Linares}, {Shahbaz}, \& {Casares}}]{2018ApJ...859...54L}
{Linares}, M., {Shahbaz}, T., \& {Casares}, J. 2018, \apj, 859, 54, \dodoi{10.3847/1538-4357/aabde6}

\bibitem[{Lindseth(2023)}]{lindseth2023gamma}
Lindseth, S.~S. 2023, Master's thesis, NTNU.
\newblock \url{https://ntnuopen.ntnu.no/ntnu-xmlui/handle/11250/3073819}

\bibitem[{{Linnell Nemec} \& {Nemec}(1985)}]{1985AJ.....90.2317L}
{Linnell Nemec}, A.~F., \& {Nemec}, J.~M. 1985, \aj, 90, 2317, \dodoi{10.1086/113936}

\bibitem[{{Lomb}(1976)}]{1976Ap&SS..39..447L}
{Lomb}, N.~R. 1976, \apss, 39, 447, \dodoi{10.1007/BF00648343}

\bibitem[{{McCully} {et~al.}(2018){McCully}, {Volgenau}, {Harbeck}, {Lister}, {Saunders}, {Turner}, {Siiverd}, \& {Bowman}}]{2018SPIE10707E..0KM}
{McCully}, C., {Volgenau}, N.~H., {Harbeck}, D.-R., {et~al.} 2018, in Society of Photo-Optical Instrumentation Engineers (SPIE) Conference Series, Vol. 10707, Software and Cyberinfrastructure for Astronomy V, ed. J.~C. {Guzman} \& J.~{Ibsen}, 107070K, \dodoi{10.1117/12.2314340}

\bibitem[{{Merloni} {et~al.}(2024){Merloni}, {Lamer}, {Liu}, {Ramos-Ceja}, {Brunner}, {Bulbul}, {Dennerl}, {Doroshenko}, {Freyberg}, {Friedrich}, {Gatuzz}, {Georgakakis}, {Haberl}, {Igo}, {Kreykenbohm}, {Liu}, {Maitra}, {Malyali}, {Mayer}, {Nandra}, {Predehl}, {Robrade}, {Salvato}, {Sanders}, {Stewart}, {Tub{\'\i}n-Arenas}, {Weber}, {Wilms}, {Arcodia}, {Artis}, {Aschersleben}, {Avakyan}, {Aydar}, {Bahar}, {Balzer}, {Becker}, {Berger}, {Boller}, {Bornemann}, {Br{\"u}ggen}, {Brusa}, {Buchner}, {Burwitz}, {Camilloni}, {Clerc}, {Comparat}, {Coutinho}, {Czesla}, {Dannhauer}, {Dauner}, {Dauser}, {Dietl}, {Dolag}, {Dwelly}, {Egg}, {Ehl}, {Freund}, {Friedrich}, {Gaida}, {Garrel}, {Ghirardini}, {Gokus}, {Gr{\"u}nwald}, {Grandis}, {Grotova}, {Gruen}, {Gueguen}, {H{\"a}mmerich}, {Hamaus}, {Hasinger}, {Haubner}, {Homan}, {Ider Chitham}, {Joseph}, {Joyce}, {K{\"o}nig}, {Kaltenbrunner}, {Khokhriakova}, {Kink}, {Kirsch}, {Kluge}, {Knies}, {Krippendorf}, {Krumpe}, {Kurpas}, {Li}, {Liu}, {Locatelli}, {Lorenz}, {M{\"u}ller},
  {Magaudda}, {Mannes}, {McCall}, {Meidinger}, {Michailidis}, {Migkas}, {Mu{\~n}oz-Giraldo}, {Musiimenta}, {Nguyen-Dang}, {Ni}, {Olechowska}, {Ota}, {Pacaud}, {Pasini}, {Perinati}, {Pires}, {Pommranz}, {Ponti}, {Poppenhaeger}, {P{\"u}hlhofer}, {Rau}, {Reh}, {Reiprich}, {Roster}, {Saeedi}, {Santangelo}, {Sasaki}, {Schmitt}, {Schneider}, {Schrabback}, {Schuster}, {Schwope}, {Seppi}, {Serim}, {Shreeram}, {Sokolova-Lapa}, {Starck}, {Stelzer}, {Stierhof}, {Suleimanov}, {Tenzer}, {Traulsen}, {Tr{\"u}mper}, {Tsuge}, {Urrutia}, {Veronica}, {Waddell}, {Willer}, {Wolf}, {Yeung}, {Zainab}, {Zangrandi}, {Zhang}, {Zhang}, \& {Zheng}}]{2024A&A...682A..34M}
{Merloni}, A., {Lamer}, G., {Liu}, T., {et~al.} 2024, \aap, 682, A34, \dodoi{10.1051/0004-6361/202347165}

\bibitem[{Nedreaas(2024)}]{nedreaas2024spidercat}
Nedreaas, I.~F. 2024, Master's thesis, NTNU.
\newblock \url{https://ntnuopen.ntnu.no/ntnu-xmlui/handle/11250/3143780}

\bibitem[{{Nolan} {et~al.}(2012){Nolan}, {Abdo}, {Ackermann}, {Ajello}, {Allafort}, {Antolini}, {Atwood}, {Axelsson}, {Baldini}, {Ballet}, {Barbiellini}, {Bastieri}, {Bechtol}, {Belfiore}, {Bellazzini}, {Berenji}, {Bignami}, {Blandford}, {Bloom}, {Bonamente}, {Bonnell}, {Borgland}, {Bottacini}, {Bouvier}, {Brandt}, {Bregeon}, {Brigida}, {Bruel}, {Buehler}, {Burnett}, {Buson}, {Caliandro}, {Cameron}, {Campana}, {Ca{\~n}adas}, {Cannon}, {Caraveo}, {Casandjian}, {Cavazzuti}, {Ceccanti}, {Cecchi}, {{\c{C}}elik}, {Charles}, {Chekhtman}, {Cheung}, {Chiang}, {Chipaux}, {Ciprini}, {Claus}, {Cohen-Tanugi}, {Cominsky}, {Conrad}, {Corbet}, {Cutini}, {D'Ammando}, {Davis}, {de Angelis}, {DeCesar}, {DeKlotz}, {De Luca}, {den Hartog}, {de Palma}, {Dermer}, {Digel}, {Silva}, {Drell}, {Drlica-Wagner}, {Dubois}, {Dumora}, {Enoto}, {Escande}, {Fabiani}, {Falletti}, {Favuzzi}, {Fegan}, {Ferrara}, {Focke}, {Fortin}, {Frailis}, {Fukazawa}, {Funk}, {Fusco}, {Gargano}, {Gasparrini}, {Gehrels}, {Germani}, {Giebels}, {Giglietto},
  {Giommi}, {Giordano}, {Giroletti}, {Glanzman}, {Godfrey}, {Grenier}, {Grondin}, {Grove}, {Guillemot}, {Guiriec}, {Gustafsson}, {Hadasch}, {Hanabata}, {Harding}, {Hayashida}, {Hays}, {Hill}, {Horan}, {Hou}, {Hughes}, {Iafrate}, {Itoh}, {J{\'o}hannesson}, {Johnson}, {Johnson}, {Johnson}, {Johnson}, {Kamae}, {Katagiri}, {Kataoka}, {Katsuta}, {Kawai}, {Kerr}, {Kn{\"o}dlseder}, {Kocevski}, {Kuss}, {Lande}, {Landriu}, {Latronico}, {Lemoine-Goumard}, {Lionetto}, {Llena Garde}, {Longo}, {Loparco}, {Lott}, {Lovellette}, {Lubrano}, {Madejski}, {Marelli}, {Massaro}, {Mazziotta}, {McConville}, {McEnery}, {Mehault}, {Michelson}, {Minuti}, {Mitthumsiri}, {Mizuno}, {Moiseev}, {Mongelli}, {Monte}, {Monzani}, {Morselli}, {Moskalenko}, {Murgia}, {Nakamori}, {Naumann-Godo}, {Norris}, {Nuss}, {Nymark}, {Ohno}, {Ohsugi}, {Okumura}, {Omodei}, {Orlando}, {Ormes}, {Ozaki}, {Paneque}, {Panetta}, {Parent}, {Perkins}, {Pesce-Rollins}, {Pierbattista}, {Pinchera}, {Piron}, {Pivato}, {Porter}, {Racusin}, {Rain{\`o}}, {Rando}, {Razzano},
  {Razzaque}, {Reimer}, {Reimer}, {Reposeur}, {Ritz}, {Rochester}, {Romani}, {Roth}, {Rousseau}, {Ryde}, {Sadrozinski}, {Salvetti}, {Sanchez}, {Saz Parkinson}, {Sbarra}, {Scargle}, {Schalk}, {Sgr{\`o}}, {Shaw}, {Shrader}, {Siskind}, {Smith}, {Spandre}, {Spinelli}, {Stephens}, {Strickman}, {Suson}, {Tajima}, {Takahashi}, {Takahashi}, {Tanaka}, {Thayer}, {Thayer}, {Thompson}, {Tibaldo}, {Tibolla}, {Tinebra}, {Tinivella}, {Torres}, {Tosti}, {Troja}, {Uchiyama}, {Vandenbroucke}, {Van Etten}, {Van Klaveren}, {Vasileiou}, {Vianello}, {Vitale}, {Waite}, {Wallace}, {Wang}, {Werner}, {Winer}, {Wood}, {Wood}, {Wood}, {Yang}, \& {Zimmer}}]{2012ApJS..199...31N}
{Nolan}, P.~L., {Abdo}, A.~A., {Ackermann}, M., {et~al.} 2012, \apjs, 199, 31, \dodoi{10.1088/0067-0049/199/2/31}

\bibitem[{{Papitto} {et~al.}(2013){Papitto}, {Ferrigno}, {Bozzo}, {Rea}, {Pavan}, {Burderi}, {Burgay}, {Campana}, {di Salvo}, {Falanga}, {Filipovi{\'c}}, {Freire}, {Hessels}, {Possenti}, {Ransom}, {Riggio}, {Romano}, {Sarkissian}, {Stairs}, {Stella}, {Torres}, {Wieringa}, \& {Wong}}]{2013Natur.501..517P}
{Papitto}, A., {Ferrigno}, C., {Bozzo}, E., {et~al.} 2013, \nat, 501, 517, \dodoi{10.1038/nature12470}

\bibitem[{Perez {et~al.}(2023)Perez, Bogdanov, Halpern, \& Gajjar}]{Perez_2023}
Perez, K.~I., Bogdanov, S., Halpern, J.~P., \& Gajjar, V. 2023, The Astrophysical Journal, 952, 150, \dodoi{10.3847/1538-4357/acdc23}

\bibitem[{{Radhakrishnan} \& {Srinivasan}(1982)}]{1982CSci...51.1096R}
{Radhakrishnan}, V., \& {Srinivasan}, G. 1982, Current Science, 51, 1096.
\newblock \url{https://ui.adsabs.harvard.edu/abs/1982CSci...51.1096R}

\bibitem[{{Ray} {et~al.}(2014){Ray}, {Belfiore}, {Saz Parkinson}, {Polisensky}, {Ransom}, {Romani}, {Hessels}, {Razzano}, {Bhattacharyya}, {Roy}, {Cognard}, \& {Pulsar Search Consortium}}]{2014AAS...22314007R}
{Ray}, P.~S., {Belfiore}, A.~M., {Saz Parkinson}, P., {et~al.} 2014, in American Astronomical Society Meeting Abstracts, Vol. 223, American Astronomical Society Meeting Abstracts \#223, 140.07.
\newblock \url{https://ui.adsabs.harvard.edu/abs/2014AAS...22314007R}

\bibitem[{{Rimoldini} {et~al.}(2022){Rimoldini}, {Eyer}, {Audard}, {Barblan}, {Carnerero}, {Clementini}, {De Ridder}, {Distefano}, {Faigler}, {Garofalo}, {Gavras}, {Gomel}, {Holl}, {Jevardat de Fombelle}, {Kruszy{\'n}ska}, {Lanzafame}, {Lebzelter}, {Leccia}, {Lecoeur-Ta{\"\i}bi}, {Mazeh}, {Molinaro}, {Mowlavi}, {Muraveva}, {Nienartowicz}, {Panahi}, {Raiteri}, {Ripepi}, {Rybicki}, {Trabucchi}, {Wyrzykowski}, \& {Zucker}}]{2022gdr3.reptE..10R}
{Rimoldini}, L., {Eyer}, L., {Audard}, M., {et~al.} 2022, {Gaia DR3 documentation Chapter 10: Variability}, Gaia DR3 documentation, European Space Agency; Gaia Data Processing and Analysis Consortium.
\newblock \url{https://ui.adsabs.harvard.edu/abs/2022gdr3.reptE..10R}

\bibitem[{Roberts(2012)}]{roberts2012surrounded}
Roberts, M.~S. 2012, Proceedings of the International Astronomical Union, 8, 127, \dodoi{10.1017/S174392131202337X}

\bibitem[{{Romani}(2015)}]{2015ApJ...812L..24R}
{Romani}, R.~W. 2015, \apjl, 812, L24, \dodoi{10.1088/2041-8205/812/2/L24}

\bibitem[{{Romani} \& {Sanchez}(2016)}]{2016ApJ...828....7R}
{Romani}, R.~W., \& {Sanchez}, N. 2016, \apj, 828, 7, \dodoi{10.3847/0004-637X/828/1/7}

\bibitem[{{Romani} \& {Shaw}(2011)}]{2011ApJ...743L..26R}
{Romani}, R.~W., \& {Shaw}, M.~S. 2011, \apjl, 743, L26, \dodoi{10.1088/2041-8205/743/2/L26}

\bibitem[{{Salvetti} {et~al.}(2015){Salvetti}, {Mignani}, {De Luca}, {Delvaux}, {Pallanca}, {Belfiore}, {Marelli}, {Breeveld}, {Greiner}, {Becker}, \& {Pizzocaro}}]{2015ApJ...814...88S}
{Salvetti}, D., {Mignani}, R.~P., {De Luca}, A., {et~al.} 2015, \apj, 814, 88, \dodoi{10.1088/0004-637X/814/2/88}

\bibitem[{{Scargle}(1982)}]{1982ApJ...263..835S}
{Scargle}, J.~D. 1982, \apj, 263, 835, \dodoi{10.1086/160554}

\bibitem[{{Schinzel} {et~al.}(2017){Schinzel}, {Petrov}, {Taylor}, \& {Edwards}}]{2017ApJ...838..139S}
{Schinzel}, F.~K., {Petrov}, L., {Taylor}, G.~B., \& {Edwards}, P.~G. 2017, \apj, 838, 139, \dodoi{10.3847/1538-4357/aa6439}

\bibitem[{{Skrutskie} {et~al.}(2006){Skrutskie}, {Cutri}, {Stiening}, {Weinberg}, {Schneider}, {Carpenter}, {Beichman}, {Capps}, {Chester}, {Elias}, {Huchra}, {Liebert}, {Lonsdale}, {Monet}, {Price}, {Seitzer}, {Jarrett}, {Kirkpatrick}, {Gizis}, {Howard}, {Evans}, {Fowler}, {Fullmer}, {Hurt}, {Light}, {Kopan}, {Marsh}, {McCallon}, {Tam}, {Van Dyk}, \& {Wheelock}}]{2006AJ....131.1163S}
{Skrutskie}, M.~F., {Cutri}, R.~M., {Stiening}, R., {et~al.} 2006, \aj, 131, 1163, \dodoi{10.1086/498708}

\bibitem[{{Smith} {et~al.}(2023){Smith}, {Abdollahi}, {Ajello}, {Bailes}, {Baldini}, {Ballet}, {Baring}, {Bassa}, {Gonzalez}, {Bellazzini}, {Berretta}, {Bhattacharyya}, {Bissaldi}, {Bonino}, {Bottacini}, {Bregeon}, {Bruel}, {Burgay}, {Burnett}, {Cameron}, {Camilo}, {Caputo}, {Caraveo}, {Cavazzuti}, {Chiaro}, {Ciprini}, {Clark}, {Cognard}, {Corongiu}, {Orestano}, {Crnogorcevic}, {Cuoco}, {Cutini}, {D'Ammando}, {de Angelis}, {DeCesar}, {De Gaetano}, {de Menezes}, {Deneva}, {de Palma}, {Di Lalla}, {Dirirsa}, {Di Venere}, {Dom{\'\i}nguez}, {Dumora}, {Fegan}, {Ferrara}, {Fiori}, {Fleischhack}, {Flynn}, {Franckowiak}, {Freire}, {Fukazawa}, {Fusco}, {Galanti}, {Gammaldi}, {Gargano}, {Gasparrini}, {Giacchino}, {Giglietto}, {Giordano}, {Giroletti}, {Green}, {Grenier}, {Guillemot}, {Guiriec}, {Gustafsson}, {Harding}, {Hays}, {Hewitt}, {Horan}, {Hou}, {Jankowski}, {Johnson}, {Johnson}, {Johnston}, {Kataoka}, {Keith}, {Kerr}, {Kramer}, {Kuss}, {Latronico}, {Lee}, {Li}, {Li}, {Limyansky}, {Longo}, {Loparco}, {Lorusso},
  {Lovellette}, {Lower}, {Lubrano}, {Lyne}, {Maan}, {Maldera}, {Manchester}, {Manfreda}, {Marelli}, {Mart{\'\i}-Devesa}, {Mazziotta}, {McEnery}, {Mereu}, {Michelson}, {Mickaliger}, {Mitthumsiri}, {Mizuno}, {Moiseev}, {Monzani}, {Morselli}, {Negro}, {Nemmen}, {Nieder}, {Nuss}, {Omodei}, {Orienti}, {Orlando}, {Ormes}, {Palatiello}, {Paneque}, {Panzarini}, {Parthasarathy}, {Persic}, {Pesce-Rollins}, {Pillera}, {Poon}, {Porter}, {Possenti}, {Principe}, {Rain{\`o}}, {Rando}, {Ransom}, {Ray}, {Razzano}, {Razzaque}, {Reimer}, {Reimer}, {Renault-Tinacci}, {Romani}, {S{\'a}nchez-Conde}, {Parkinson}, {Scotton}, {Serini}, {Sgr{\`o}}, {Shannon}, {Sharma}, {Shen}, {Siskind}, {Spandre}, {Spinelli}, {Stappers}, {Stephens}, {Suson}, {Tabassum}, {Tajima}, {Tak}, {Theureau}, {Thompson}, {Tibolla}, {Torres}, {Valverde}, {Venter}, {Wadiasingh}, {Wang}, {Wang}, {Wang}, {Weltevrede}, {Wood}, {Yan}, {Zaharijas}, {Zhang}, \& {Zhu}}]{2023ApJ...958..191S}
{Smith}, D.~A., {Abdollahi}, S., {Ajello}, M., {et~al.} 2023, \apj, 958, 191, \dodoi{10.3847/1538-4357/acee67}

\bibitem[{{Stellingwerf}(1978)}]{1978ApJ...224..953S}
{Stellingwerf}, R.~F. 1978, \apj, 224, 953, \dodoi{10.1086/156444}

\bibitem[{{Stringer} {et~al.}(2021){Stringer}, {Breton}, {Clark}, {Voisin}, {Kennedy}, {Mata S{\'a}nchez}, {Shahbaz}, {Dhillon}, {van Kerkwijk}, \& {Marsh}}]{2021MNRAS.507.2174S}
{Stringer}, J.~G., {Breton}, R.~P., {Clark}, C.~J., {et~al.} 2021, \mnras, 507, 2174, \dodoi{10.1093/mnras/stab2167}

\bibitem[{{Swihart} {et~al.}(2022){Swihart}, {Strader}, {Aydi}, {Chomiuk}, {Dage}, {Kawash}, {Sokolovsky}, \& {Ferrara}}]{2022ApJ...926..201S}
{Swihart}, S.~J., {Strader}, J., {Aydi}, E., {et~al.} 2022, \apj, 926, 201, \dodoi{10.3847/1538-4357/ac4ae4}

\bibitem[{Swihart {et~al.}(2021)Swihart, Strader, Aydi, Chomiuk, Dage, \& Shishkovsky}]{swihart2021discovery}
Swihart, S.~J., Strader, J., Aydi, E., {et~al.} 2021, The Astrophysical Journal, 909, 185

\bibitem[{Swihart {et~al.}(2022)Swihart, Strader, Chomiuk, Aydi, Sokolovsky, Ray, \& Kerr}]{Swihart_2022}
Swihart, S.~J., Strader, J., Chomiuk, L., {et~al.} 2022, The Astrophysical Journal, 941, 199, \dodoi{10.3847/1538-4357/aca2ac}

\bibitem[{{Swihart} {et~al.}(2019){Swihart}, {Strader}, {Chomiuk}, \& {Shishkovsky}}]{2019ApJ...876....8S}
{Swihart}, S.~J., {Strader}, J., {Chomiuk}, L., \& {Shishkovsky}, L. 2019, \apj, 876, 8, \dodoi{10.3847/1538-4357/ab125e}

\bibitem[{{Tauris} {et~al.}(2013){Tauris}, {Sanyal}, {Yoon}, \& {Langer}}]{2013A&A...558A..39T}
{Tauris}, T.~M., {Sanyal}, D., {Yoon}, S.~C., \& {Langer}, N. 2013, \aap, 558, A39, \dodoi{10.1051/0004-6361/201321662}

\bibitem[{{Tody}(1986)}]{1986SPIE..627..733T}
{Tody}, D. 1986, in Society of Photo-Optical Instrumentation Engineers (SPIE) Conference Series, Vol. 627, Instrumentation in astronomy VI, ed. D.~L. {Crawford}, 733, \dodoi{10.1117/12.968154}

\bibitem[{{Turchetta} {et~al.}(2023){Turchetta}, {Linares}, {Koljonen}, \& {Sen}}]{2023MNRAS.525.2565T}
{Turchetta}, M., {Linares}, M., {Koljonen}, K., \& {Sen}, B. 2023, \mnras, 525, 2565, \dodoi{10.1093/mnras/stad2435}

\bibitem[{{van den Heuvel} \& {van Paradijs}(1988)}]{1988Natur.334..227V}
{van den Heuvel}, E.~P.~J., \& {van Paradijs}, J. 1988, \nat, 334, 227, \dodoi{10.1038/334227a0}

\bibitem[{{Wadiasingh} {et~al.}(2018){Wadiasingh}, {Venter}, {Harding}, {B{\"o}ttcher}, \& {Kilian}}]{2018ApJ...869..120W}
{Wadiasingh}, Z., {Venter}, C., {Harding}, A.~K., {B{\"o}ttcher}, M., \& {Kilian}, P. 2018, \apj, 869, 120, \dodoi{10.3847/1538-4357/aaed43}

\bibitem[{{Webb} {et~al.}(2020){Webb}, {Coriat}, {Traulsen}, {Ballet}, {Motch}, {Carrera}, {Koliopanos}, {Authier}, {de la Calle}, {Ceballos}, {Colomo}, {Chuard}, {Freyberg}, {Garcia}, {Kolehmainen}, {Lamer}, {Lin}, {Maggi}, {Michel}, {Page}, {Page}, {Perea-Calderon}, {Pineau}, {Rodriguez}, {Rosen}, {Santos Lleo}, {Saxton}, {Schwope}, {Tom{\'a}s}, {Watson}, \& {Zakardjian}}]{2020A&A...641A.136W}
{Webb}, N.~A., {Coriat}, M., {Traulsen}, I., {et~al.} 2020, \aap, 641, A136, \dodoi{10.1051/0004-6361/201937353}

\bibitem[{{Weber} {et~al.}(2016){Weber}, {Granzer}, \& {Strassmeier}}]{2016SPIE.9910E..0NW}
{Weber}, M., {Granzer}, T., \& {Strassmeier}, K.~G. 2016, in Society of Photo-Optical Instrumentation Engineers (SPIE) Conference Series, Vol. 9910, Observatory Operations: Strategies, Processes, and Systems VI, ed. A.~B. {Peck}, R.~L. {Seaman}, \& C.~R. {Benn}, 99100N, \dodoi{10.1117/12.2232251}

\bibitem[{{Wijnands} \& {van der Klis}(1998)}]{Wijnands98}
{Wijnands}, R., \& {van der Klis}, M. 1998, \nat, 394, 344, \dodoi{10.1038/28557}

\bibitem[{{Yao} {et~al.}(2017){Yao}, {Manchester}, \& {Wang}}]{2017ApJ...835...29Y}
{Yao}, J.~M., {Manchester}, R.~N., \& {Wang}, N. 2017, \apj, 835, 29, \dodoi{10.3847/1538-4357/835/1/29}

\end{thebibliography}
\bibliographystyle{aasjournal_arxiv}



\appendix
\restartappendixnumbering
\section{COBIPULSE fields of view} \label{sec:appA}
\begin{figure*}[ht!]
\gridline{\leftfig{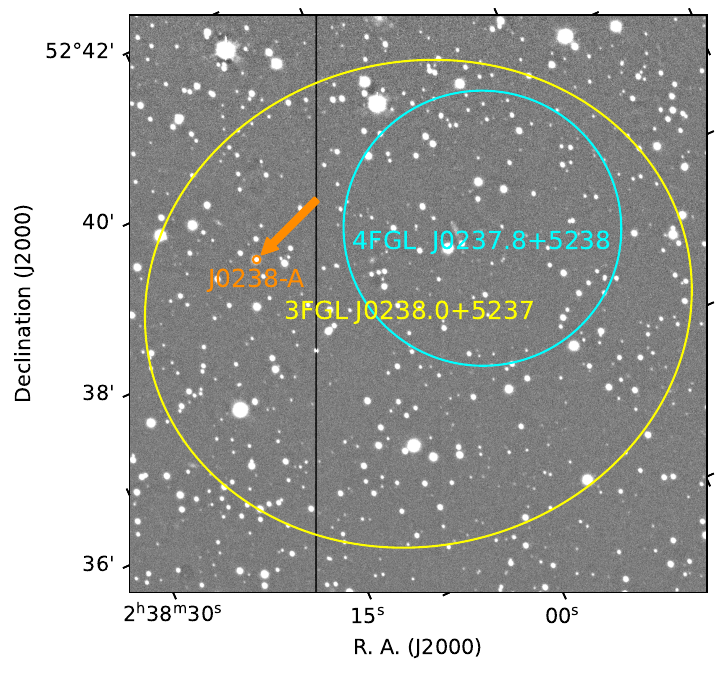}{0.43\textwidth}{}
          \leftfig{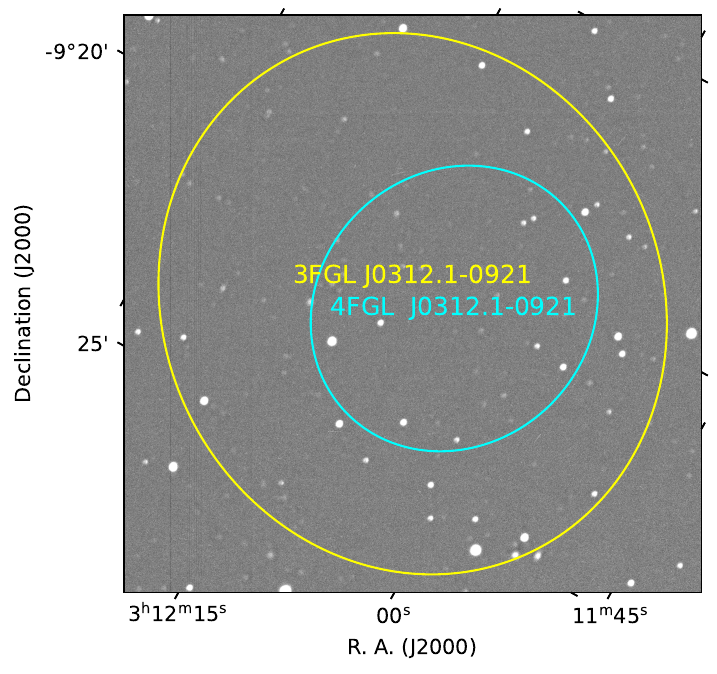}{0.43\textwidth}{}
          }
\gridline{\leftfig{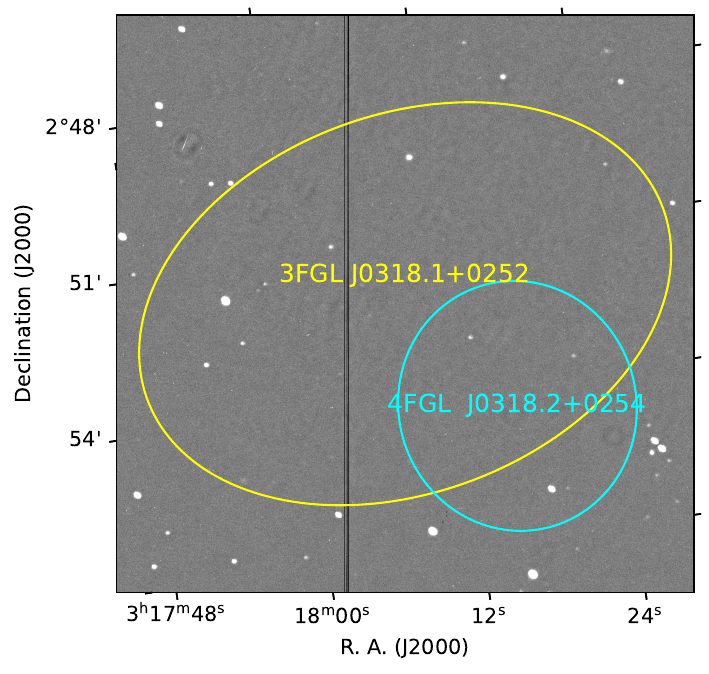}{0.43\textwidth}{}
          \leftfig{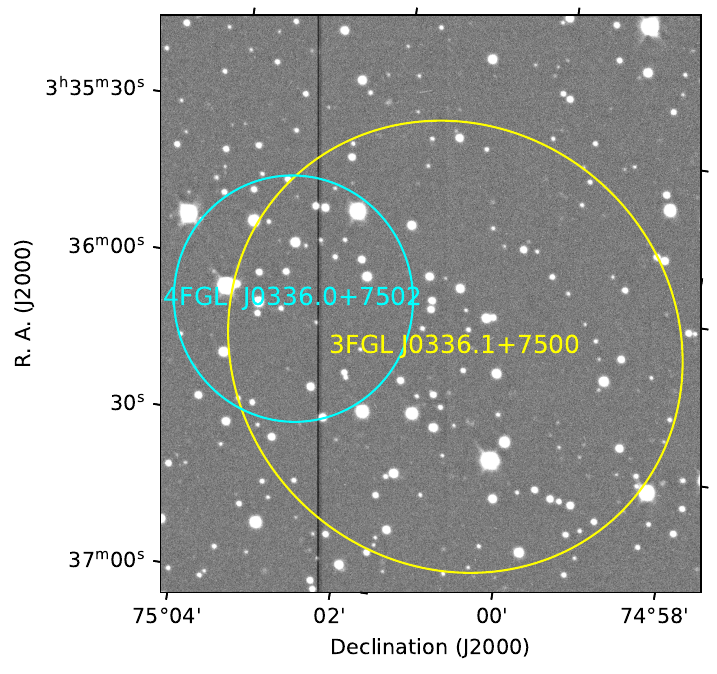}{0.43\textwidth}{}
          }
\gridline{\leftfig{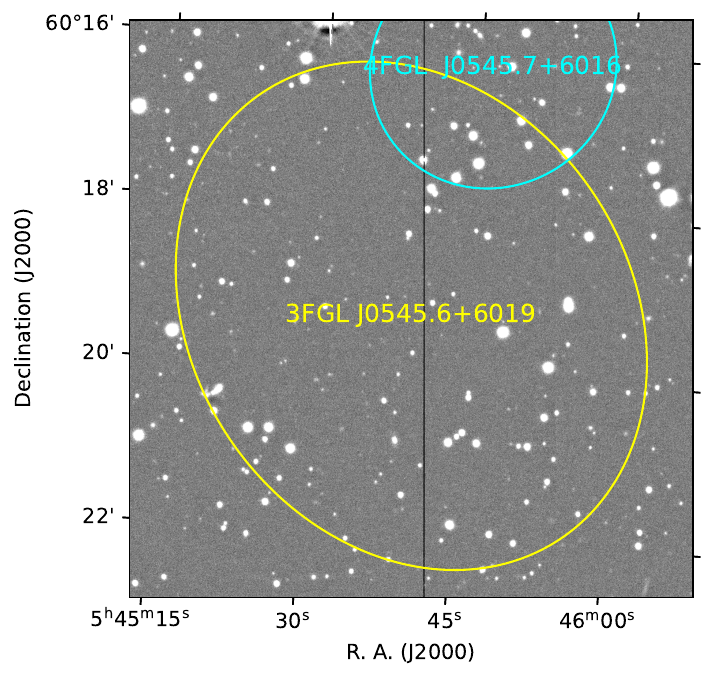}{0.43\textwidth}{}
          \leftfig{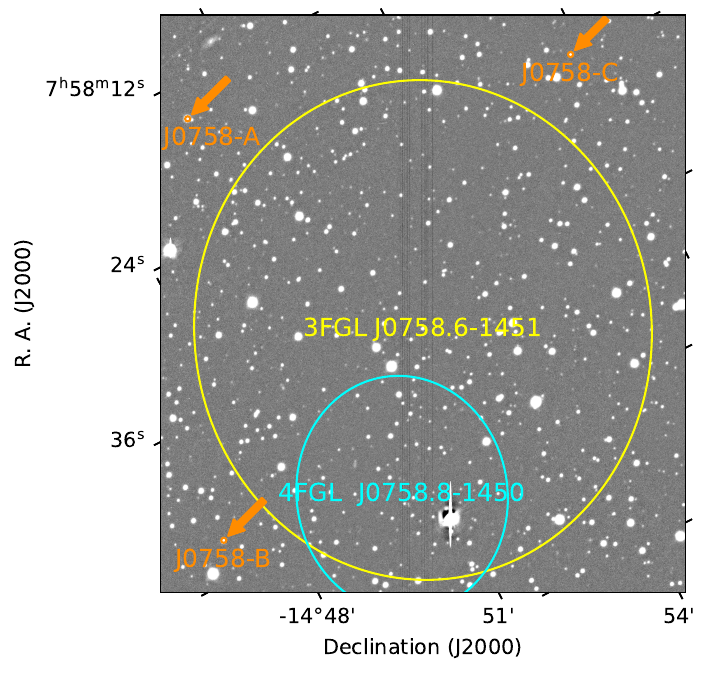}{0.43\textwidth}{}
          }
\caption{Fermi-3FGL/4FGL fields of view observed with STELLA/WiFSIP ($1.2$ $\mathrm{m}$).\label{fig:STELLAFoVs}}
\end{figure*}
\begin{figure*}[ht!]
\gridline{\leftfig{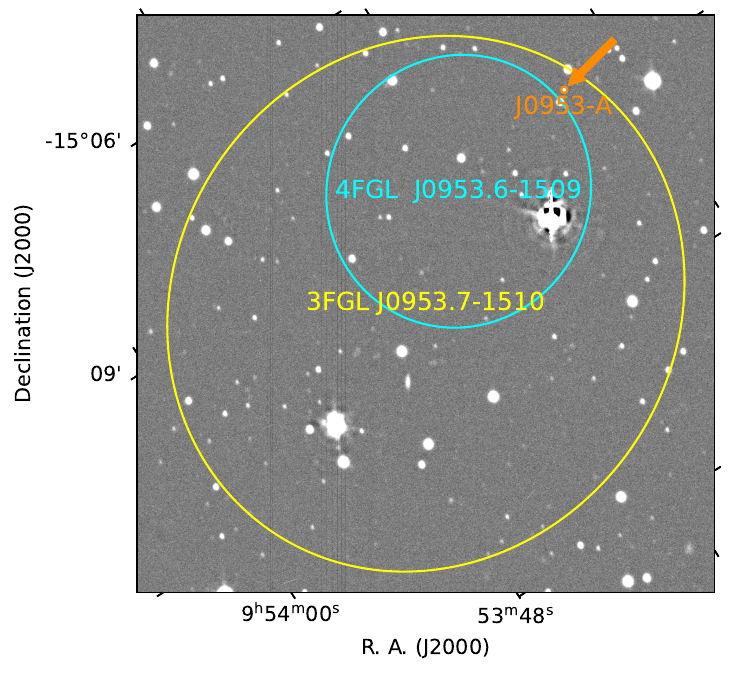}{0.44\textwidth}{}
          \leftfig{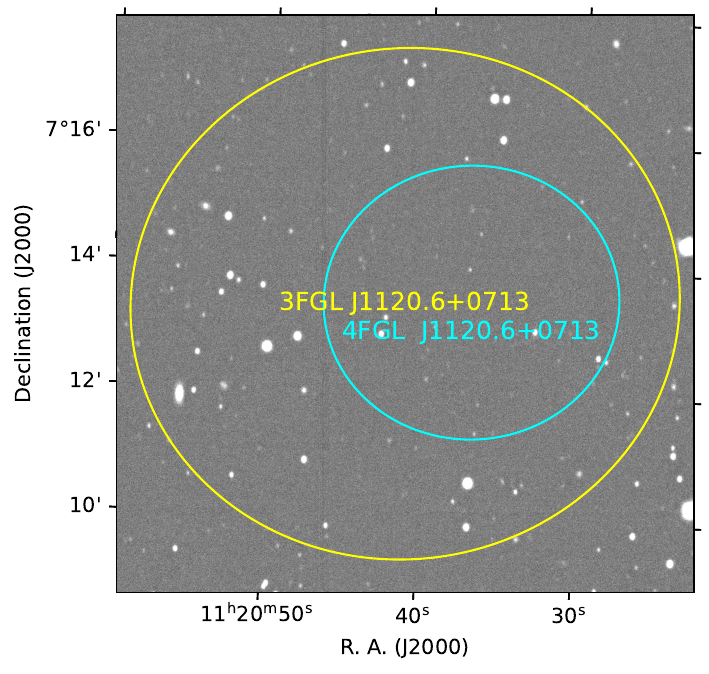}{0.43\textwidth}{}
          }
\gridline{\leftfig{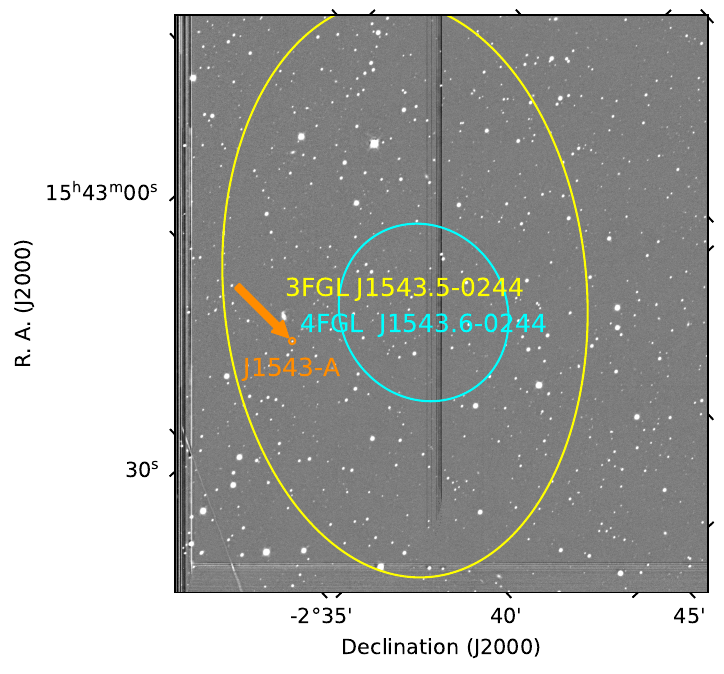}{0.43\textwidth}{}
          \leftfig{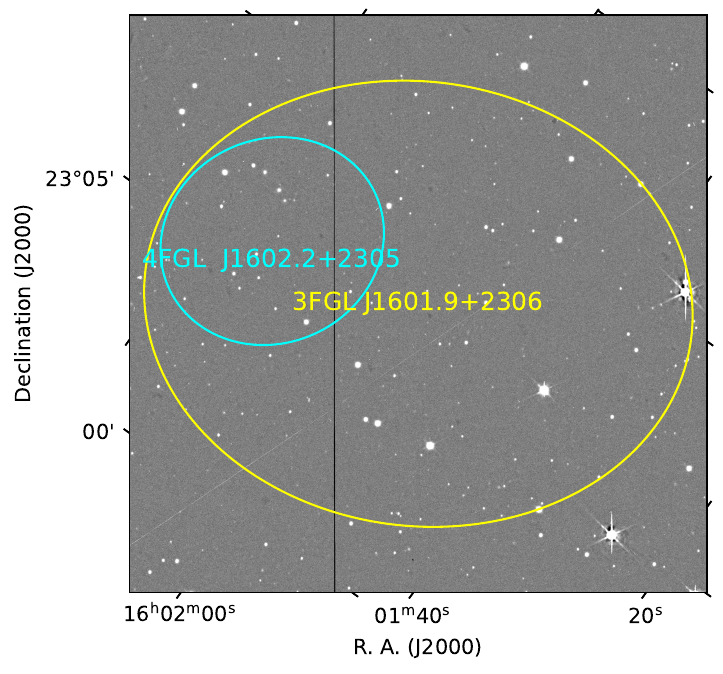}{0.43\textwidth}{}
          }
\gridline{\leftfig{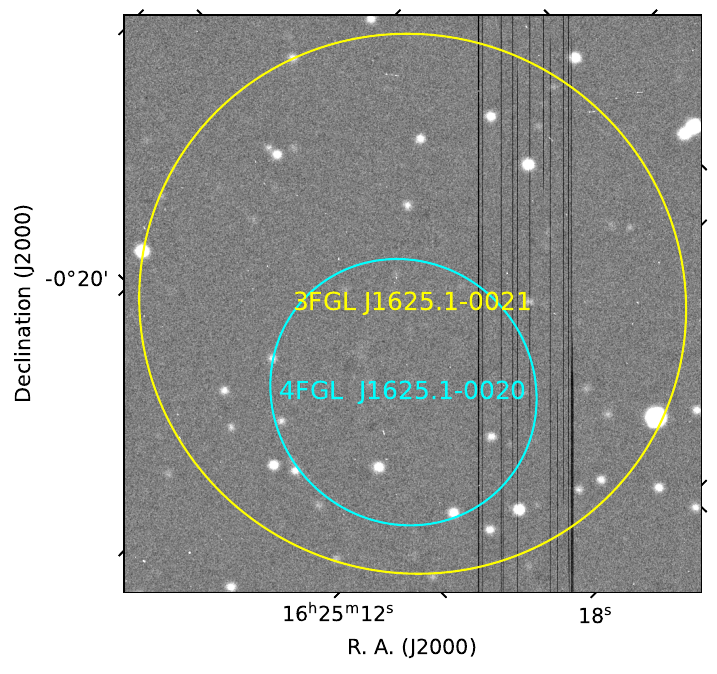}{0.43\textwidth}{}
          \leftfig{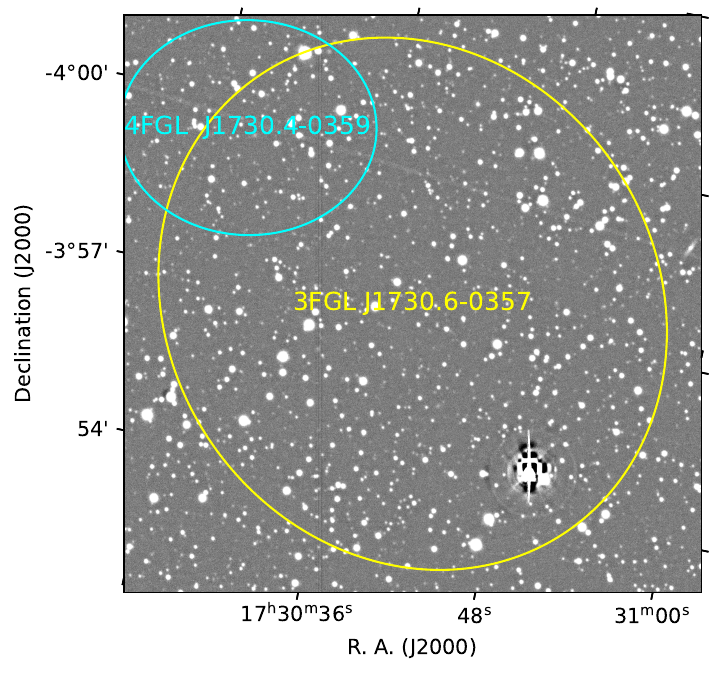}{0.43\textwidth}{}
          }
\caption{Continued.\label{fig:STELLAFoVs_cont1}}
\end{figure*}
\begin{figure*}[ht!]
\gridline{\leftfig{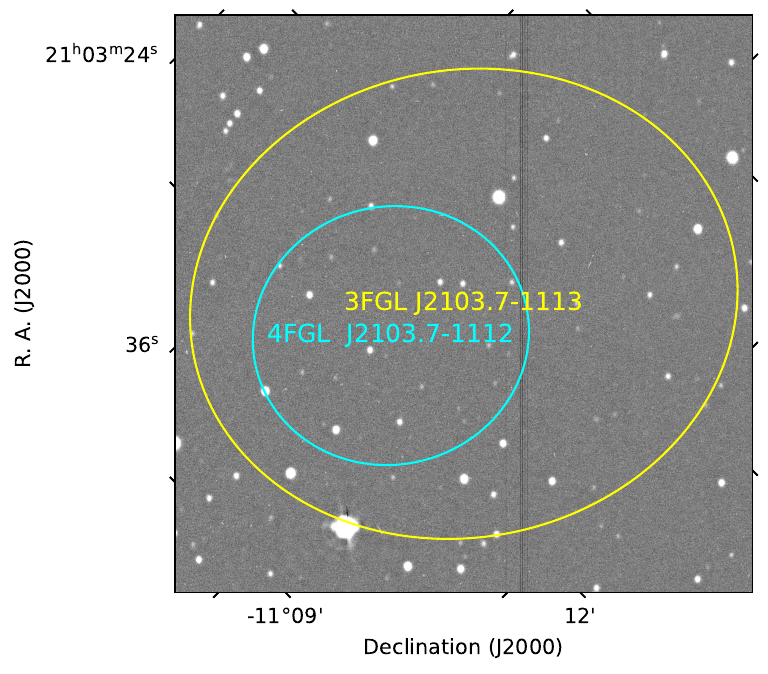}{0.47\textwidth}{}
          \leftfig{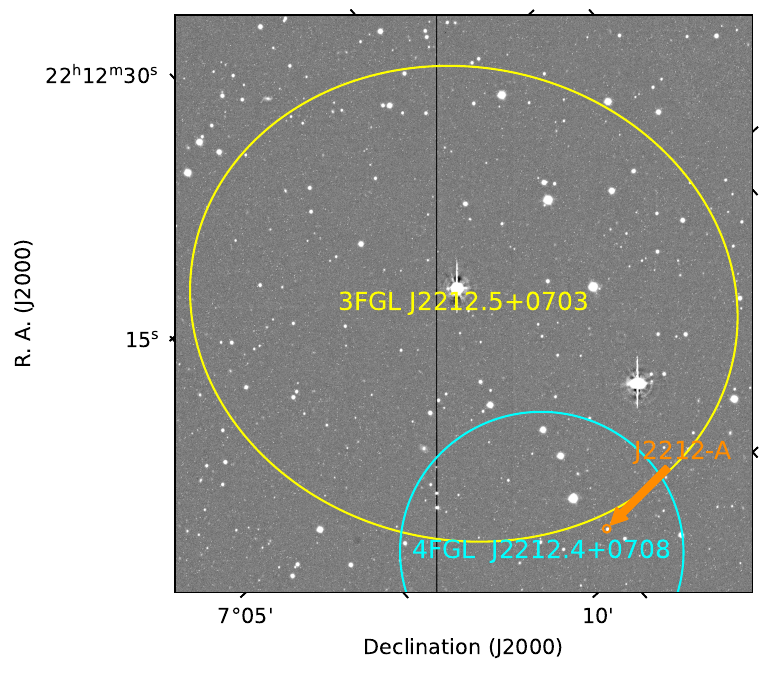}{0.47\textwidth}{}
          }
\gridline{\leftfig{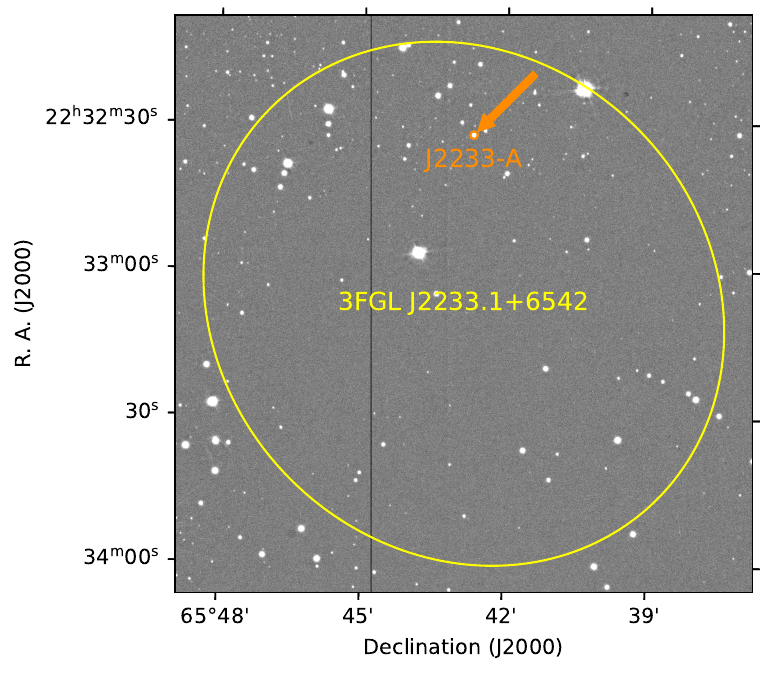}{0.47\textwidth}{}
          \leftfig{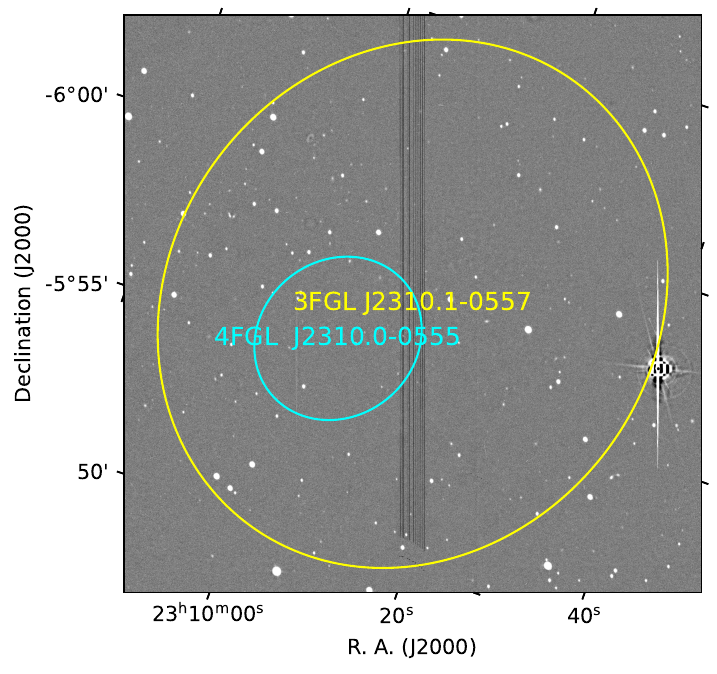}{0.44\textwidth}{}
          }
\caption{Continued.\label{fig:STELLAFoVs_cont2}}
\end{figure*}
\begin{figure*}[ht!]
\gridline{\leftfig{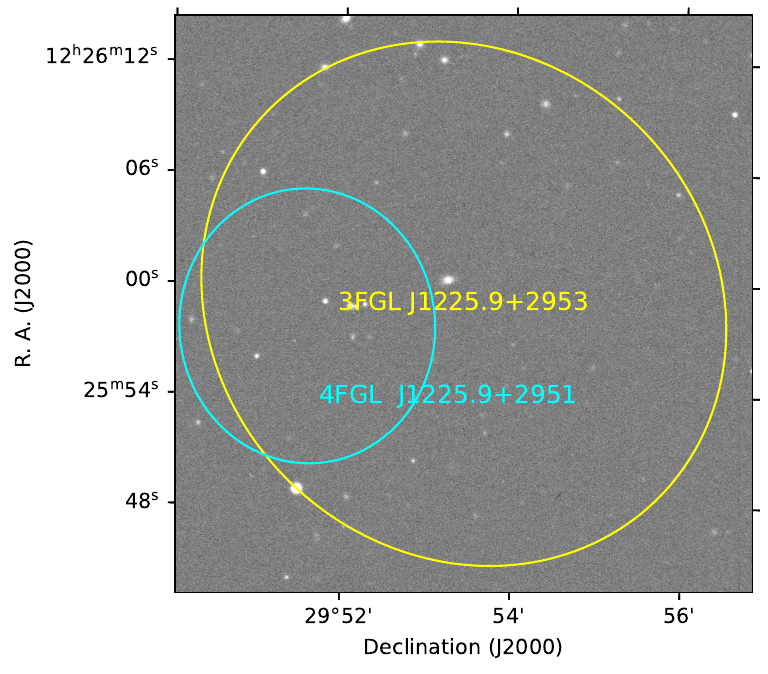}{0.5\textwidth}{}
          \fig{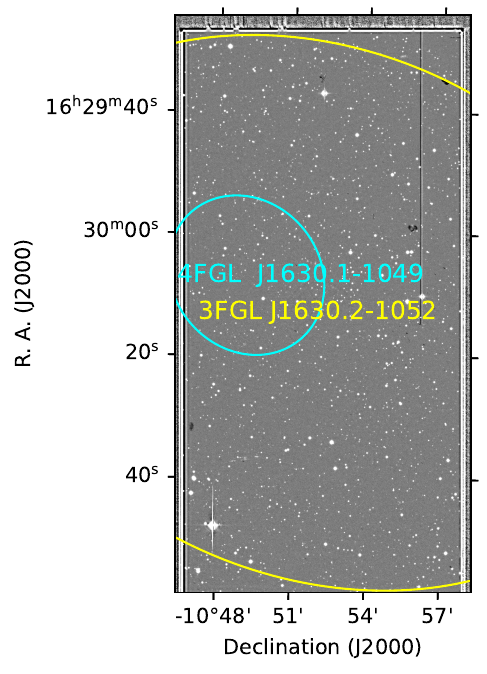}{0.32\textwidth}{}
          }
\gridline{\leftfig{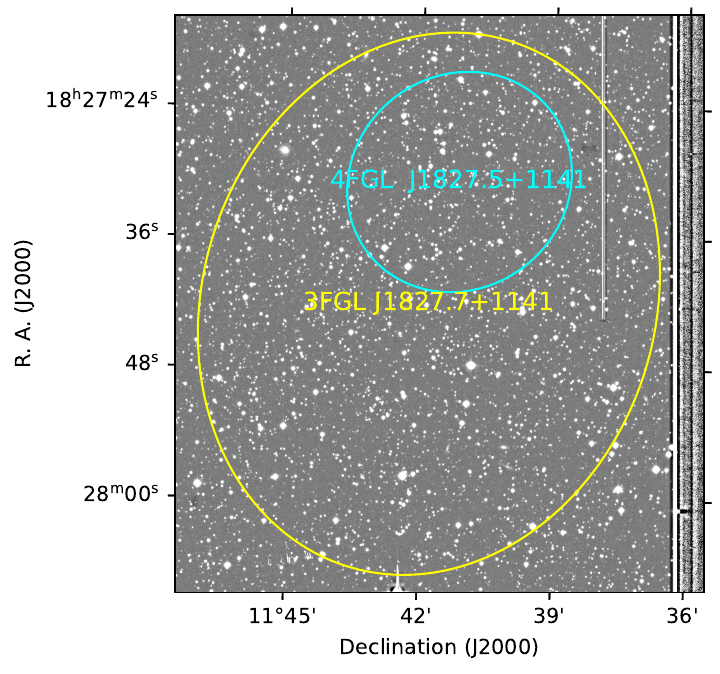}{0.5\textwidth}{}
          \leftfig{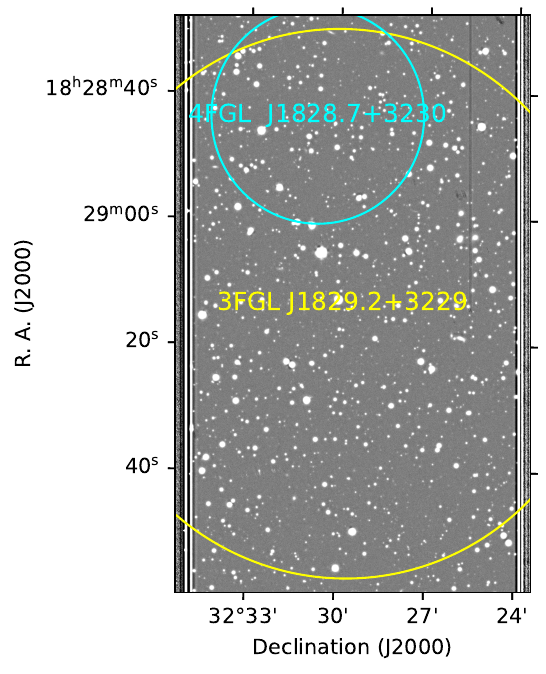}{0.38\textwidth}{}
          }
\caption{Fermi-3FGL/4FGL fields of view observed with INT/WFC ($2.5$ $\mathrm{m}$).\label{fig:INTFoVs}}
\end{figure*}
\begin{figure*}[ht!]
\gridline{\leftfig{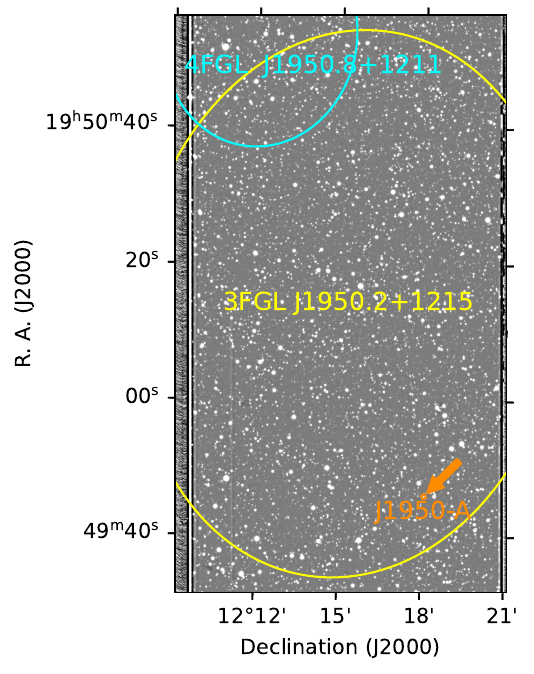}{0.5\textwidth}{}
          }
\caption{Continued.\label{fig:INTFoVs_cont1}}
\end{figure*}
\begin{figure*}[ht!]
\gridline{\leftfig{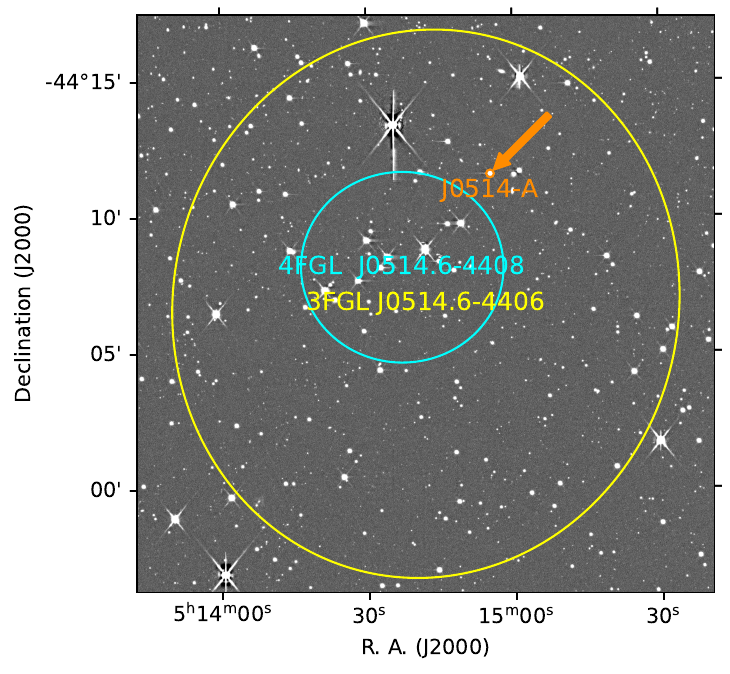}{0.42\textwidth}{}
          \leftfig{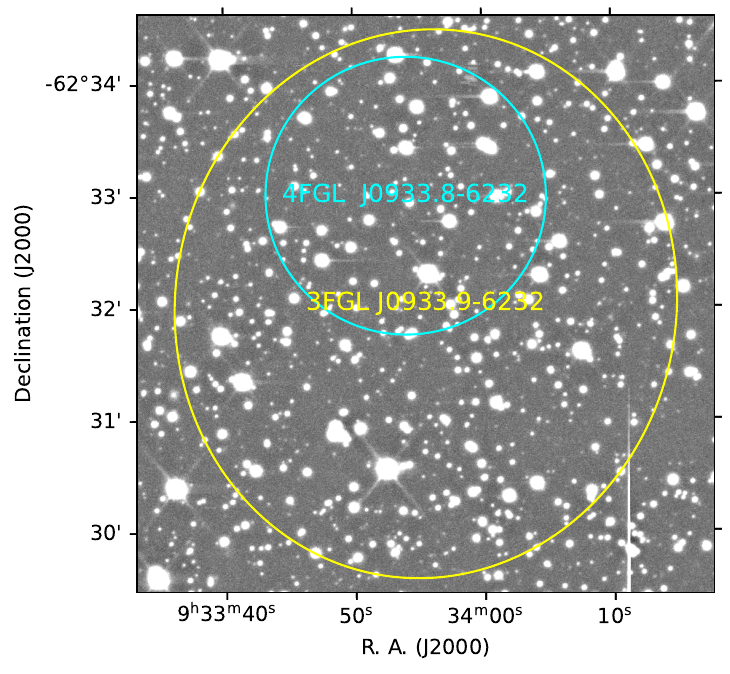}{0.42\textwidth}{}
          }
\gridline{\leftfig{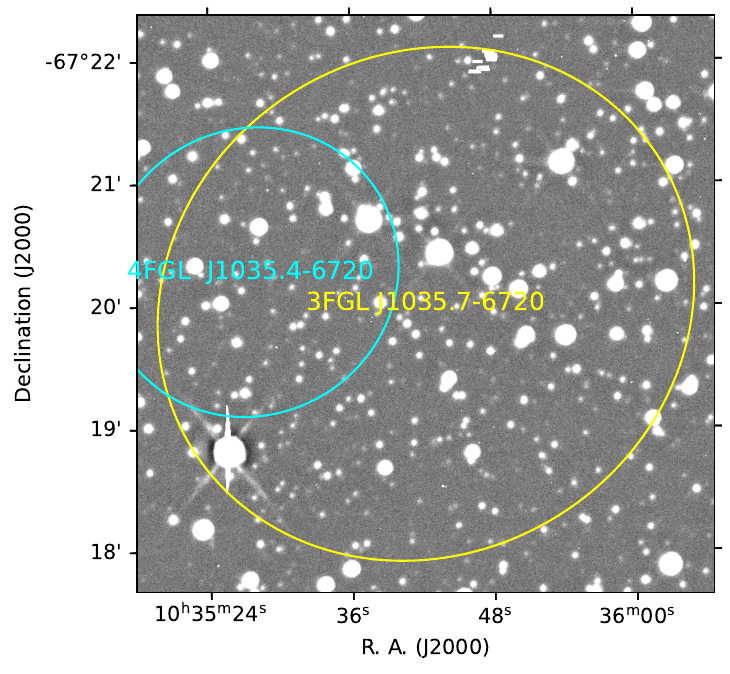}{0.42\textwidth}{}
          \leftfig{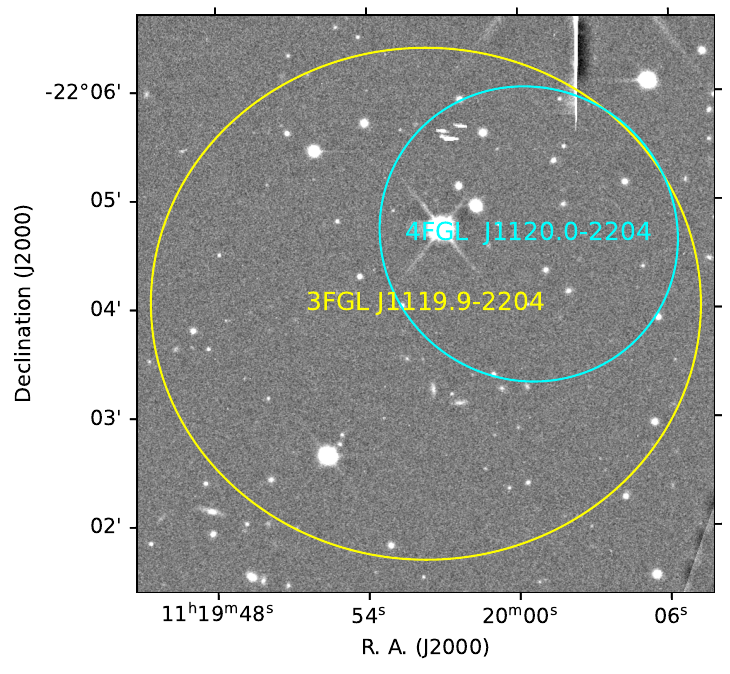}{0.42\textwidth}{}
          }
\gridline{\leftfig{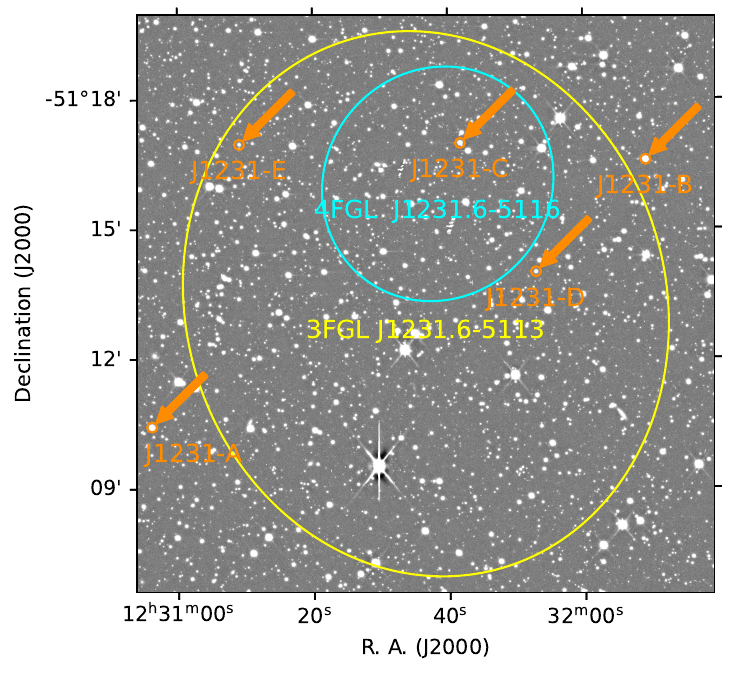}{0.42\textwidth}{}
          \leftfig{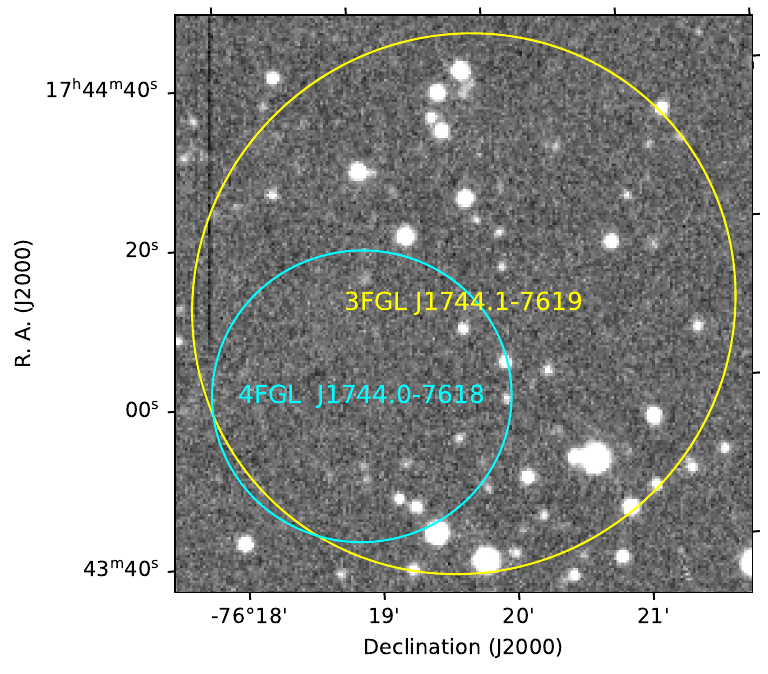}{0.44\textwidth}{}
          }
\caption{Fermi-3FGL/4FGL fields of view observed with LCO/Sinistro ($1$ $\mathrm{m}$) and LCO/SBIG ($0.4$ $\mathrm{m}$).\label{fig:LCOFoVs}}
\end{figure*}
\begin{figure*}[ht!]
\gridline{\leftfig{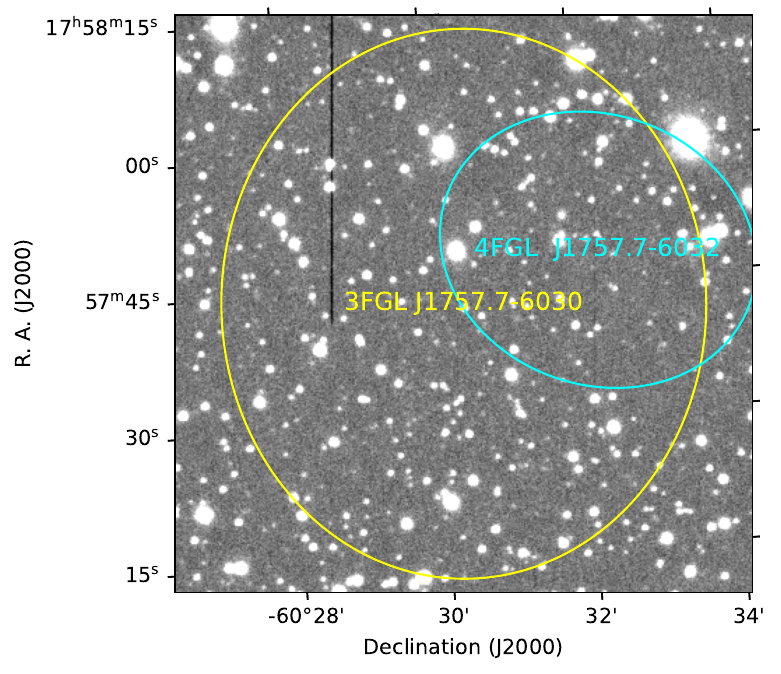}{0.42\textwidth}{}
          \leftfig{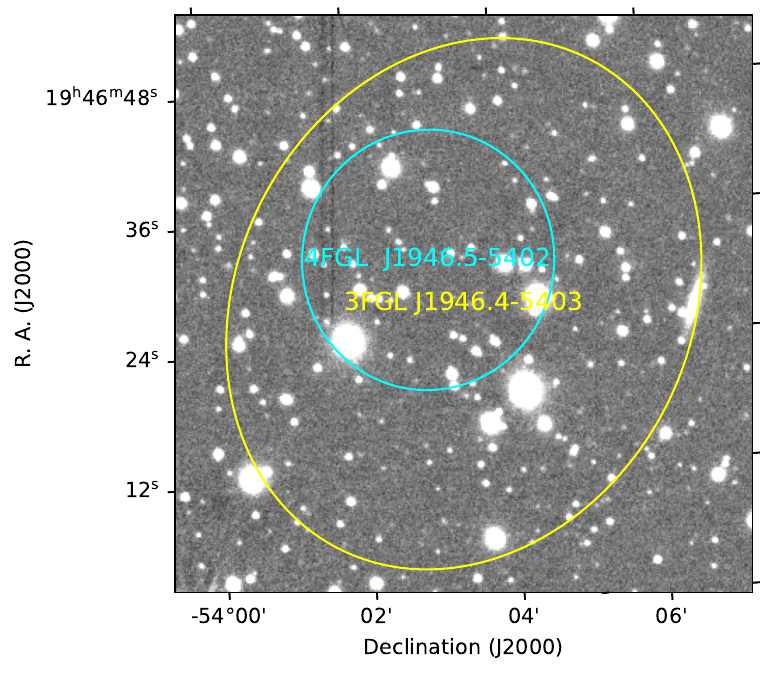}{0.42\textwidth}{}
          }
\gridline{\leftfig{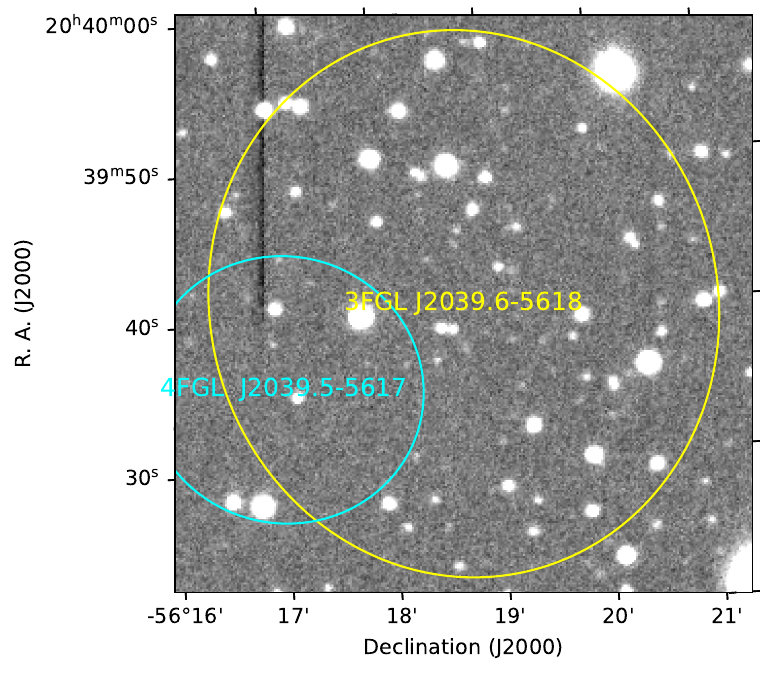}{0.42\textwidth}{}
          }
\caption{Continued.\label{fig:LCOFoVs_cont1}}
\end{figure*}

In Figures \ref{fig:STELLAFoVs}-\ref{fig:STELLAFoVs_cont2}, \ref{fig:INTFoVs}-\ref{fig:INTFoVs_cont1}, and \ref{fig:LCOFoVs}-\ref{fig:LCOFoVs_cont1} we present the combined \textit{r'}-band images for each field observed with the STELLA, INT, and LCO optical telescopes, respectively. We do not include the 3FGL~J0737.2$-$3233, 3FGL~J2117.6+3725, and 3FGL~J2221.6+6507 fields since they are already shown in Section \ref{sec:results}. We plot the 3FGL $95\%$ error ellipses in yellow, the associated 4FGL $95\%$ error ellipses in cyan, and the 17 periodic variables
that we did not attribute to spiders in orange (see Appendix \ref{sec:appC} for details).
\section{Selection of optical variables} \label{sec:appB}
\restartappendixnumbering
\begin{figure*}[ht!]
\gridline{\leftfig{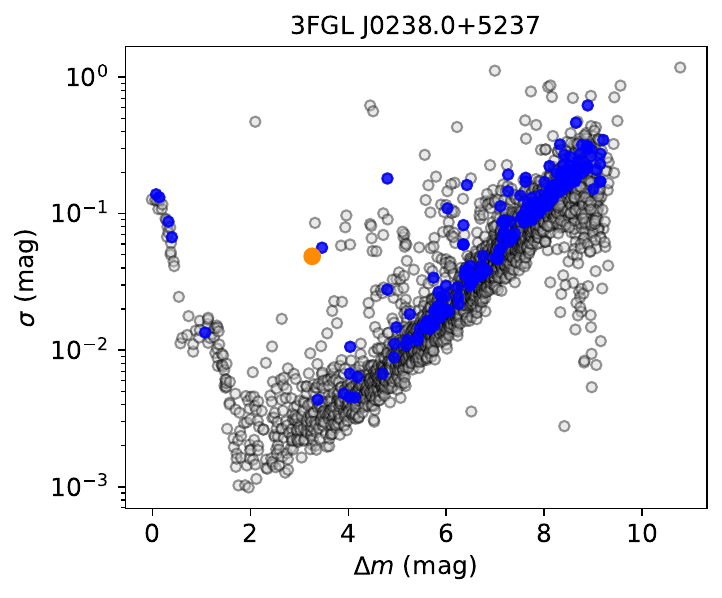}{0.48\textwidth}{}
          \leftfig{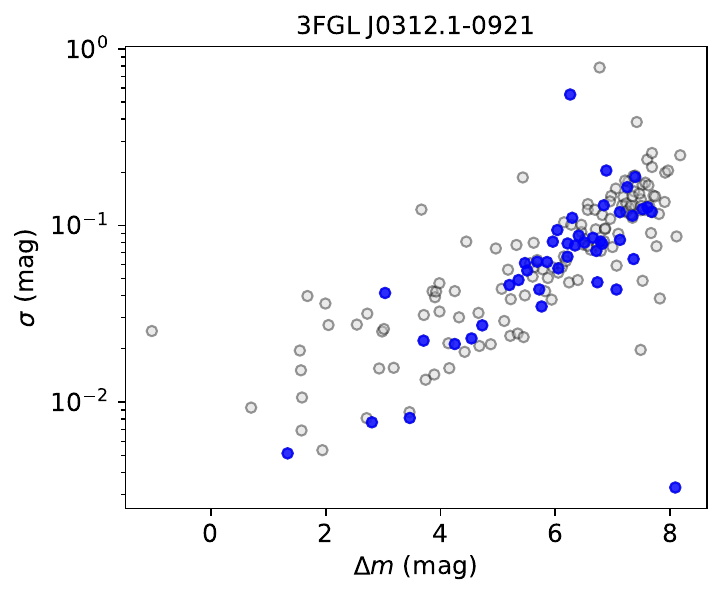}{0.48\textwidth}{}
          }
\gridline{\leftfig{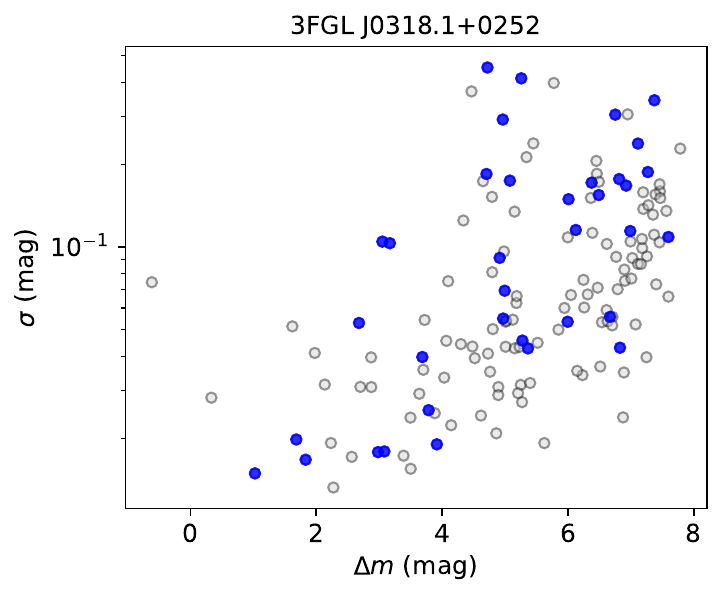}{0.48\textwidth}{}
          \leftfig{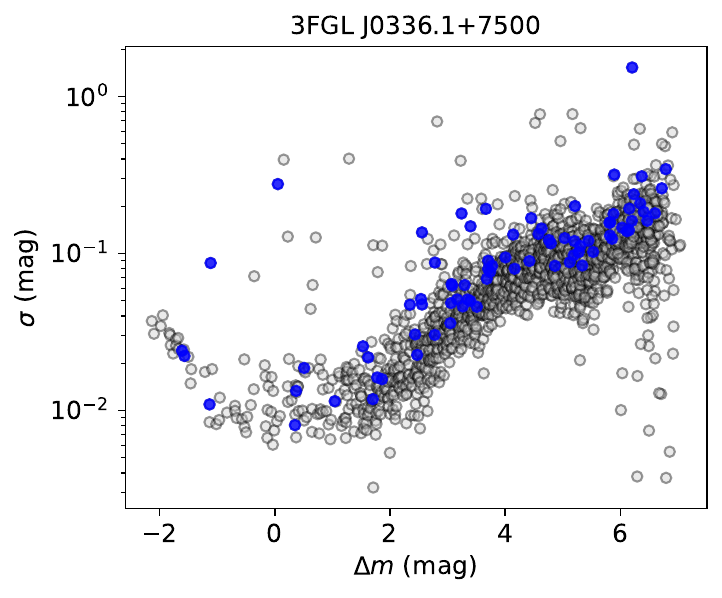}{0.48\textwidth}{}
          }
\gridline{\leftfig{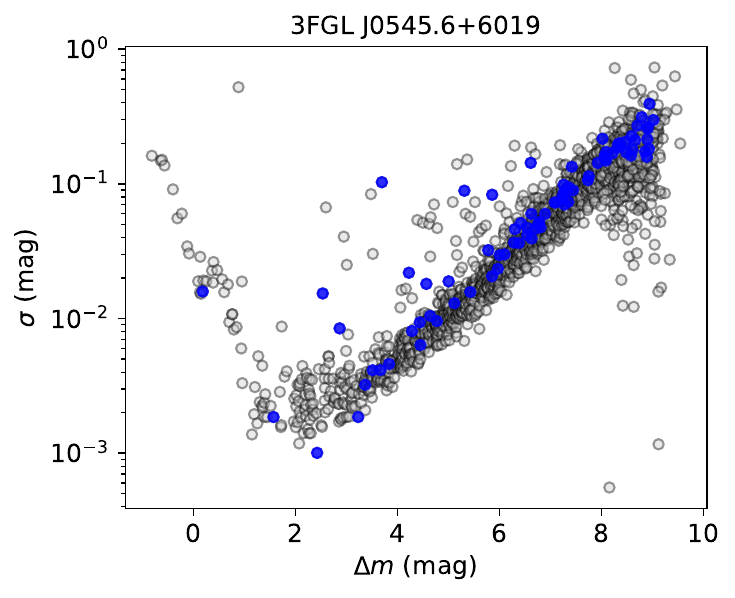}{0.48\textwidth}{}
          \leftfig{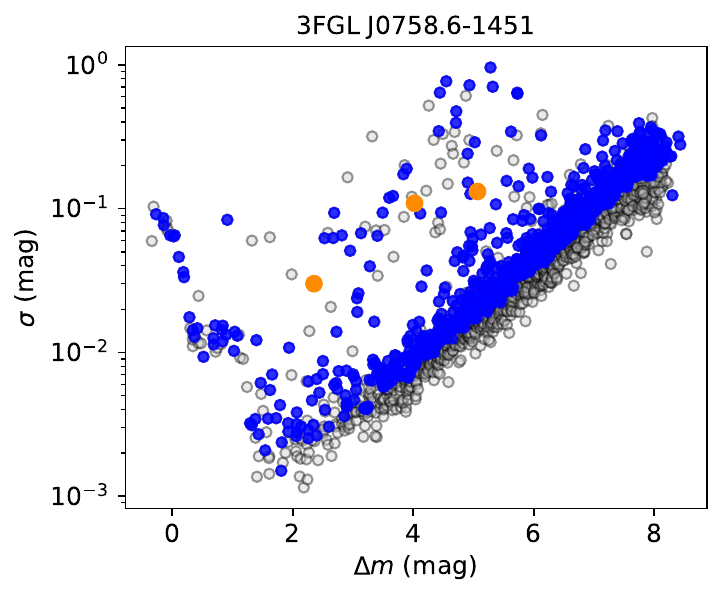}{0.48\textwidth}{}
          }
\caption{$\sigma$ vs $\Delta m$ plots for each of the 33 COBIPULSE fields of view.\label{fig:sigmavsdm}}
\end{figure*}
\begin{figure*}[ht!]
\gridline{\leftfig{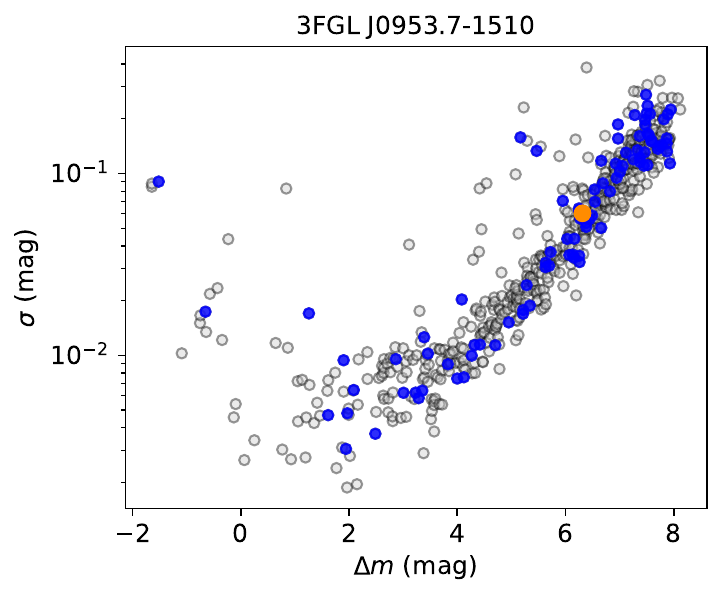}{0.48\textwidth}{}
          \leftfig{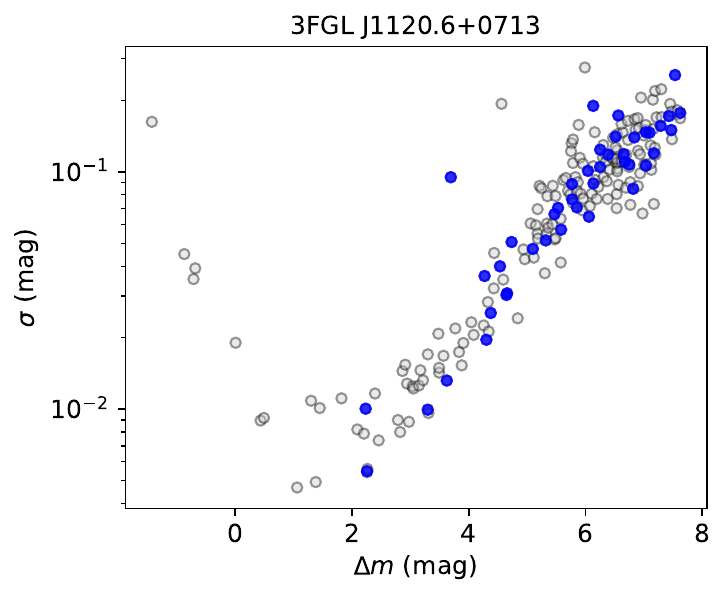}{0.48\textwidth}{}
          }
\gridline{\leftfig{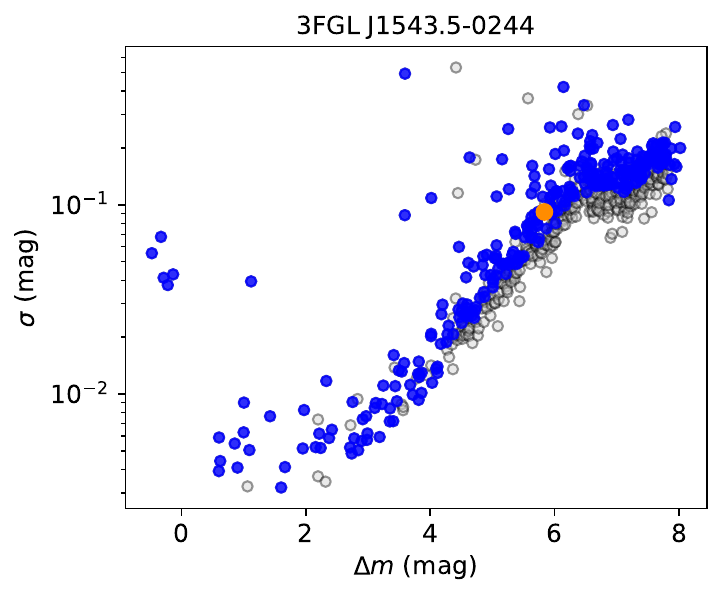}{0.48\textwidth}{}
          \leftfig{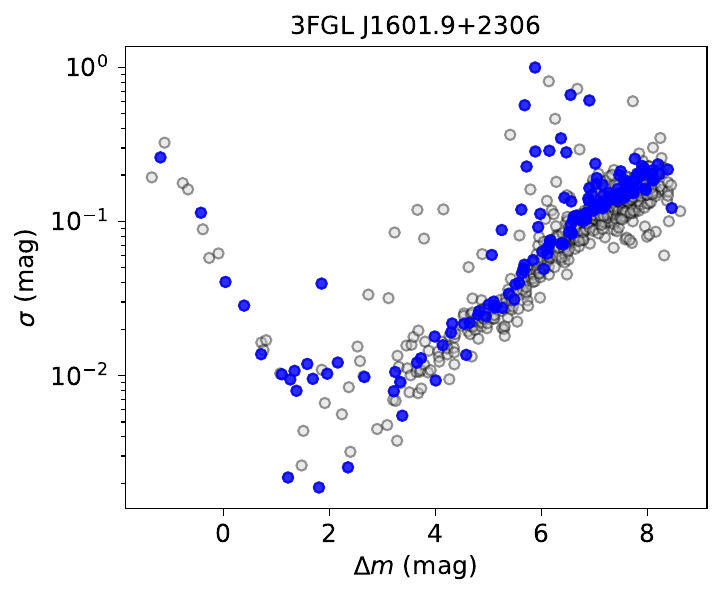}{0.48\textwidth}{}
          }
\gridline{\leftfig{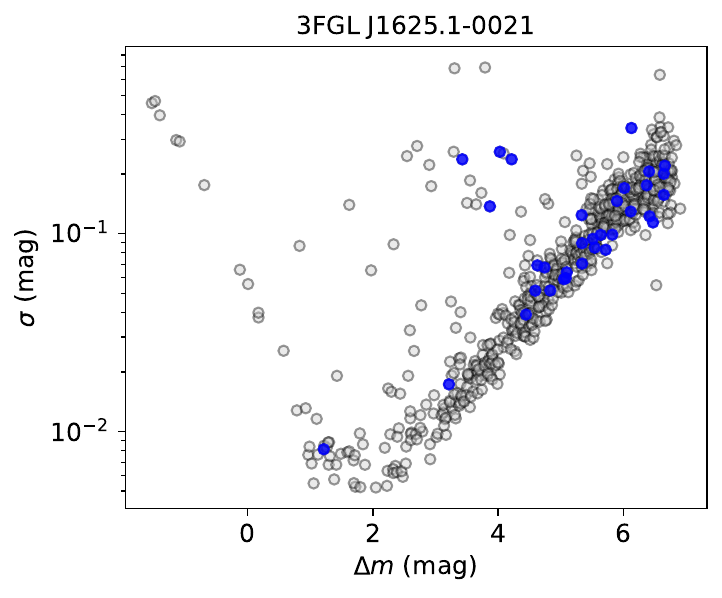}{0.48\textwidth}{}
          \leftfig{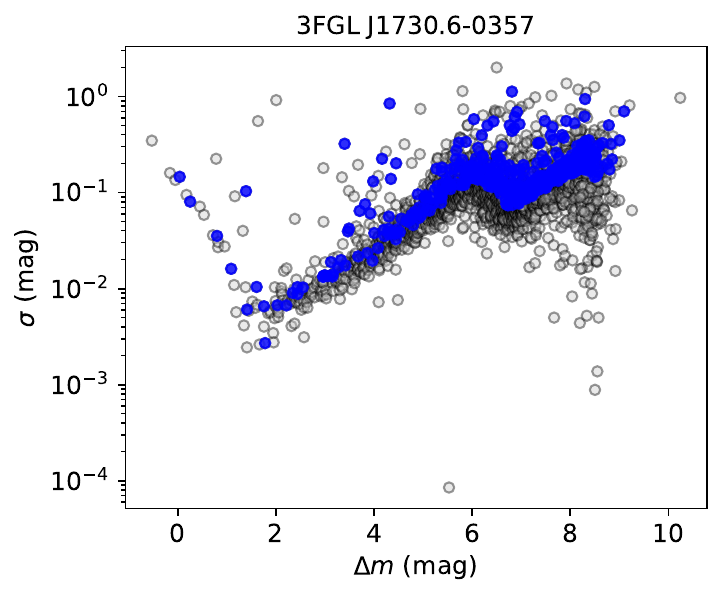}{0.48\textwidth}{}
          }
\caption{Continued.\label{fig:sigmavsdm_cont1}}
\end{figure*}
\begin{figure*}[ht!]
\gridline{\leftfig{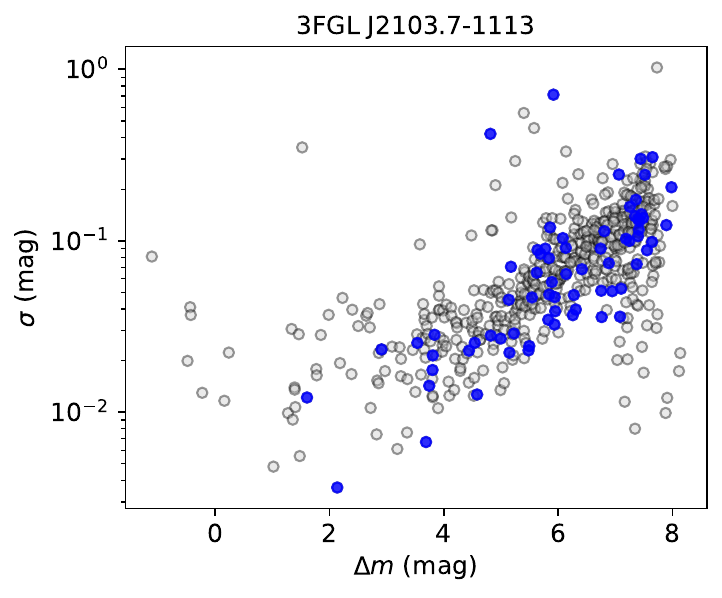}{0.48\textwidth}{}
          \leftfig{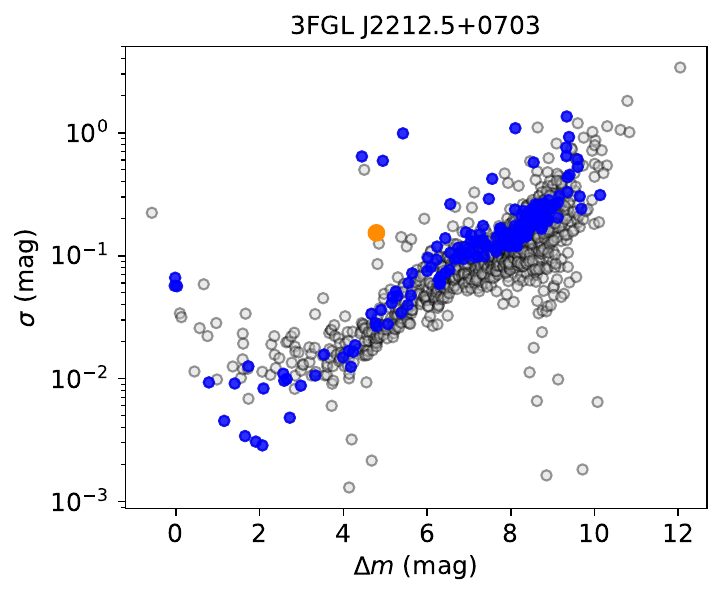}{0.48\textwidth}{}
          }
\gridline{\leftfig{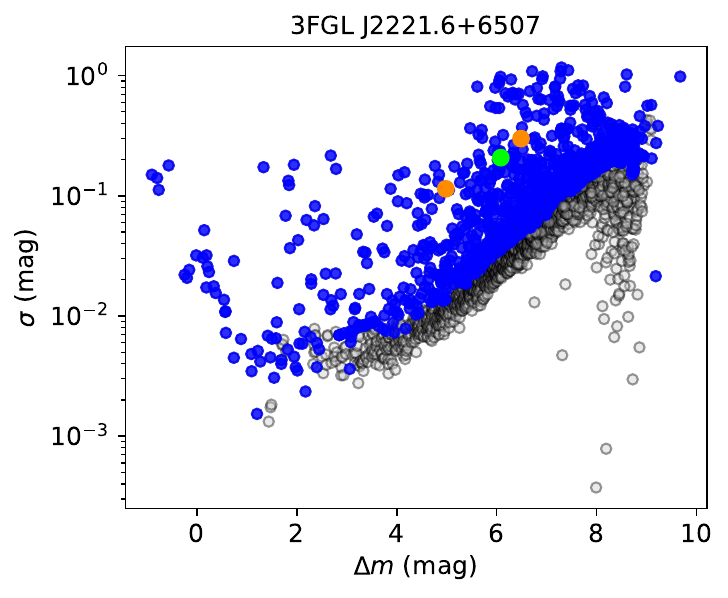}{0.48\textwidth}{}
          \leftfig{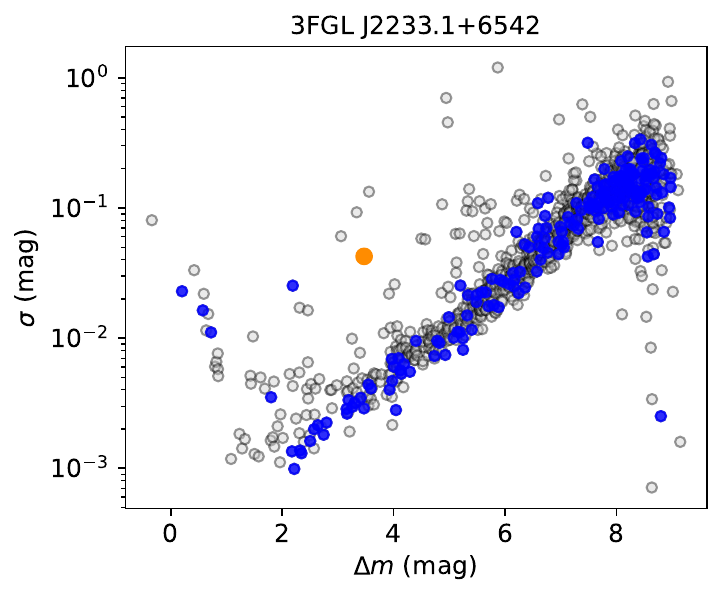}{0.48\textwidth}{}
          }
\gridline{\leftfig{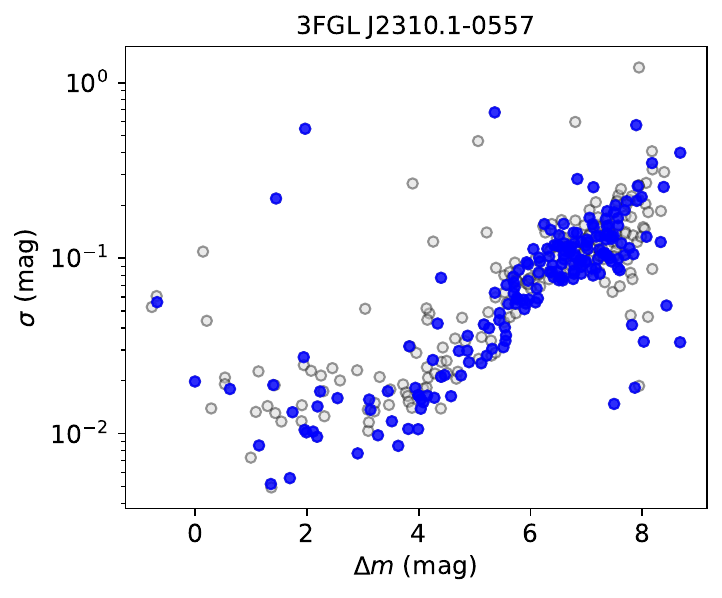}{0.48\textwidth}{}
          \leftfig{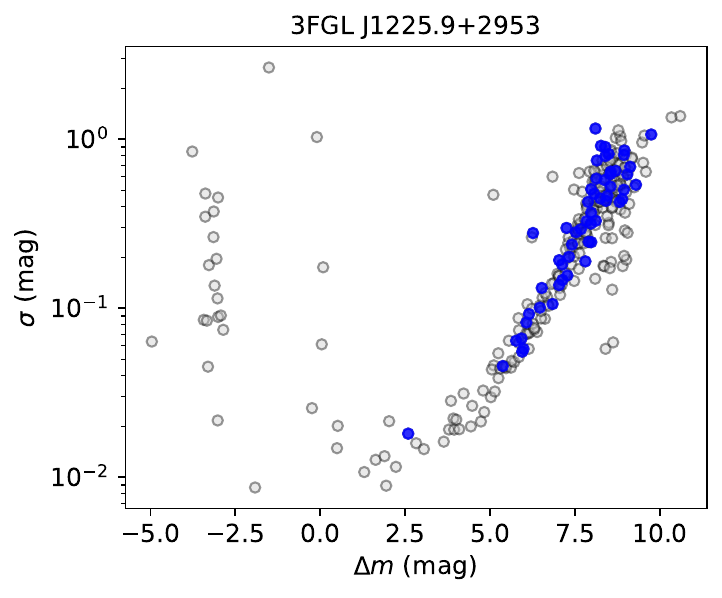}{0.48\textwidth}{}
          }
\caption{Continued.\label{fig:sigmavsdm_cont2}}
\end{figure*}
\begin{figure*}[ht!]
\gridline{\leftfig{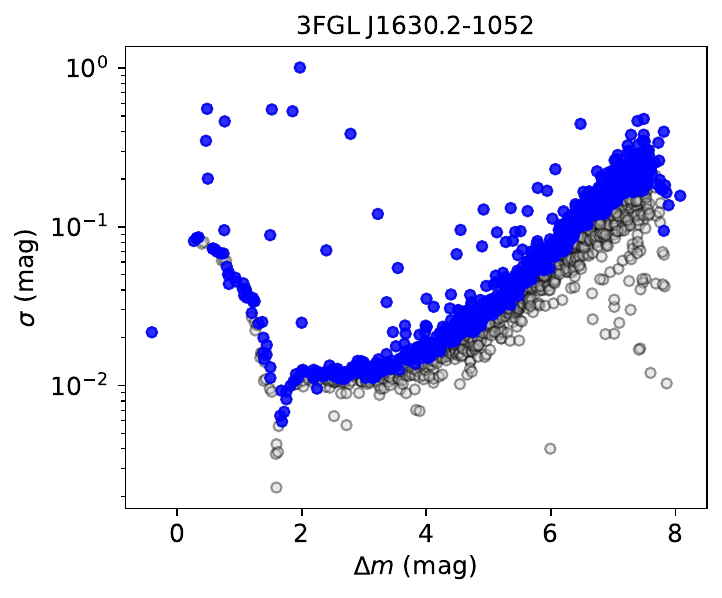}{0.48\textwidth}{}
          \leftfig{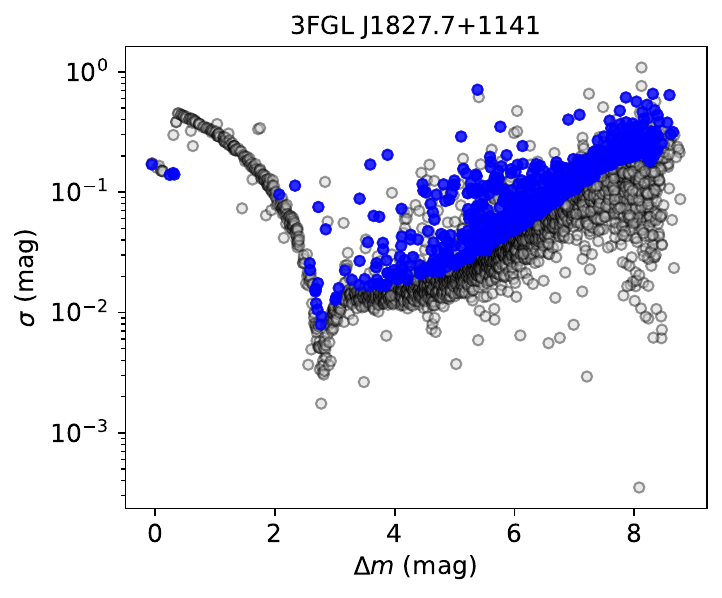}{0.48\textwidth}{}
          }
\gridline{\leftfig{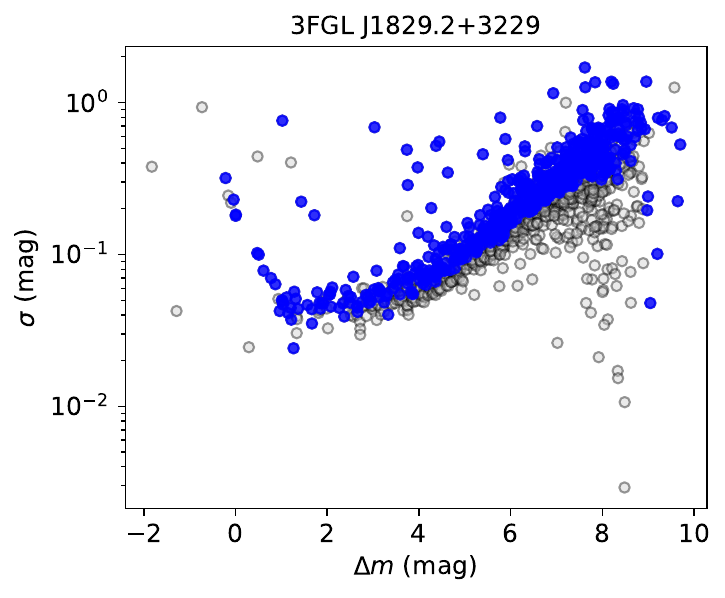}{0.48\textwidth}{}
          \leftfig{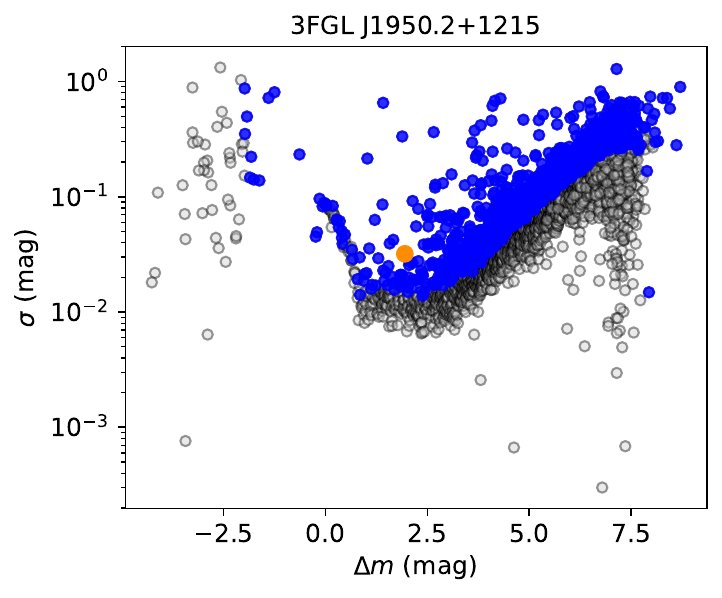}{0.48\textwidth}{}
          }
\gridline{\leftfig{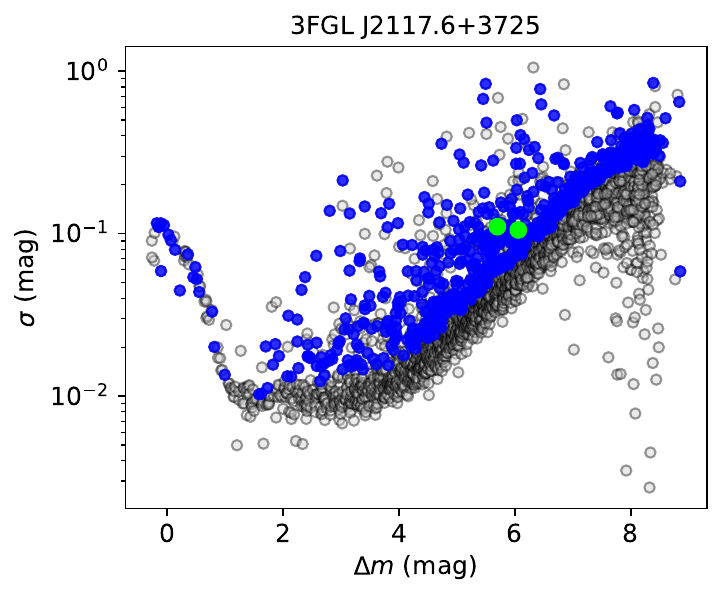}{0.48\textwidth}{}
          \leftfig{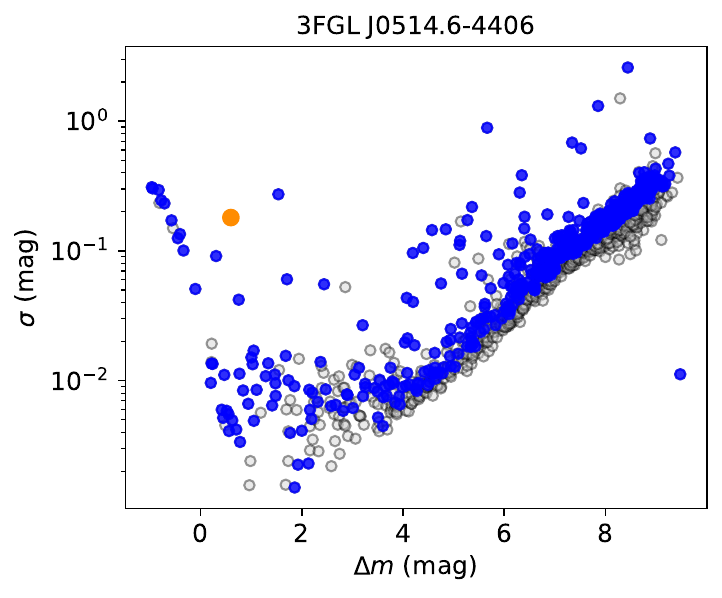}{0.48\textwidth}{}
          }
\caption{Continued.\label{fig:sigmavsdm_cont3}}
\end{figure*}
\begin{figure*}[ht!]
\gridline{\leftfig{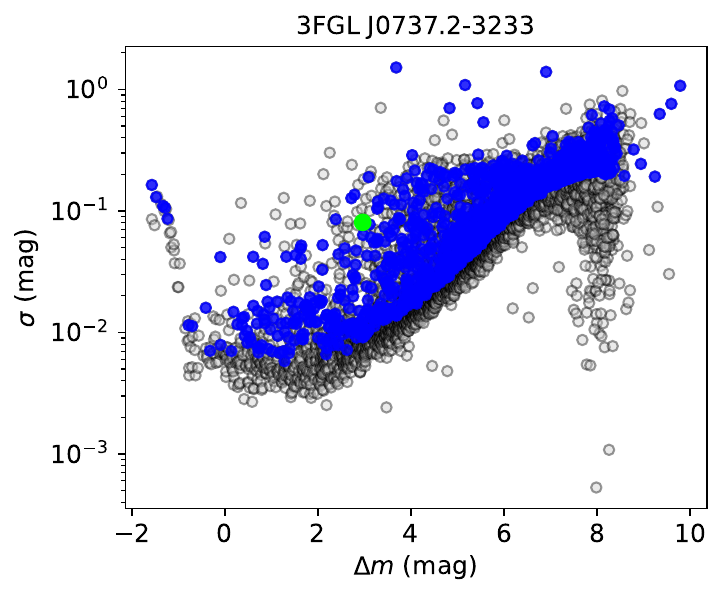}{0.48\textwidth}{}
          \leftfig{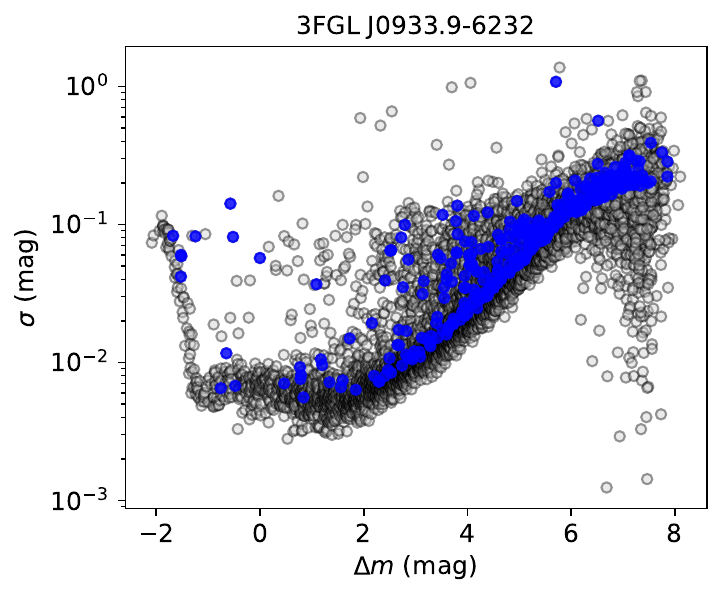}{0.48\textwidth}{}
          }
\gridline{\leftfig{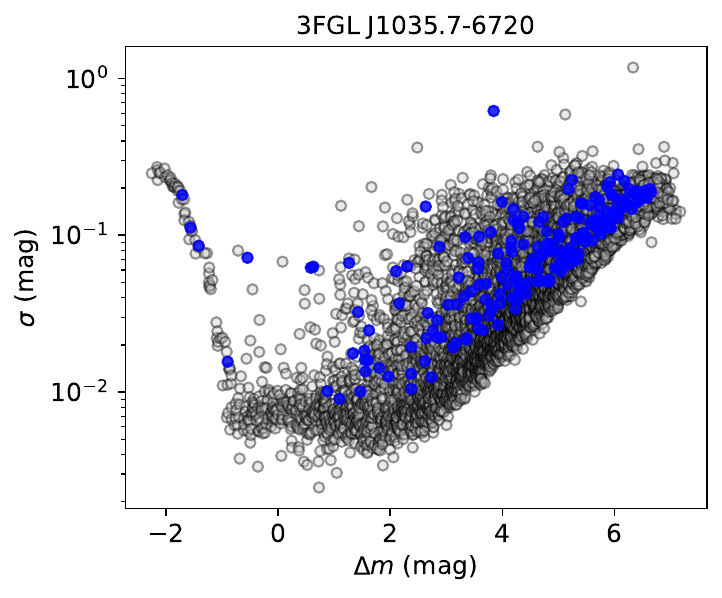}{0.48\textwidth}{}
          \leftfig{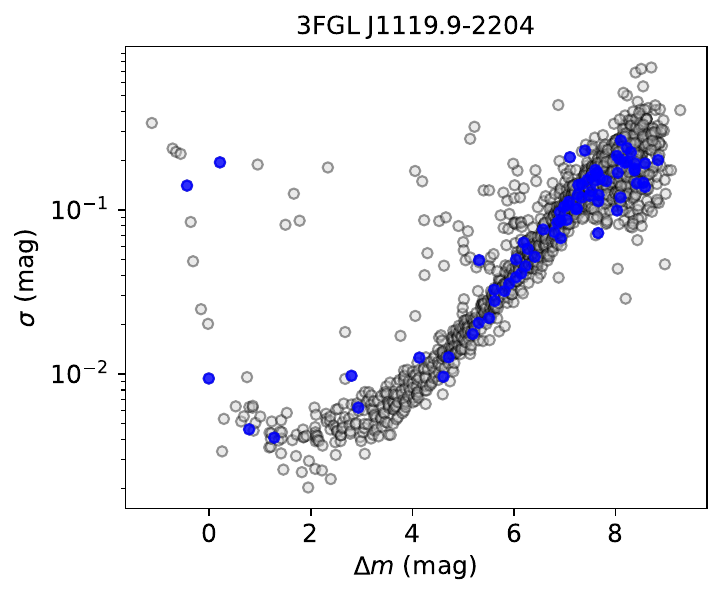}{0.48\textwidth}{}
          }
\gridline{\leftfig{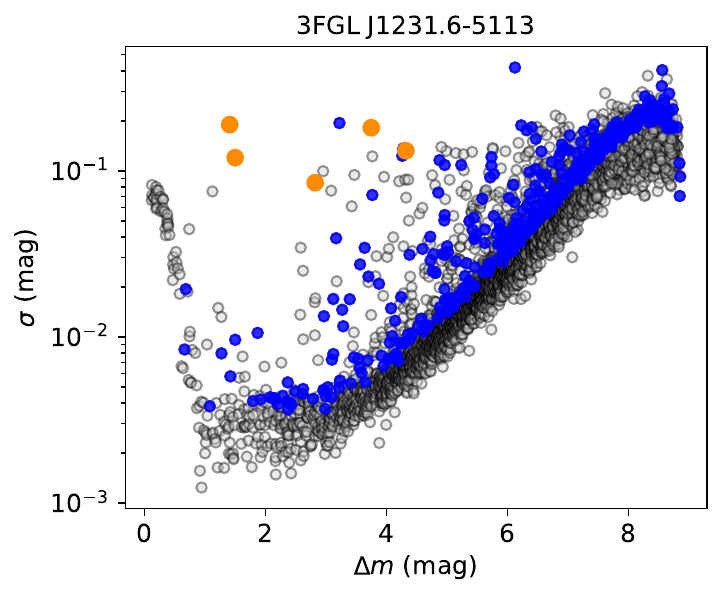}{0.48\textwidth}{}
          \leftfig{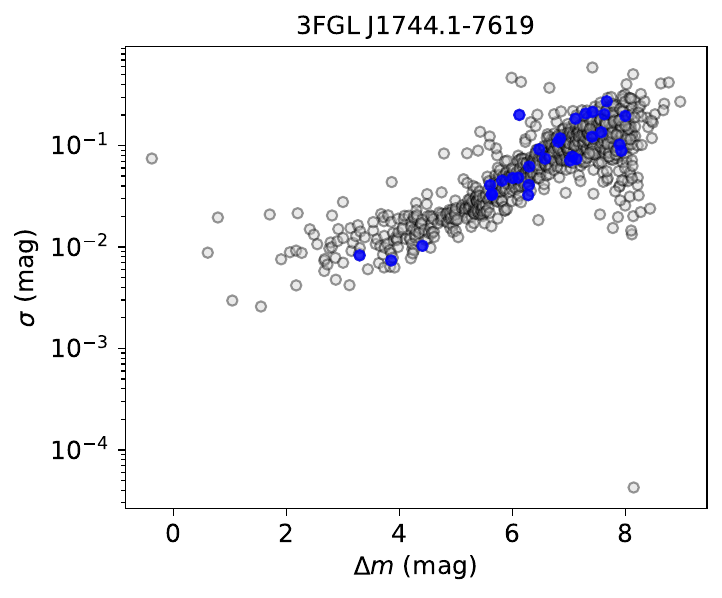}{0.48\textwidth}{}
          }
\caption{Continued.\label{fig:sigmavsdm_cont4}}
\end{figure*}
\begin{figure*}[ht!]
\gridline{\leftfig{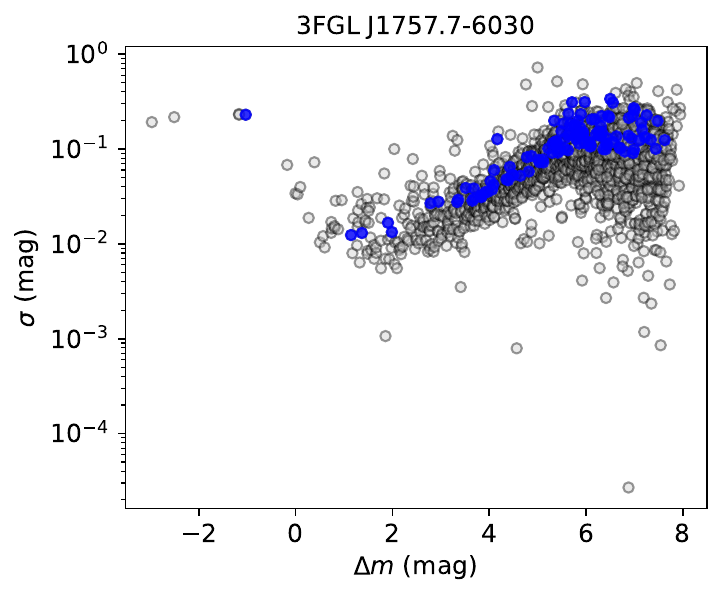}{0.48\textwidth}{}
          \leftfig{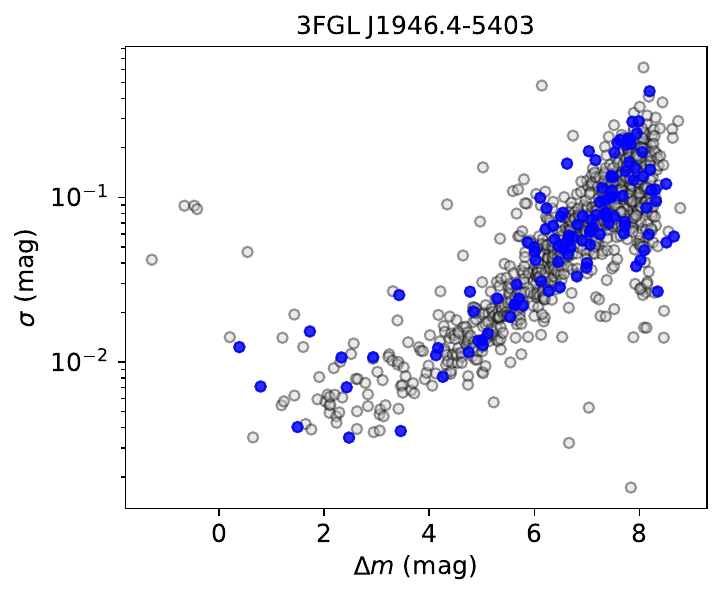}{0.48\textwidth}{}
          }
\gridline{\leftfig{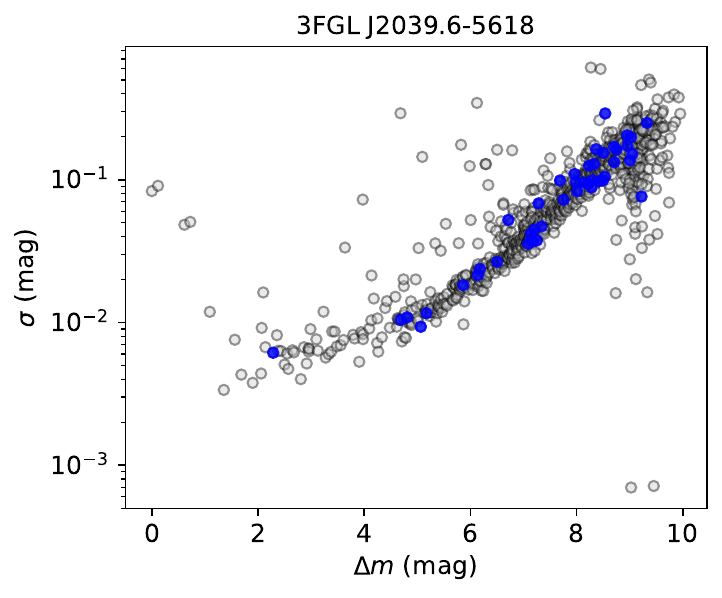}{0.48\textwidth}{}
          }
\caption{Continued.\label{fig:sigmavsdm_cont5}}
\end{figure*}

In Figures \ref{fig:sigmavsdm}-\ref{fig:sigmavsdm_cont5}, we report the light curve standard deviation ($\sigma$) vs. differential magnitude ($\Delta m$) plots for all the optical sources detected separately for each of the 33 COBIPULSE fields of view. We indicate in blue all the ``photometric variables" (see Section \ref{subsec:variablesandperiods}),
the optical periodic variables not associated to spiders in orange, and the variables classified as spider candidates in green. Most of these plots show a ``tick" shape, with a sharp minimum in $\sigma$ at $\Delta m\approx2 \ \mathrm{mag}$. This is due to the increasing optical variability observed in fainter sources ($\Delta m\gtrsim2 \ \mathrm{mag}$), which have less accurate photometry and more scattered data points, even if they are intrinsically stable. Conversely, the brightest sources ($\Delta m\lesssim2 \ \mathrm{mag}$) show spurious variability, as they are usually close to saturating the CCD detector.

\section{Periodic variables} \label{sec:appC}
\restartappendixnumbering 
\begin{table*}
\raggedright
    \caption{Optical location, best photometric period from our analysis with uncertainty reported in brackets, and classification with corresponding period measurements from other catalogs for each periodic variable.}
    \setlength{\tabcolsep}{3.0pt}
    \begin{tabular}{lcccccc}
    \hline\hline
        Name & R.A. (J2000) & Decl. (J2000) & Error radius & Photometric period & Catalog ID\tablenotemark{a} & Catalog period\\
        3FGL & (h:m:s) & ($^{\circ}$: $'$: $''$) & ($''$) & (d) & (ID, CLASS) & (d)\\ 
        \hline
    \hline
        J0238.0$+$5237-A & 02:38:12.12 & $+$52:38:54.2 & 0.8 & 0.3800(5) & \textit{ATO} J039.5505$+$52.6484, CBF & 0.380069\\
        J0758.6$-$1451-A & 07:58:14.8 & $-$14:49:51 & 1.0 & 0.9964(5) & \textit{ATO} J119.5615$-$14.8308, LPV & 0.996184\\
        J0758.6$-$1451-B & 07:58:45.0 & $-$14:46:52 & 1.0 & 0.5351(5) & \textit{ATO} J119.6874$-$14.7812, PULSE & 0.535069\\
        J0758.6$-$1451-C & 07:58:23.8 & $-$14:56:45 & 1.0 & 0.1505(5) & \textit{ATO} J119.5991$-$14.9459, CBH & 0.165609\\
        J0953.7$-$1510-A & 09:53:28.2 & $-$15:08:52 & 1.4 & 0.0957(5) & -- & --\\
        J1543.5$-$0244-A & 15:43:27.6 & $-$02:40:31 & 1.2 & 0.3176(5) & -- & --\\
        J2212.5$+$0703-A & 22:12:31.4 & $+$07:09:08 & 1.3 & 0.7640(5) & \textit{ATO} J333.1308+07.1520, dubious & 0.764068\\
        J2221.6$+$6507-A & 22:21:40.7 & $+$65:08:25 & 1.0 & 0.5162(5) & \textit{ZTF} J222140.71+650824.6, EW & 0.5161604\\
        J2221.6$+$6507-B & 22:21:40.9 & $+$65:05:43 & 1.0 & 0.6586(5) & \textit{ZTF} J222140.94+650542.6, EW & 0.4390726\\
        J2233.1$+$6542-A & 22:32:32.47 & $+$65:42:43.0 & 0.7 & 0.4056(5) & \textit{ZTF} J223232.46+654243.0, EW & 0.4056560\\
        J1950.2$+$1215-A & 19:49:45.93 & $+$12:18:06.2 & 0.9 & 1.1663(5) & \textit{ATO} J297.4413+12.3017, DBF & 1.166389\\
        J0514.6$-$4406-A & 05:14:55.29 & $-$44:11:34.1 & 0.9 & 0.5707(5) & \textit{CRTS} J051455.2-441132, Ecl & 0.570336\\
        J1231.6$-$5113-A & 12:30:00.93 & $-$51:10:26.0 & 0.7 & 0.2953(5) & \textit{Gaia} 6079101525426807808, ECL & 0.295279\\
        J1231.6$-$5113-B & 12:32:09.18 & $-$51:16:34.6 & 0.7 & 0.2292(5) & \textit{Gaia} 6079071250193978752, ECL & 0.457214\\
        J1231.6$-$5113-C & 12:31:41.77 & $-$51:16:58.5 & 0.7 & 0.1854(5) & \textit{Gaia} 6079072800684576384, ECL & 0.370825\\
        J1231.6$-$5113-D & 12:31:52.88 & $-$51:13:59.7 & 0.7 & 0.2619(5) & \textit{Gaia} 6079073213001480192, ECL & 0.578556\\
        J1231.6$-$5113-E & 12:31:09.10 & $-$51:16:57.6 & 0.7 & 0.1221(5) & \textit{Gaia} 6079049947157294592, ECL & 0.243222\\
        \hline
    \end{tabular}
    \tablenotetext{a}{ID and classification of the periodic variables as found in catalogs (refer to the main text for details). If the field is empty, the source is considered  a newly discovered optical variable.}
    \tablecomments{Classes of optical variables attributed by the \textit{ATLAS} catalog:
     \\
     \textbf{CBF}=close binary, full period identified, contact or near-contact eclipsing binary star;
     \\
     \textbf{CBH}=close binary, half period identified, contact or near-contact eclipsing binary star;
     \\
     \textbf{DBF}=distant binary, full period identified, detached eclipsing binary star;
     \\
     \textbf{LPV}=long-period variable that does not seem to fit into any of the \textit{ATLAS} more specific categories;
     \\
     \textbf{PULSE}=pulsating star showing the classic sawtooth light curve, identified as RR Lyrae, $\delta$ Scuti stars, or Cepheids;
     \textbf{dubious}=star might not be a real variable.
     \\
     Classes of optical variables attributed by the \textit{ZTF} catalog:
     \\
     \textbf{EW}=W UMa-type eclipsing binary star.
     \\
     Classes of optical variables attributed by the \textit{CSS} catalog:
     \\
     \textbf{Ecl}=eclipsing binary star.
     \\
     Classes of optical variables attributed by the \textit{Gaia}-DR3 catalog:
     \\
     \textbf{ECL}=eclipsing binary star of type beta Persei (Algol).}
    \label{tab:periodicresults}
\end{table*}

\begin{figure*}[ht!]
\gridline{\leftfig{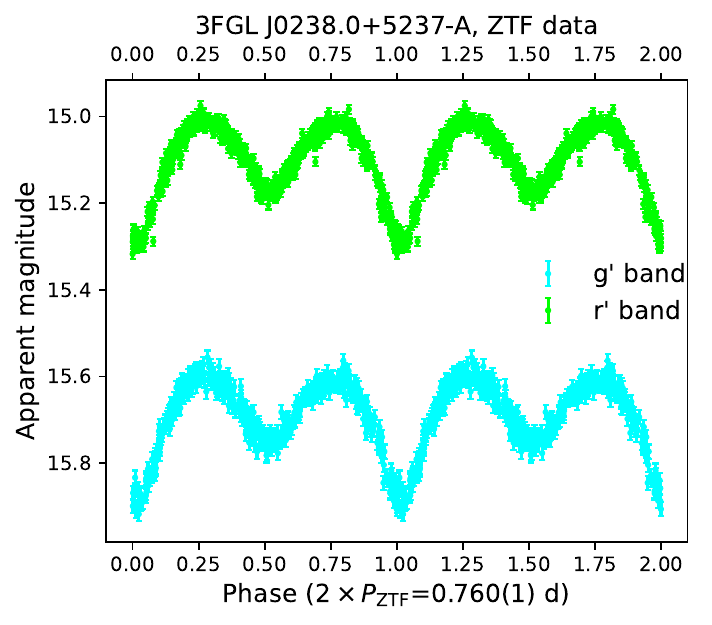}{0.46\textwidth}{}
          \leftfig{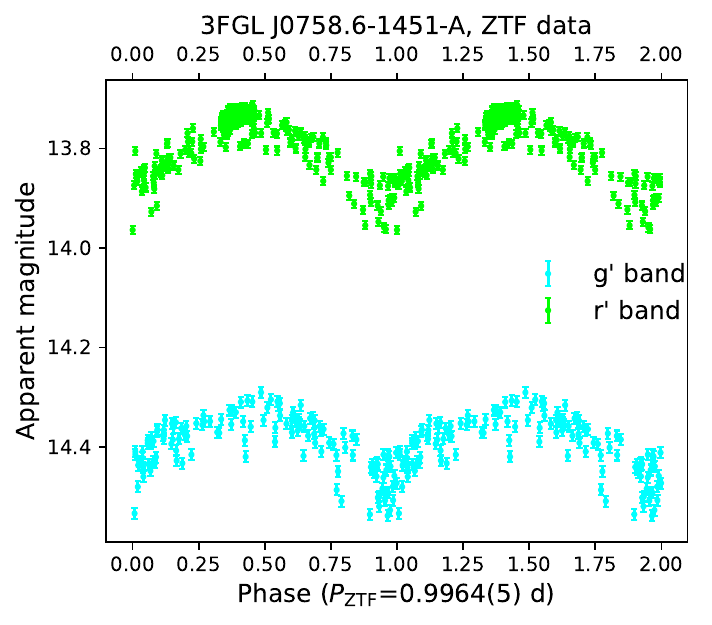}{0.46\textwidth}{}
          }
\gridline{\leftfig{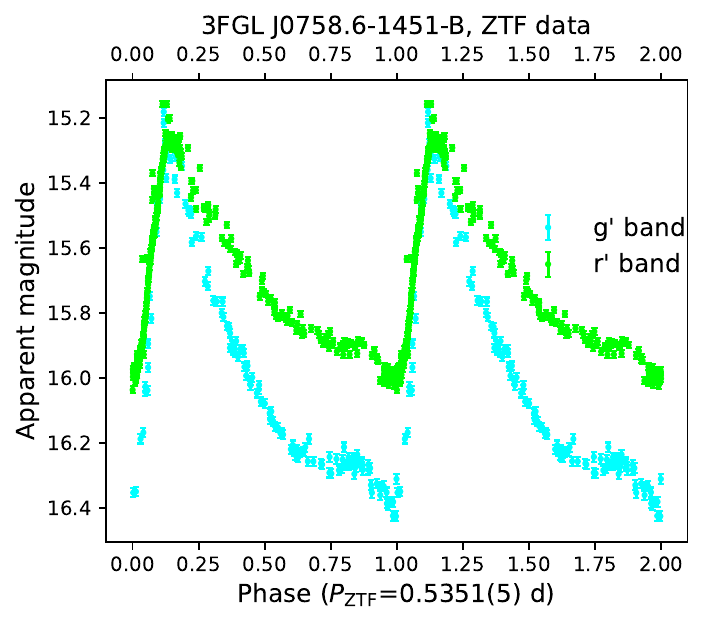}{0.46\textwidth}{}
          \leftfig{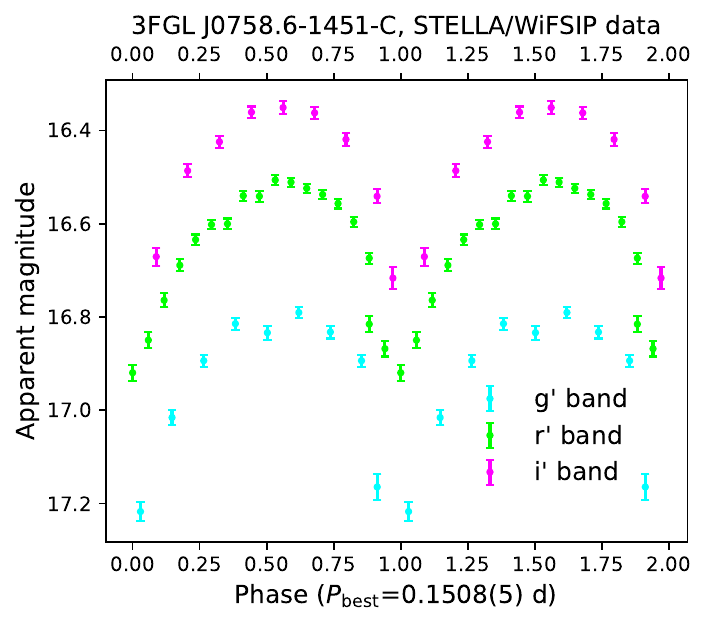}{0.46\textwidth}{}
          }
\gridline{\leftfig{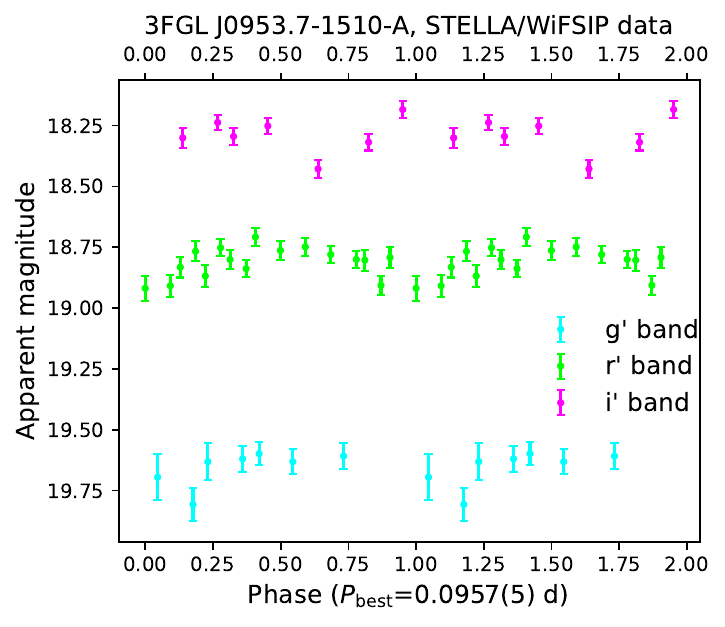}{0.46\textwidth}{}
          \leftfig{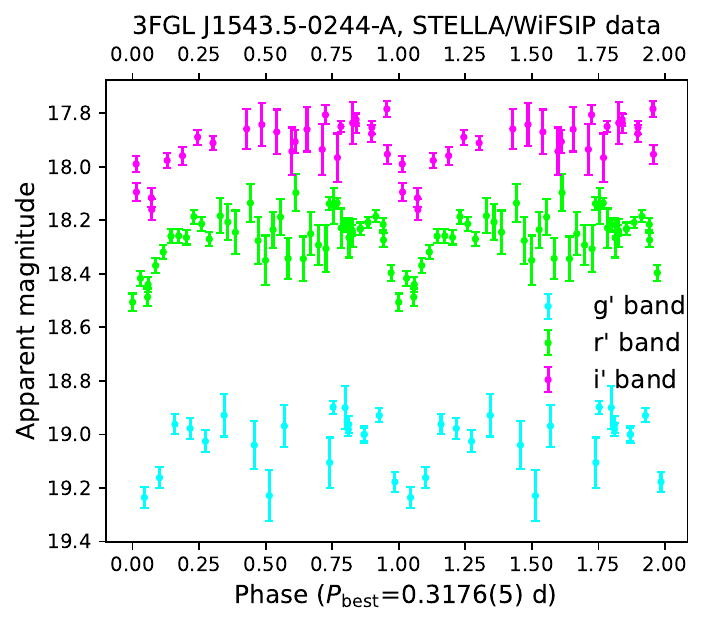}{0.46\textwidth}{}
          }
\caption{Phase-folded optical light curves for each of the 17 COBIPULSE periodic variables.\label{fig:pervarlightcurves}}
\end{figure*}
\begin{figure*}[ht!]
\gridline{\leftfig{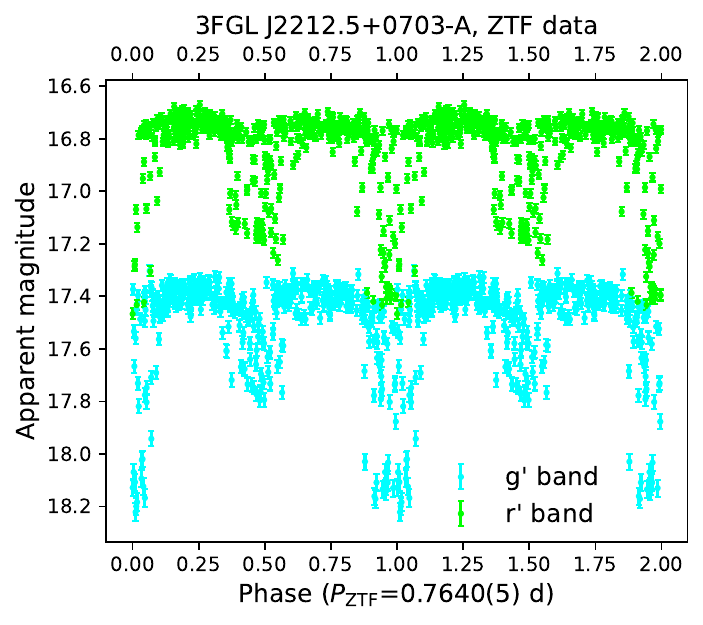}{0.46\textwidth}{}
          \leftfig{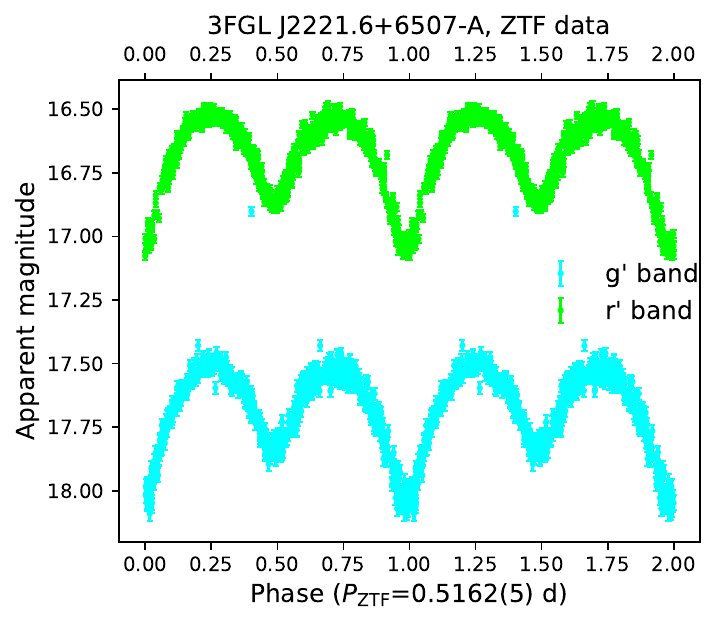}{0.46\textwidth}{}
          }
\gridline{\leftfig{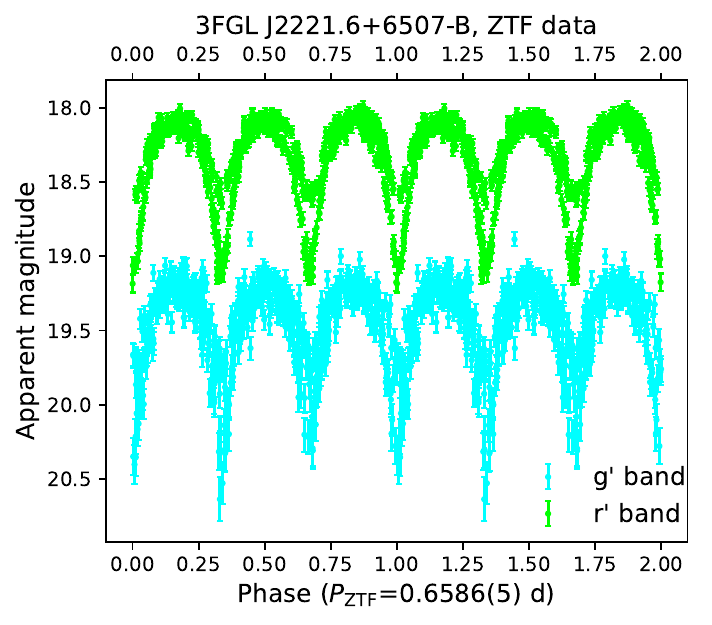}{0.46\textwidth}{}
          \leftfig{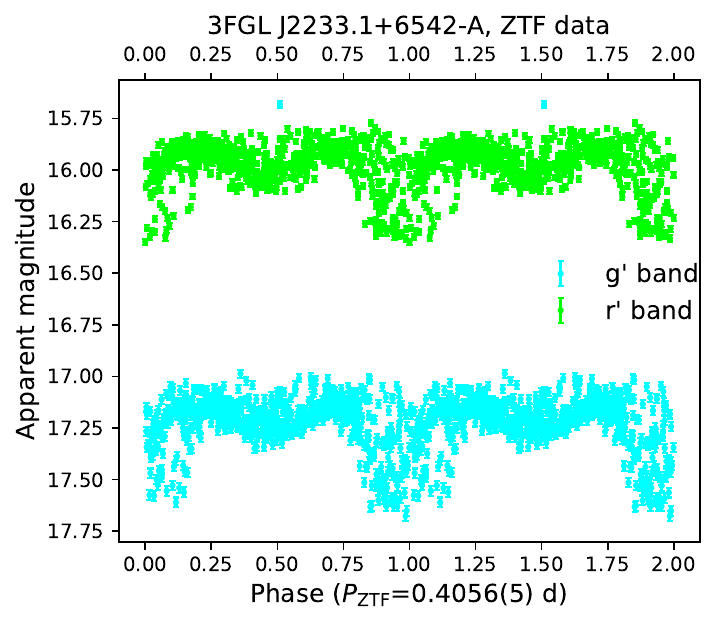}{0.46\textwidth}{}
          }
\gridline{\leftfig{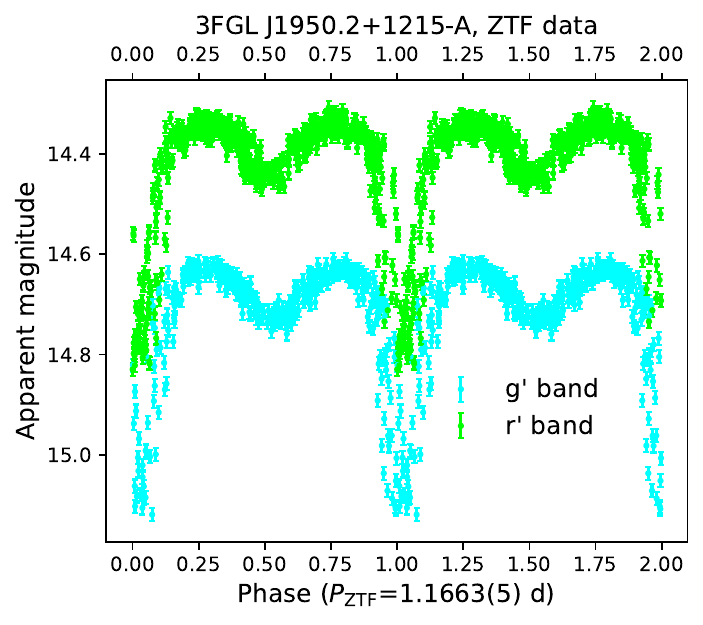}{0.46\textwidth}{}
          \leftfig{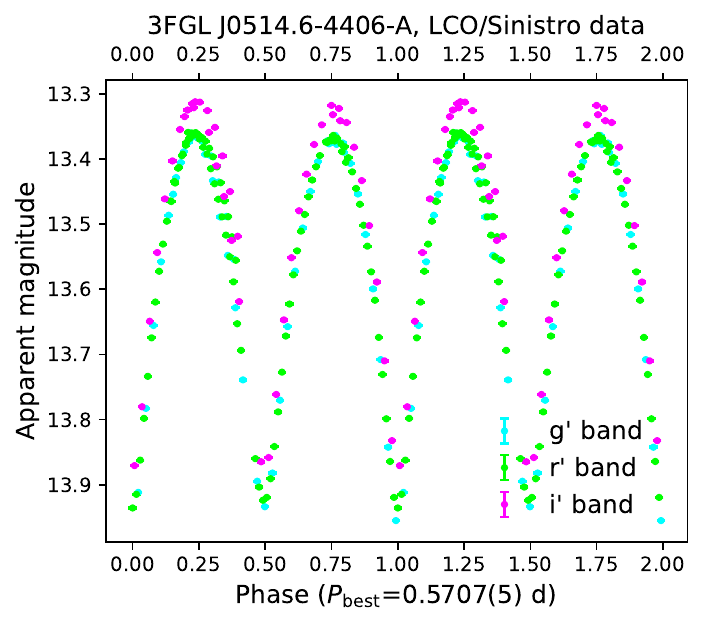}{0.46\textwidth}{}
          }
\caption{Continued.\label{fig:pervarlightcurves_cont1}}
\end{figure*}
\begin{figure*}[ht!]
\gridline{\leftfig{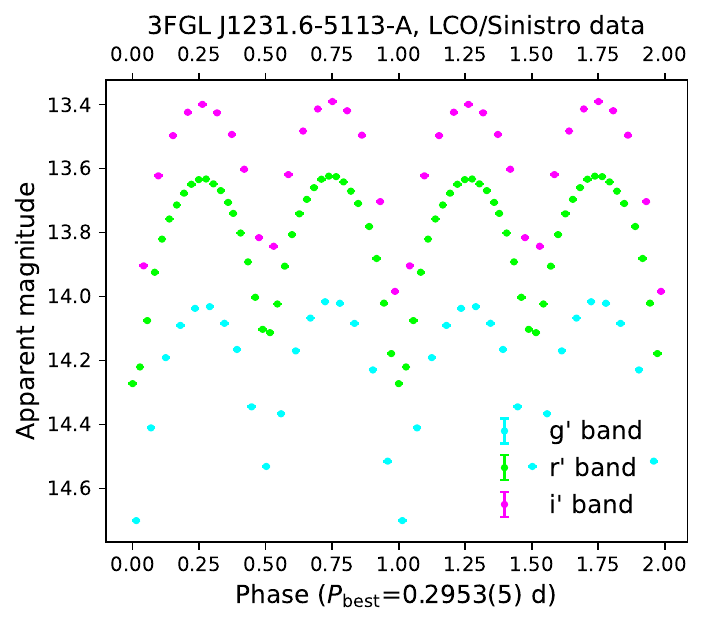}{0.46\textwidth}{}
          \leftfig{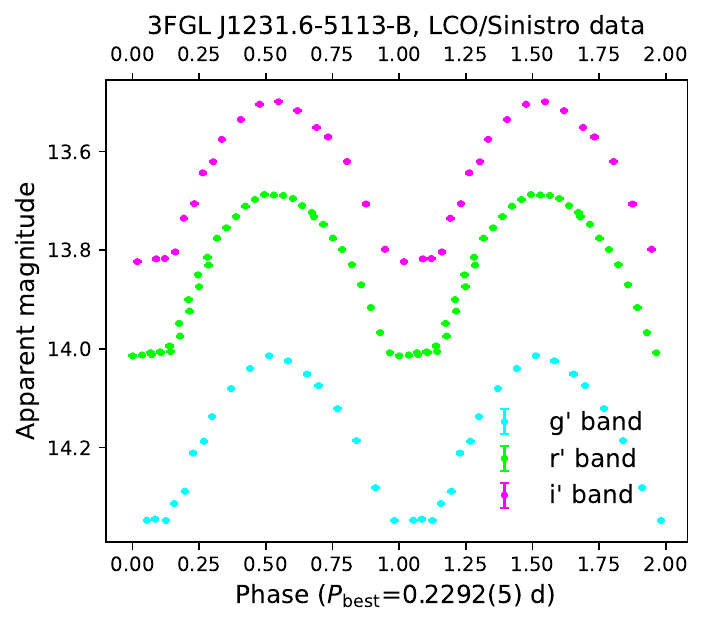}{0.46\textwidth}{}
          }
\gridline{\leftfig{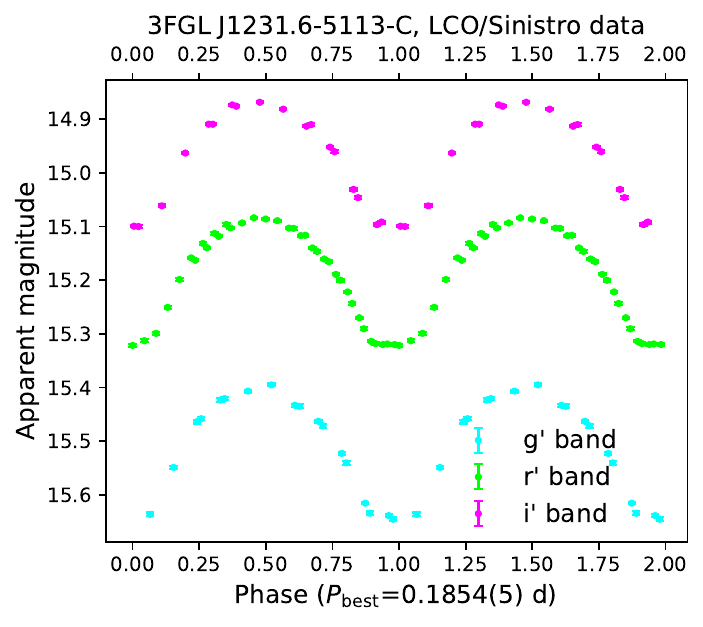}{0.46\textwidth}{}
          \leftfig{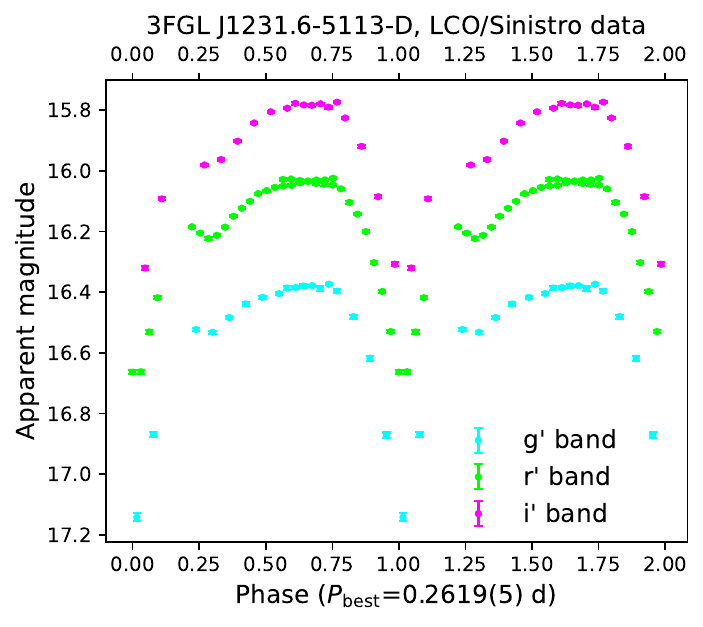}{0.46\textwidth}{}
          }
\gridline{\leftfig{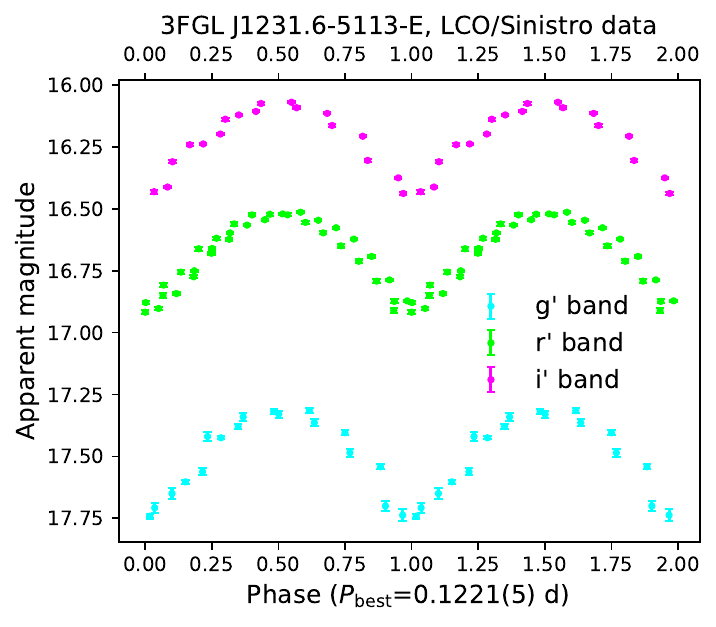}{0.46\textwidth}{}
          }
\caption{Continued.\label{fig:pervarlightcurves_cont2}}
\end{figure*}

Table \ref{tab:periodicresults} lists the optical locations and photometric periods estimated for the 17 periodic variables identified in COBIPULSE. For comparison, we also include the periods measured from other variable catalogs, if any previous identification was present from \textit{CSS} \citep{catalinasouth_2017}, \textit{ATLAS} \citep{2018AJ....156..241H}, \textit{ZTF} \citep{Chen_2020}, or \textit{Gaia}-DR3 \citep{2022gdr3.reptE..10R}. We also show in Figure \ref{fig:pervarlightcurves}-\ref{fig:pervarlightcurves_cont2} the optical light curves for each periodic variable, phase-folded with the photometric period found either from our data (STELLA, INT, or LCO) or from ZTF, in case this was needed to cover a full orbital cycle of the system.

\end{document}